\documentclass{aa}

\usepackage{amsmath}
\usepackage{hyperref}
\usepackage{xcolor}

\title{Gaps in stellar streams as a result of globular cluster fly-bys}

\subtitle{The case of Palomar~5--PREPRINT}

\author{Salvatore Ferrone
       \inst{1,2}
         \and
       Marco Montuori\inst{2}
       \and
       Paola Di Matteo\inst{1}
       \and
       Alessandra Mastrobuono-Battisti \inst{3}
       \and 
       Rodrigo Ibata \inst{4}
       \and 
       Paolo Bianchini \inst{4}
       \and
       Sergey Khoperskov \inst{5}
       \and
       Nicolas Leclerc \inst{6}
       \and 
       Clement Hottier \inst{6}
       \and 
       Eliot Stein  \inst{7}
       \and
       David Valls-Gabaud \inst{8}
       \and
       Owain N. Snaith \inst{9}
       \and
       Misha Haywood \inst{1}
       }

\institute{
    LIRA, Observatoire de Paris, Université PSL, Sorbonne Université, Université Paris Cité, CY Cergy Paris Université, CNRS,75014 Paris, France\\
  \email{salvatore.ferrone@obspm.fr}
  \and
    Dipartimento di Fisica, Universit\`a di Roma ``La Sapienza'', Piazza Aldo Moro
  \and
    Dipartimento di Fisica e Astronomia. ``Galileo Gallilei'' Università di Padova, Vicolo dell'Osservatorio 3 Padova 35122, Italy.
  \and
    Universit\'e de Strasbourg, CNRS, Observatoire astronomique de Strasbourg, UMR 7550, F-67000 Strasbourg, France
  \and
    Leibniz-Institut für Astrophysik Potsdam (AIP), An der Sternwarte 16, 14482 Potsdam, Germany
  \and
    UNIDIA, Observatoire de Paris, Université PSL, CNRS, 92190 Meudon, France
  \and
    DTIS, ONERA, Université Paris Saclay, 91123 Palaiseau, France 
  \and
    LUX, CNRS UMR 8252, Observatoire de Paris, PSL, 61 Avenue de l'Observatoire, 75014 Paris, France
  \and
    Department of Physics and Astronomy, University of Exeter, Stocker Rd, Exeter EX4 4QL, United Kingdom
  }
      
  \date{SUBMITTED TO A\&A 27 JAN 2025; Accepted XXX}

\definecolor{darkgreen}{rgb}{0.0, 0.5, 0.0}

\begin{document}

\abstract
  {Thin stellar streams, such as those resulting from the tidal disruption of globular clusters, have long been known and used as probes of the gravitational potential of our Galaxy, both its visible and dark contents. In particular, the presence of under-density regions, or gaps, along these streams is commonly interpreted as being due to the close passage of dark matter sub-halos. }
  {In this work, we investigate the perturbations induced on streams by the passage of dense stellar systems, such as globular clusters themselves, to test the possibility that they may cause the formation of gaps as well. In particular, we focus on the study of the stream of Palomar~5, a well-known globular cluster in the Galactic halo, which has particularly long tidal tails.  }
  {For this purpose, we used a particle-test code to simulate Palomar~5's tidal tails when subjected to the Galaxy's gravitational field plus its whole system of globular clusters.}
  {Our study shows that the tails of Palomar~5 can be strongly perturbed by the close passage of other clusters, in particular of NGC~2808, NGC~7078, NGC~104, and that these perturbations induce the formation of gaps in the tails.}
  {These results show that globular clusters are capable of inducing gaps in streams--as other baryonic components such as giant molecular clouds and the galactic bar have been shown to do in other works. Therefore, when searching to construct the distribution function of dark matter sub halos within the Milky Way, the gap contribution from globular clusters must be included.}

\maketitle
\nolinenumbers
\section{Introduction}

  Stellar streams are several-kpc long structures formed by the tidal disruption of globular clusters or dwarf galaxies orbiting a host galaxy. These tidal forces arise due to differential gravitational pulls across extended objects, causing stars farther from the galactic center to lag behind while those closer are pulled away. This stretching creates two tidal tails that trace the cluster's orbit, unless in the closest vicinity to the object \citep{2007ApJ...659.1212M}. Numerical predictions of this phenomenon exist since the 1970s \citep[see, for example][]{1975AJ.....80..290K}. 
  
  These predictions occurred well before the first detections of Galactic globular cluster tidal tails \citep{1995AJ....109.2553G}. Interestingly, \citet{1995AJ....109.2553G}'s detections were made nearly contemporary  with the discovery of the Sagittarius stellar stream by \citet{1994Natur.370..194I} which is the closest example of a stream emerging from a dwarf satellite currently interacting with the Milky Way. 
  
  Subsequent studies\footnote{For more subsequent observation detections of tidal debris, i.e. globular cluster stars beyond the tidal radius see the works of: \citet{1997A&A...320..776L, 2000A&A...356..127T, 2000A&A...359..907L, 2001AAS...19910906S, 2003AJ....126..815L,2011ApJ...726...47S,2018MNRAS.476.4814S,2020MNRAS.495.2222S}.} extended Grillmair's findings to other globular cluster streams, but were often limited to the detections of stars still close to the cluster tidal radius, until the discovery made by \citet{2001ApJ...548L.165O,2002AAS...200.1001O, 2003AJ....126.2385O} of long and thin tails outside the Palomar~5 globular cluster. With a mass of $1.34\pm 0.24 \times 10^4 M_{\sun}$ \citep{2019MNRAS.482.5138B}, Palomar~5 is one of the least massive globular clusters in the Galaxy. \citet{2003AJ....126.2385O} showed that its tails contain more mass than the cluster itself. The works of \citet{2006ApJ...641L..37G} and \citet{2015MNRAS.446.3297K} showed the tails to be extended for more than $20^\circ$ degrees in the sky. The discovery of its prominent tails stimulated a vigorous and successful search in the following years. New streams were discovered, mostly taking advantage of SDSS data, but also Pan-STARRS and ATLAS  \citep{2006ApJ...643L..17G, 2006ApJ...637L..29B, 2009ApJ...693.1118G, 2012ApJ...760L...6B, 2013ApJ...769L..23G, 2014ApJ...790L..10G, 2015ApJ...812L..26G, 2014MNRAS.443L..84B, 2016MNRAS.463.1759B, 2017ApJ...847..119G, 2014MNRAS.442L..85K}.

  \citet{2025NewAR.10001713B} provides a review of stellar stream astronomy, which has entered a new era since the publication of the data from the Gaia astrometric mission \citep{2016A&A...595A...1G}. Gaia's characterization of billions of stars in the Milky Way allows us to look for these structures coupling photometry, astrometry and, for the brightest stars, spectroscopy. The possibility given by Gaia to track stars with coherent movements, over the entirety of the sky, has led to the discovery of dozens of new streams. For example, \citet{2018MNRAS.477.4063M} developed the \texttt{streamfinder} algorithm and applied it across a series works \citep{2018MNRAS.481.3442M,  2018ApJ...865...85I, 2019ApJ...872..152I} to discover a multitude of streams.
  
  The possibility to couple Gaia data with spectroscopic surveys has extended the study of stellar streams beyond the quantification of their orbital properties, to a full chemical characterization. For example \citet{2019MNRAS.490.3508L} presents $S^5$, and is one example of a cross-catalog among other works \citep{2020AJ....160..181J, 2021ApJ...911..149L, 2022ApJ...928...30L, 2024MNRAS.529.2413U}. Currently, about a hundred stellar streams are known in our Galaxy. \citet{2023MNRAS.520.5225M} provides a compilation of their tracks on the sky. Interestingly, only about 20 streams are associated to known Galactic globular clusters.

  One of the interests in studying stellar streams is that they can constrain the gravitational field of their host galaxies, particularly the Milky Way. Compared to the measurement of the HI rotation curve, stellar streams offer the opportunity to investigate the potential of the host galaxy over a wide range of distances, reaching the outermost regions of the halo. For example, \citet{2011MNRAS.417..198V} demonstrated how stellar streams can infer the mass and scale parameters of dark matter halos using various amounts of observational data, from basic right ascension and declination to full six-dimensional phase space information. \citet{2018ApJ...867..101B} reviewed this concept from an information-theoretic point of view, identifying which orbits and configurations of stellar streams yield the most information about the Galactic potential. However, the use of single streams for constraining the potential led to some ambiguous and not converging results over time. For example, \citet{2010ApJ...718.1128L} made use of the Sagittarius stream to infer that the dark matter halo of our Galaxy has a triaxial shape \citep[but see also][]{2004MNRAS.351..643H, 2005ApJ...619..800J, 2005ApJ...619..807L}; \citet{2016ApJ...833...31B} concluded that the dark matter halo of our Galaxy is nearly spherical at the distances of the Palomar~5 and GD-1 streams. The two contrasting results could also show that at one distance is the halo is triaxial and at another spherical--highlighting the need for streams at different distances to map the halo shape. Recently, \citet{2024ApJ...967...89I} constructed a Milky Way model by applying a Monte Carlo Markov Chain (MCMC) fitting procedure. This method identified the set of potential parameters in an axisymmetric model of the Milky Way that best reproduces all observed stellar streams.


  Beyond the global visible and dark mass distribution streams can be used also to infer the granularity of the dark matter, that is the mass and density of the sub-halos populating our Galaxy. According to simulations by \citet{2008MNRAS.391.1685S}, $\Lambda$~Cold Dark Matter ($\Lambda$CDM) predicts that galaxies grow hierarchically, with dark matter clumps forming at a wide range of masses and sizes. These clumps, or \textit{sub-halos}, are predicted to follow a mass distribution with a power-law slope slightly shallower than -2.0. To date, the smallest observed dark matter halo was detected through gravitational lensing in an Einstein ring with a mass of $10^8$ solar masses by \citet{2012Natur.481..341V}. However, some models predict that dark matter clumps could exist down to at least the mass of Earth-like planets \citep[see][and discussion in \citet{2021arXiv211101148A}]{2005JCAP...08..003G}. 

  \citet{2002MNRAS.332..915I} first suggested that dark matter sub-halos could influence stellar streams by diffusing their orbital elements. Later, \citet{2012ApJ...748...20C} expanded this idea, proposing that sub-halos could create gaps in stellar streams during flyby encounters, where a sub-halo approaches closely enough to a segment of a stream and significantly changes the orbits of the closest stars.  \citet{2012ApJ...760...75C, 2013ApJ...768..171C} applied this idea and looked for the presence of gaps in the well-known Palomar~5 and GD-1 streams, concluding that the density variations found in their streams were consistent with expectations from $\Lambda$CDM models. \citet{2019ApJ...880...38B} provided further observational evidence for this idea, identifying a gap and a spur in the GD-1 stream that they could not explain by known objects, such as globular clusters, and which they suggested was due to the close passage of a dark matter sub-halo. Interestingly enough, as stated in \citet{2019ApJ...880...38B}, the recovered properties of this sub-halo (mass and size) were denser than those with the $\Lambda$CDM mass-size relationship presented in \cite{2017MNRAS.466.4974M}. The above-mentioned works are only a few of a much more extensive literature that has explored the impact of dark matter sub-halos in simulated streams \citep{2016ApJ...828L..10H, 2021MNRAS.507.1999H, 2021JCAP...10..043B, 2024arXiv240402953H, 2024arXiv241021174N}, or searched for their traces in observed streams \citep{2016MNRAS.460.2711T, 2017MNRAS.470...60E, 2020ApJ...889...70B, 2020ApJ...892L..37B}.

  In contrast, a more limited number of works have been interested in understanding whether other structures, e.g. from baryonic matter, can cause variations in the density of streams and gaps that can be assimilated and confused with those produced by dark matter sub-halos.  Among these works, it is worth mentioning the results of \citet{2017NatAs...1..633P} that suggests how the presence of the bar at the center of the Galaxy can perturb the characteristics of a stream such as Palomar~5 and generate gaps along its tail. That the Galactic bar could have an influence on streams morphology was also discussed by \citet{2016MNRAS.460..497H} and \citet{2016ApJ...824..104P}, in the case of the Ophiuchus stream \citep{2014MNRAS.443L..84B}. Besides the Galactic bar, also giant molecular clouds can produce gaps in stellar streams, as shown by \citet{2016MNRAS.463L..17A}. All these works thus seem to indicate that baryonic structures can play an important role in tail morphology. In this context,  an extensive numerical study, specifically focused on modeling the tails of Palomar~5, under the influence of the Galactic bar, spiral arms, giant molecular clouds and globular clusters, has been realized by \citet{2019MNRAS.484.2009B}, who concluded that both the influence of the bar and that of the giant molecular clouds can leave imprints in Palomar~5 tidal tails similar to those left by dark matter sub-halos, while the effect of globular clusters was found to be negligible. 
  
  Few studies have specifically investigated the effect of globular clusters on stellar streams. Here's an improved version with clearer phrasing and better flow:

  \citet{2017MNRAS.470...60E} concluded that the observed density variations in the tails of Palomar~5 could not be attributed to the passage of globular clusters. Their analysis focused on the characteristics of the observed gaps and involved constraining properties of the progenitor using reconstructive modeling. By trial and error, they identified a specific configuration of masses, sizes, impact parameters, times of impact, and relative velocities for two perturbers that successfully reproduced the observed density distribution. However, their method does not perform full forward modeling of the entire globular cluster system on Palomar~5's stream, as we do in this work. While they suggest that the impact rate of globular clusters is likely less significant—given their lower abundance compared to the expected population of dark matter sub-halos—-they do not explore this aspect in detail. However, they do state that it is indeed an avenue for future investigation. 
  
  More recently \citet{2022ApJ...941..129D} have examined the possibility that the gaps found in the GD-1 stellar stream could be due to the close passage of globular clusters, concluding that this scenario is highly unlikely. These first works seem to suggest that the impact of globular clusters on stellar streams is negligible. This does not necessarily need to be the general case, especially for streams of clusters as Palomar~5, which are confined in the inner 20~kpc of the Galaxy, where many other globular clusters do also orbit. For example, \citet{2023A&A...678A..69I} showed that a globular cluster can even collide with other clusters, which implies that cluster stream collisions should happen much more frequently since streams are far more extended than clusters. 
  
  In the present study we wish to fill this gap in the literature of numerically modeling cluster-stream interactions. We seek to quantify the importance of the effects of passing globular clusters in the vicinity of streams, to understand whether these systems can also be effective - and how often - in altering the distribution of stars in the tails and producing under-dense regions or gaps. To this end, on the following pages we will present the results of simulations of the streams of Palomar~5, subject to the gravitational interaction with the set of 165 Galactic globular clusters for which positions and velocities are known to date, and for which orbits can therefore be reconstructed \citep{2021MNRAS.505.5957B}. We chose to simulate Palomar~5 because it is a typical halo cluster with extended tails, and because it is a cluster for which the effect of baryonic structures on its stream has already been studied. As we will show, and in tension with previous claims, the close passage of other clusters with the stream of Palomar~5 is not a rare event, and indeed, in the 50 simulations we ran, we find the formation of numerous gaps, for an average of 1.5 gaps per simulation, generated by 18 different clusters among the whole system of Galactic globular clusters.

  \begin{figure*}
    \centering
    \includegraphics[width=\linewidth]{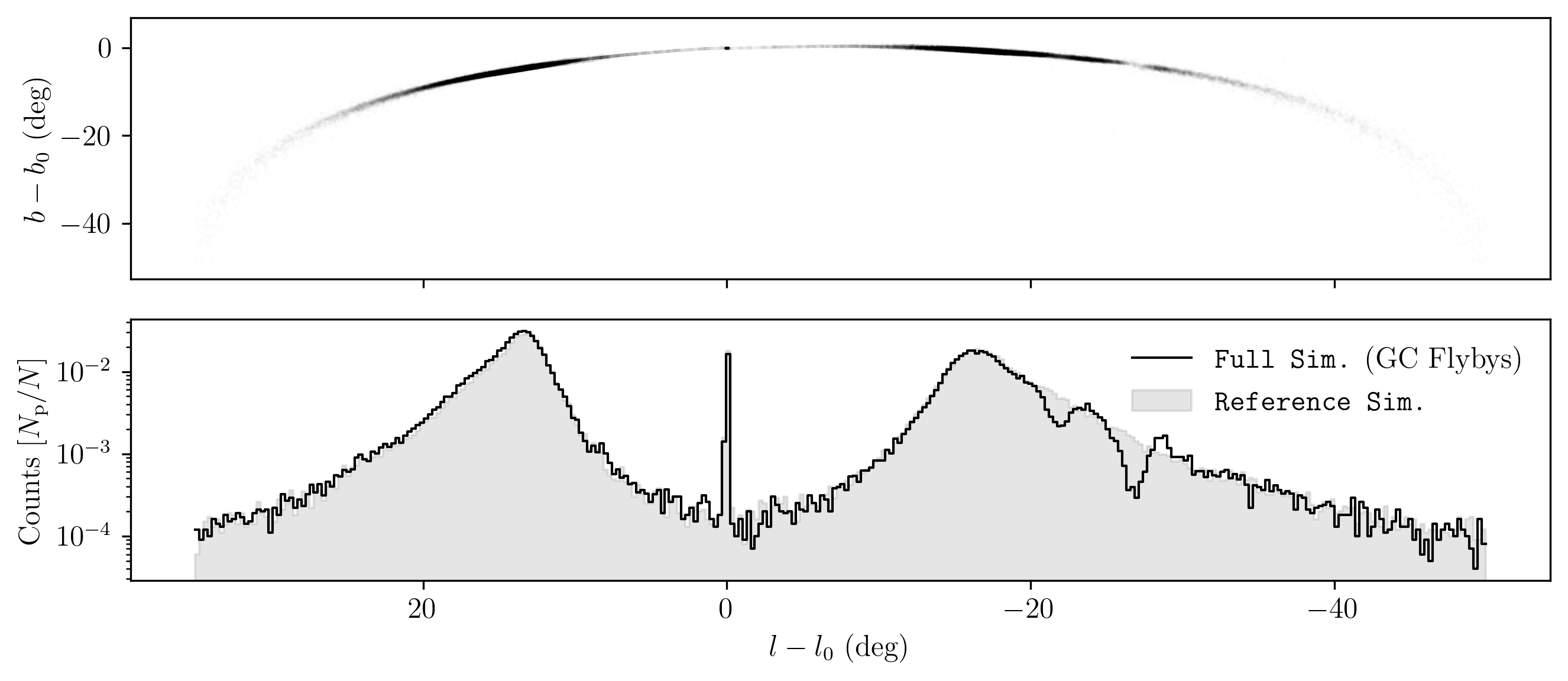}
    \caption{Simulated Palomar~5 stream created by modeling the host cluster as a Plummer sphere disrupting within an axis-symmetric Galactic potential plus the gravitational effect of 164 other galactic globular clusters. The top panel shows the distribution of star-particles that escaped the cluster due to tidal forces. The bottom panel shows the 1D density profile marginalized over longitude. The gray fill shows a reference simulation where the same conditions to produce the stream are used but the effect of other globular clusters are not included. The large central peak in density are of the particles still bound to Palomar~5. $\ell_0,b_0$ are located Palomar~5's cluster's center of mass. Two large gaps are present and are due to the passage of two globular clusters. $N_p$ indicates the number of particles in a bin while $N$ is the total number of particles which is 100,000.}
    \label{fig:stream_on_sky}
    \end{figure*}

\section{Methods}
  The most accurate model of the Palomar~5 stream would involve full modeling of the internal dynamics of the cluster. This would mean computing $N$-body interactions with a $\mathcal{O}(N^2)$ computation time, stellar evolution, supernovae, an initial mass distribution, treatment of binary star systems, etc. \citep[for such an example, see][]{2021NatAs...5..957G, 2016MNRAS.458.1450W}. Instead, we opt for solving the \textit{restricted-three body problem}, or as known as the \texttt{particle-test method} as we did for \citet{2023A&A...673A..44F}, which we describe here for completeness. As demonstrated by \citet{2012A&A...546L...7M}, although the restricted three-body problem neglects the internal evolution of the cluster, it still reproduces very similar stream properties. This is because the model captures key extra-cluster physics, such as disk shocking and epicyclic stripping.

  \subsection{Numerical Methodology}
    We begin by extracting positions in the sky, proper motions, line-of-sight velocities and distances, as well as masses and half-mass radii, of 165 globular clusters from the Galactic globular cluster catalog by \cite{2021MNRAS.505.5957B}.\footnote{The Baumgardt catalog has been assembled across a series of works, see: \cite{2020PASA...37...46B,2019MNRAS.482.5138B,2018MNRAS.478.1520B}. The catalog can be found on the World Wide Web at \href{https://people.smp.uq.edu.au/HolgerBaumgardt/globular/}{https://people.smp.uq.edu.au/HolgerBaumgardt/globular/}.} We then convert the initial conditions from sky coordinates into a Galactocentric reference frame, by adopting a velocity for the local standard of rest of $v_{\text{LSR}} = 240$~km~s$^{-1}$ and a peculiar velocity of the Sun equal to $(U_\odot, V_\odot, W_\odot)=(11.1, 12.24, 7.25)$~km~s$^{-1}$, as reported by \citet{2012MNRAS.427..274S}.  The position of the Sun was set to $(x_\odot,y_\odot,z_\odot) = (-8.34,0,0.027)$~kpc, vertical position above the disk was taken from \citet{2001ApJ...553..184C} and the distance of the Sun to the Galactic center was taken from \citet{2014ApJ...783..130R}. These transformations were performed using \texttt{astropy} \citep{2013A&A...558A..33A}.

    For the Galactic potential, we used the second model from \citet{2017A&A...598A..66P}, a superposition of a thin disk, thick disk, and dark matter halo, with masses and scale lengths provided in Table~1 of \citet{2023A&A...673A..44F}. This model is time-independent throughout our simulations. This model satisfies a series of observational constraints such as local solar stellar density, the Galactic rotation curve, etc., similarly to other Galactic models such as \texttt{MWpotential2014} from \citet{2015ApJS..216...29B} and \texttt{McMillian2017} from \citet{2017MNRAS.465...76M}. However, to balance data volume and computation time, we use only one Galactic potential, which should suffice. \citet{2021MNRAS.505.5978V} found that only a few outer globular clusters are strongly affected by different potential models. Generally, kinematic uncertainties are the dominant factor in differences between orbital solutions per cluster. Similarly, \citet{2024MNRAS.528.5189G} generated a globular cluster mass-loss catalog using seven different potential models and found their debris distributions to be rather model-independent, similarly to \citet{2023A&A...673A..44F}. While the clusters' exact positions in time may depend on the model, we assert that interaction rates and stream formation are largely independent the choice of Galactic potential model.
    
    Lastly, we select an integration time of 5~Gyr as a compromise between maximizing interaction statistics and adhering to the assumption that the galaxy can be modeled as a time-independent, constant mass distribution. \citet{2023A&A...673A.152I} analyzed the orbits of the Galactic cluster population using the same initial conditions as in this work, within five live Milky Way-like potentials from IllustrisTNG \citep{2018MNRAS.473.4077P}. They found that in all sampled potentials, orbital changes remain minimal over 5~Gyr, becoming significant only at earlier look-back times when the host galaxy had either significantly less mass or was undergoing a merger event.

    \subsubsection*{ \texttt{Full} simulations}
    There is a primary methodological departure from \citet{2023A&A...673A..44F}.  In that work, globular clusters evolved under the gravitational effect of the Galaxy only, while now we include also the effect of all the other Galactic globular clusters, by taking into account the direct $N$-body interactions among them. To quantify these interactions, all clusters are represented by Plummer spheres, each with mass and half-mass radii as reported in the Baumgardt catalog \citep{2021MNRAS.505.5957B}. For the remainder of this paper, we refer to these simulations, where the globular clusters interactions are taken into account as \texttt{full} simulations. \\
    For these simulations, we proceed into two steps:
      \begin{enumerate}
          \item First, starting from the Galactocentric positions and velocities of all 165 Galactic globular clusters, we integrate their orbits back in time for 5~Gyr under the influence of the Galaxy itself and their mutual influence. In the backward integration, the system of equations of motion for the globular clusters are thus: 
          \begin{equation}
          \ddot{\vec{r}}_i = -\nabla \Phi(R_i,z_i) + \left.\sum_{j\neq i}^{N_{GC}} \frac{Gm_j}{\left(|\vec{r}_j - \vec{r}_i|^2 + b_j^2\right)^{3/2}}\right. \left(\vec{r}_j - \vec{r}_i\right),
          \end{equation}\label{eq:GCNBody} 
          \noindent where $\vec{r}$ indicates the galactocentric position vector, the index $i$ indicates the globular cluster of interest with $R_i,z_i$ being its galactocentric positions in cylindrical coordinates; the index $j$ indicates the other globular clusters that are summed over. $N_{GC}$ is the total number of globular clusters which in this study is 165, $m_j$ is the mass of the j-th cluster in the sample, $b_j$ is its Plummer scale radius, and $\vec{r_j}$ is its galactocentric position. $\Phi$ represents the same Galactic smooth potential that we discussed previously\citep[][Model~II, in the present case]{2017A&A...598A..66P}. Note that the masses and sizes of the globular clusters are kept constant in these simulations and not allowed to vary with time. This equally implies that we do not take into account the internal evolution of the globular clusters in these models. We discuss the implications of the modeling limitations in the Discussion.
      
        \item Once the positions and velocities of the entire globular cluster system 5~Gyr ago are found, we sample Palomar~5 with 100,000 particles from a Plummer distribution taking the mass and half-mass radius from the Baumgardt catalog. We then integrate the evolution of these particles forward in time to the present day, by taking into account that each particle feels the gravitational potential of the Galaxy, that of its host cluster, and that of all the other clusters in the Galaxy. Note that we do not take into account self-gravity among particles. The particles experience the gravitational field yet do not contribute to it, which is a common assumption made in Galactic dynamics since the mass of an individual star is negligible compared to the mass of the larger dynamical system. The equation of motion of a generic particle among the 100,000 which populate Palomar~5 is thus: 
        \begin{equation}
          \ddot{\vec{r}}_p = -\nabla \Phi(R_p,z_p) + \left.\sum_{j}^{N_{GC}} \frac{Gm_j}{\left(|\vec{r}_j(t) - \vec{r}_p|^2 + b_j^2\right)^{3/2}}\right. \left(\vec{r}_j(t)- \vec{r}_p\right),
          \end{equation} \label{eq:equation_of_motion_particle} where the index $p$ represents one of the 100,000 particles of interest, $\vec{r_p}$ being its position, and $j$ indexes over the globular clusters as in Eq.~\ref{eq:GCNBody}. Note that in Eq.~\ref{eq:equation_of_motion_particle} the positions of the globular clusters are time-dependent since they are being loaded during this step and not computed, unlike Eq.~\ref{eq:GCNBody}. 

      \end{enumerate}

        The procedure described so far has been repeated 50 times, each time generating a new set of initial conditions, given the uncertainties on proper motions, line-of-sight velocities, distances to the Sun, and masses, of all clusters, as reported in the Baumgardt catalog. We handle these uncertainties through a Monte-Carlo approach, by sampling them with a Gaussian distribution, and by taking into account the covariance term between the proper motions. For the first simulation, we use the most probable values for the initial conditions. The uncertainties are sampled for all globular clusters. Additionally, for each re-sampling of Palomar~5's mass, we also resample the distribution of the 100,000 star-particles.

      During the integration, intermediate snapshots were saved to facilitate the analysis of stellar streams and the effects of cluster impacts. Specifically, for each of the 50 realizations of the Palomar~5 stream, we saved $5000$ intermediate time-steps which is a temporal resolution of 1 million years. We provide the parameters that specify our data volume in Table~\ref{tab:data_volume}. Using single precision floating point numbers, the size of our simulations is approximately:
    \begin{equation} \label{eq:data_volume_estimate}
      N_p \times N_{\textrm{ts}}\times N_{\textrm{phase}}\times N_{\textrm{sampling}} \times 4~\textrm{bytes}\approx 600~\textrm{Gb}.
    \end{equation}

    \begin{table}[h]
      \centering
      \begin{tabular}{|c|c|c|c|}
          \hline
          $N_p$ & $N_{\textrm{ts}}$ & $N_{\textrm{phase}}$ & $N_{\textrm{sampling}}$ \\
          \hline
          $100000$ & $5000$ & $6$ & $50$ \\
          \hline
      \end{tabular}
      \caption{The parameters determining the data volume of this experiment. where $N_p$  is the number of particles, $N_{\textrm{ts}}$ is the number of time-steps saved, $N_{\textrm{phase}}$ is the number of phase space coordinates, and $N_{\textrm{sampling}}$ is the number of Monte-Carlo samplings of the initial conditions.}
      \label{tab:data_volume}
    \end{table}

  \subsubsection*{ \texttt{Reference} simulations}

    To quantify the impact of globular cluster passages to the density of Palomar~5 stream, we performed a second set of simulations, which we refer to as the \texttt{reference} simulations in this paper.  These \texttt{reference} simulations use the same 50 sets of initial conditions as the \texttt{full} simulations, the same Galactic potential but exclude mutual interactions between globular clusters. The approach adopted for this second set of simulations is thus equivalent to that adopted already in \citet{2023A&A...673A..44F}. In Eq.~\ref{eq:GCNBody} only the gradient of the Galactic potential  is considered. In Eq.~\ref{eq:equation_of_motion_particle}, of the second term on the right-side of the equation, only the influence of Palomar~5's Plummer sphere on Palomar~5's particles is considered. In other words, sum iterates over only one globular cluster, the host. All the interactions with the other clusters in the Galaxy are indeed omitted.

  \subsubsection*{Numerical Stability}

    The integration was conducted using a leapfrog integrator, chosen for its ability to preserve phase-space volume and conserve the Hamiltonian with each integration step. This method is preferable to, for instance, a Runge-Kutta scheme, which can introduce non-physical and significant numerical errors in systems that require long-term stability and energy conservation. One drawback of the leapfrog integrator is that it requires a uniform time-step across the entire computation, leading to unnecessary computations for a particle after it escapes form the host cluster. However, energy conservation and phase-space volume preservation are paramount when modeling stellar streams. The time step was therefore set to be small enough to conserve energy for the most interior particles within the cluster---ensuring that a higher mass loss did not arise from numerical error. For these simulations, a time-step of 10,000 years was selected to maintain adequate energy conservation, with a median variation of $10^{-12} \frac{\Delta E}{E_0}$, where $E_0$ is a particle's initial energy, and $\Delta E$ is the difference between its final and initial energy. 
    
    We also checked the \textit{reverse integrability} of the globular cluster system for the \texttt{reference} simulations. By reverse integrability, we mean the integrator's capability to track the cluster backward in time and then re-integrate it forward along the same trajectory. Integrating point masses in a static axis-symmetric potential conserves $L_z$ and $E$, which create regular periodic and non-chaotic orbits. Therefore, any drift would arise from purely numerical error. We selected a time stamp for which the drift in the final position after forward integration, compared to the initial position from the backward integration, was consistently at least two orders of magnitude smaller than the Plummer scale radius used for Palomar~5. This high precision ensures that no fictitious numerical forces influenced the system, preventing any artificial mass loss or retention of star particles.

\section{Results}

  \subsection{Overview}

    \begin{figure*}
      \centering
      \includegraphics[width=\linewidth]{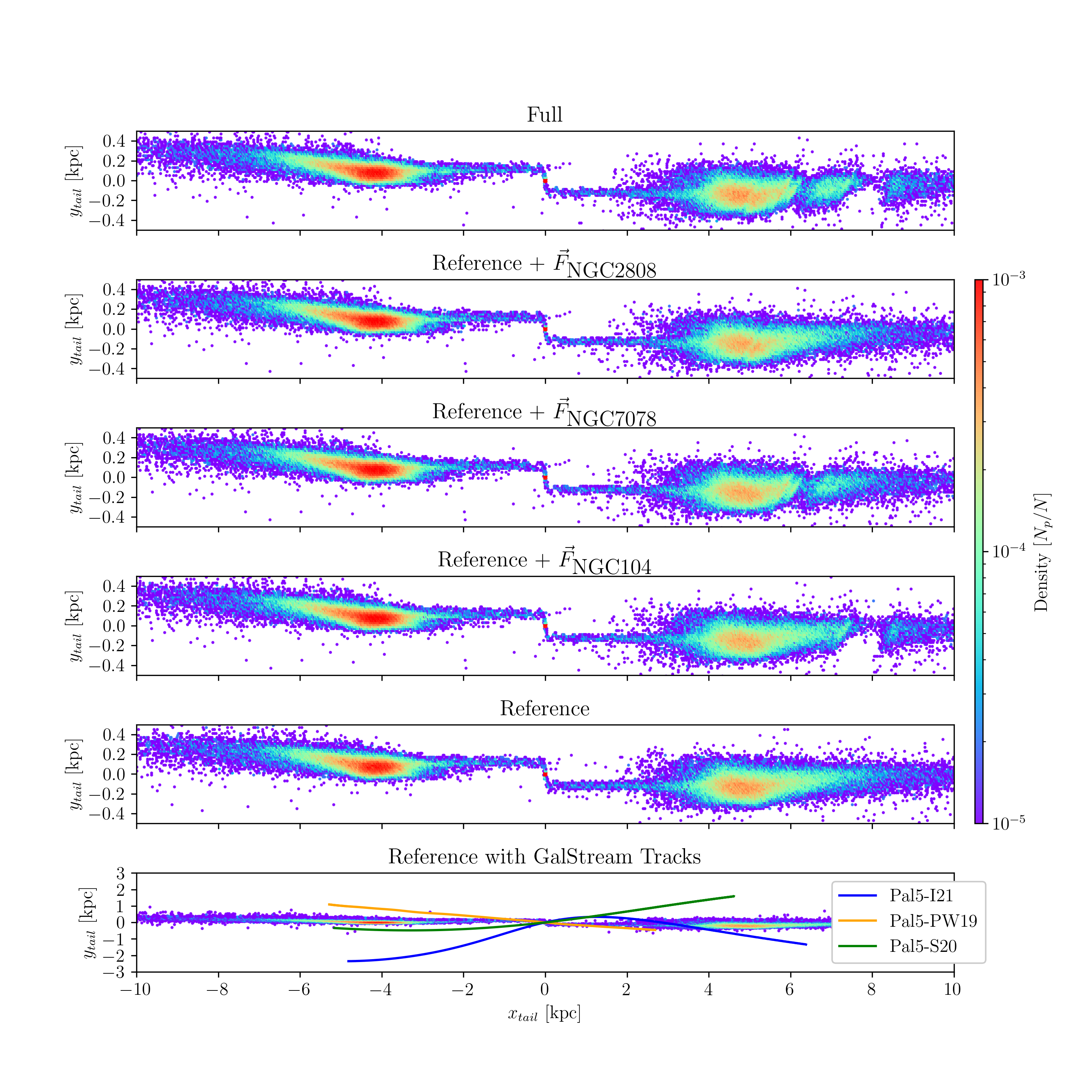}
      \caption{Density maps of Palomar~5's stream in the tail coordinate system. The color scale represents normalized particle counts (total: 100,000). The top panel shows the \texttt{full} simulation with three gaps on the stream's right-hand side. The next three panels depict simulations with identical initial conditions but exclude the gravitational influence of all clusters except those forming a given gap. The \texttt{Reference} simulation omits the influence from other globular clusters. The bottom panel compares Palomar~5's observed stream length to the \texttt{Reference} simulation, using the same Monte Carlo realization as Fig.~\ref{fig:stream_on_sky} and \texttt{Sampling~009} of Fig.~\ref{fig:gallery2}.}
      \label{fig:decomposition}
    \end{figure*} 
    
    \begin{figure*}
      \centering
      \includegraphics[width=\linewidth]{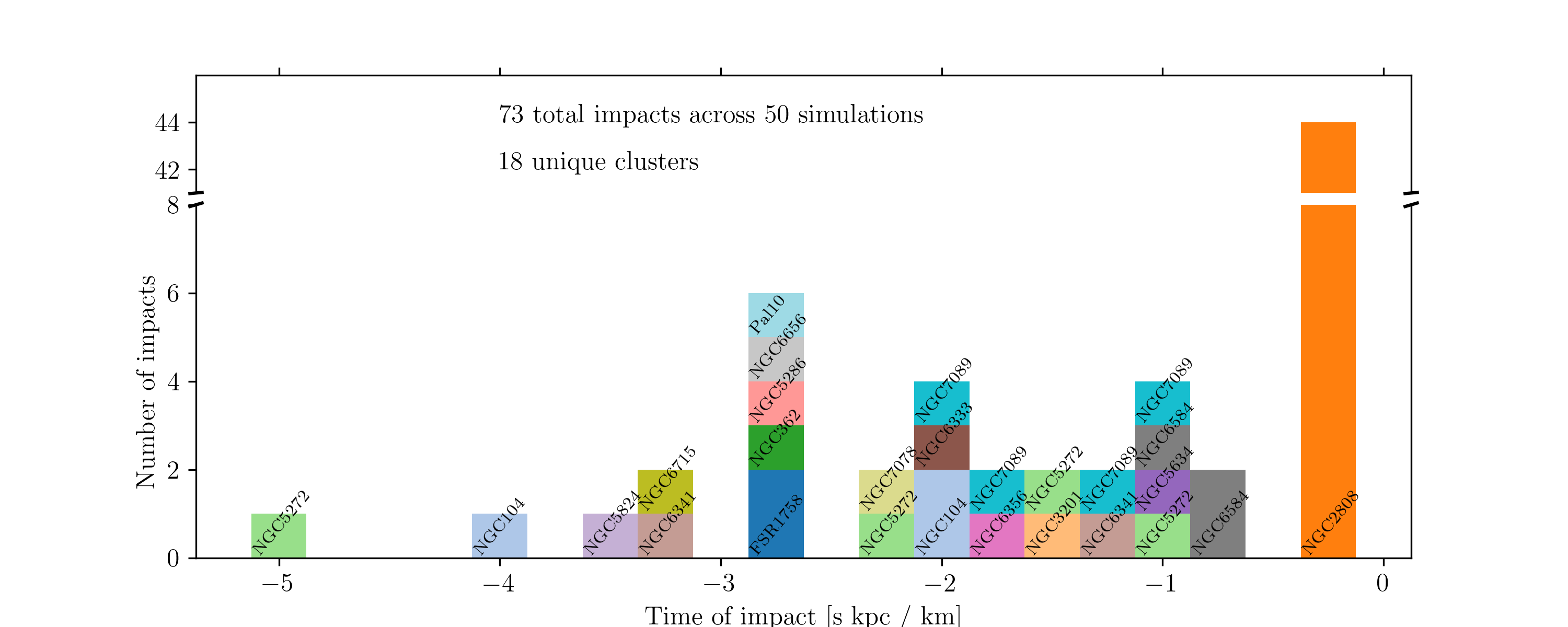}
      \caption{The simulation time at which the impacts occurred for all gap causing flybys summed over all 50 simulations. Each perturbing cluster is label and is color consistent. The time axis is given in simulation units, with 1~s~kpc~km~$^{-1}$ corresponding to roughly 1~Gyr. Note that the y-axis is broken to accommodate the large number of encounters from NGC2808, without overshadowing the other interactions.}
      \label{fig:histogram_impact_time}
    \end{figure*}

    
    The presence of the other globular clusters has a clear effect on the properties of Palomar~5 stream. A particularly clear example of this effect is shown in Fig.~\ref{fig:stream_on_sky} which we selected for its prominent gaps. Two of these gaps are clearly visible in Galactic coordinates and become even more apparent when marginalizing over latitude to reconstruct the 1D density profile of the stream, as a function of longitude.     

    Regarding the shape of the density distribution, the central peak corresponds to the still-intact globular cluster whose stars have not yet escaped. The density peaks in the stream are of the same order as the cluster itself. This is inconsistent with reality, the cluster's peak density should be higher than the stream's peak density. Of course, this discrepancy is a result of our modeling choices. We use the present day mass and radius for Palomar~5 for the whole duration of the simulation. Consequently the system is modeled as being less dense than it should have been. In turn, our simulations have a strong initial mass loss which adds to the amplitudes of the profile density peaks of Fig.~\ref{fig:stream_on_sky}. This inaccuracy is acceptable for the scope of this work. First is that Palomar~5's tails indeed have more mass than the cluster itself. \citet{2017ApJ...842..120I} reports that there could be three times as much mass in Palomar~5's tails as the cluster itself. Secondly, the exact form of the density distribution is less important than having a population of particles present that can probe a cluster flyby event.
    
    To make a more quantitative comparison between the \texttt{reference} and \texttt{full} simulations, we work in the tail coordinate system, in which the stream is aligned with the cluster's orbit. The definition of this coordinate system is based on the work of \citet{2004AJ....127.2753D} and is shown in Fig.~\ref{fig:TailCoordinates}. Briefly, in this system the $x_{\textrm{tail}}$ coordinate represents the position of a particle along the orbit relative to the globular cluster. Positive values of $x_{\textrm{tail}}$ are ahead of the cluster while negative $x_{\textrm{tail}}$ is behind the cluster. The $y_{\textrm{tail}}$ coordinate measures the particle's distance within the orbital plane, where positive values indicate that the particle is farther from the Galactic center and negative values indicate that it is closer. Fig.~\ref{fig:decomposition} shows a comparison between one of the 50 realizations of Palomar~5 stream taking into account the gravitational interactions with all other globular clusters in the Galaxy (top panel) and omitting them (bottom panel). This comparison clearly shows the presence of two wide ($\sim$100~pc and $\sim$1~kpc) gaps in the leading tail and of a more subtle under-density at $x_{\textrm{tail}}\sim 5$~kpc (we refer the reader to Appendix~\ref{sec:gap_detection} for a detailed description of the under-densities and gaps detection method). 
    
    To find out which globular clusters were responsible of creating these gaps, and when close passages took place, we estimated the gravitational acceleration along the orbit of Palomar~5 and represented it in the ($t$, $\tau$) space. $t$ is the simulation time and $\tau$ is indicates how long it will take for Palomar~5 to reach, or how  long ago it passed, a given point on its orbit. The use of $\tau$ is advantageous because the growth of the stream is approximately linear in $\tau$. On the other hand, in physical space streams on eccentric orbits indeed grow in time yet the growth rate is modulated with periodic expansion and contraction depending on the orbital phase \citep[see the top panel of Fig.~5.][for an example]{2016MNRAS.457.3817S}. Adopting this time space, and reporting the gravitational acceleration along Palomar~5 orbit in this space, the identification of the globular clusters which produced the perturbation, and the time at which it occurred becomes straight forward. We refer the reader to Appendix~\ref{sec:Perturber_Identification} for an in-depth description of the procedure. In this way, we could identify that the clusters responsible for creating gaps in the simulation of Palomar~5 stream reported in Fig.~\ref{fig:decomposition} are NGC~2808, NGC~7078 and NGC~104, and that their close passages occurred 200~Myr, -1.9 and -2.1~Gyr ago, respectively. 
    
    To further probe that the three clusters above are responsible of producing the gaps observed in this simulation, we conducted additional experiments by including only the perturbation of each of these three clusters at a time, and neglecting the gravitational perturbations of all the other clusters in the Galaxy. The results of these simulations are reported in the middle panels of Fig.~\ref{fig:decomposition} and clearly show that NGC~2808, NGC~7078 and NGC~104 are the clusters responsible for creating the under-dense regions observed in the leading tails of Palomar~5. It is worth noting that the times at which the passages of these clusters occurred, according to our analysis, are in agreement with the observed width of the corresponding gaps: the encounter with NGC~2808 being very recent (only 200~Myr ago), its induced gap is still very thin, because it takes time for a perturbation to grow into an extended gap, as it is the case for those induced by the passages of  NGC~7078 and NGC~104, which occurred in earlier times. It is also interesting to emphasize that gaps as thin as those generated by the passage of NGC~2808, 200~Myr ago, can be detected by working in the tail coordinate system and by making a comparative analysis (\texttt{full} versus \texttt{reference} simulations): they are so thin that they cannot be directly identified in Galactic coordinates (see Fig.~\ref{fig:stream_on_sky}).\\

    The analysis presented in Fig.~\ref{fig:decomposition} has been repeated for the whole set of simulations, and is reported in Appendix~\ref{sec:gallery_of_gaps} for completeness. For all of the streams, morphologically speaking, the only changes appear to be the existence of gaps or not. We do not observe a thickening of the streams due to an increased velocity dispersion. 
    
    From this analysis we can derive a statistical view of: 1) the number of gaps generated on the Palomar~5 stream by the whole system of Galactic globular clusters; 2) the clusters which generated these gaps and the time history of these perturbations. From this, we can then quantify: 3) the properties of the perturbers (their masses, sizes and their orbital parameters) as well as 4) the impact geometry of the encounters, which allows us to understand which encounters are more favorable to generate gaps in the Palomar~5 stream. In the following, we will present results of this analysis, by starting from addressing points 1) and 2).

  \subsection{The history and statistics of gap creations in Palomar~5 stream}\label{sect:history}

    By applying the methodology described above to the whole set of simulations, we can reconstruct the history of close passages of Galactic globular clusters to Palomar~5's stream, in the last 5~Gyr, which -- we remind the reader -- is the time interval investigated in our simulations. The results of this analysis are reported in Fig.~\ref{fig:histogram_impact_time} and in Table~\ref{tab:gaps_per_perturber}. NGC~2808 impacts Palomar~5's stream about 200 Myr ago in 44 out of 50 simulations, creating a small gap at a similar position to the one reported in Fig.~\ref{fig:decomposition} in each case. Since this interaction occurred less than one orbital period ago, despite the uncertainties the orbital solutions remain nontheless similar and thus produce consistent results across all simulations. However, as we continue to turn back time further, the uncertainties in the initial conditions allow the for the various orbital solutions to diverge from one another. Thus, in one configuration a cluster can impact the stream at a given time and yet at the same moment in a different set of initial conditions, it could be on the other side of the galaxy.

    \begin{table}[h]
      \centering
      \begin{tabular}{|ll|ll|ll|}
      \hline
      NGC2808 & 44 & NGC7089 & 5 & NGC5272 & 4 \\
      NGC6584 & 3 & NGC6341 & 2 & NGC6656 & 2 \\
      NGC104 & 2 & NGC3201 & 1 & NGC5634 & 1 \\
      NGC5286 & 1 & NGC362 & 1 & NGC5824 & 1 \\
      NGC6356 & 1 & NGC6333 & 1 & NGC6715 & 1 \\
      FSR1758 & 1 & NGC7078 & 1 & Pal10 & 1 \\
      \hline
      \end{tabular}
      \caption{The number of gaps created by each perturber across all 50 simulations of Palomar~5's tidal tail. These data are color-coded and illustrated in Fig.~\ref{fig:histogram_impact_time}.}
      \label{tab:gaps_per_perturber}
    \end{table} 
    
    In total, we report the finding of 73 gaps in our 50 simulations which is on average 1.5 gaps/simulation and were created by 18 different perturbers. The distribution of number of gaps appearing per simulation is presented in Table~\ref{table:gap_distribution}. If we consider NGC~2808 to be an outlier and exclude it from the experiment, we observe an average of 0.6 gaps per simulation.  
    
    \begin{table}[h]
      \centering
        \begin{tabular}{|l|c|c|c|c|c|}
          \hline
          Number of Gaps & 0 & 1 & 2 & 3 & 4 \\
          \hline
          Number of Sims. & 4 & 25 & 16 & 4 & 1 \\
          \hline
        \end{tabular}
        \vspace{0.5cm}
      \caption{Occurrence of gaps in Palomar~5 streams,  in our simulations. More specifically, the table reports the number of simulations (second row) for which a given number of gaps (first row) is found. }\label{table:gap_distribution}
    \end{table}

    We need a more sophisticated statistic that can be compared to other simulations, so we turn to the \textit{gap creation rate} developed by \citet{2012ApJ...748...20C}. The gap creation rate is the number of gaps that appear per time and normalized by the length of the stream. For our simulations, this rate is given by: \begin{equation} \label{eq:gap_creation_rate} \mathcal{R}_{\textrm{Pal 5}} =  \frac{1}{T}\int_{0}^T l^{-1}(t) \sum_i \delta(t-t_i) dt,\end{equation}where $T$ is the total integration time, $l(t)$ is the length of the stream, and $\sum_i \delta(t-t_i)$ sums over the gap occurrences, with $i$ indexing over the number of gaps in a given simulation. Here, $\delta$ represents the Dirac delta function. This expression can be simplified to:\begin{equation}\mathcal{R}_{\textrm{Pal 5}} =  \frac{1}{T} \sum_i \frac{1}{l (t_i)}. \end{equation}
    
\begin{figure}
      \centering
      \includegraphics[width=\linewidth]{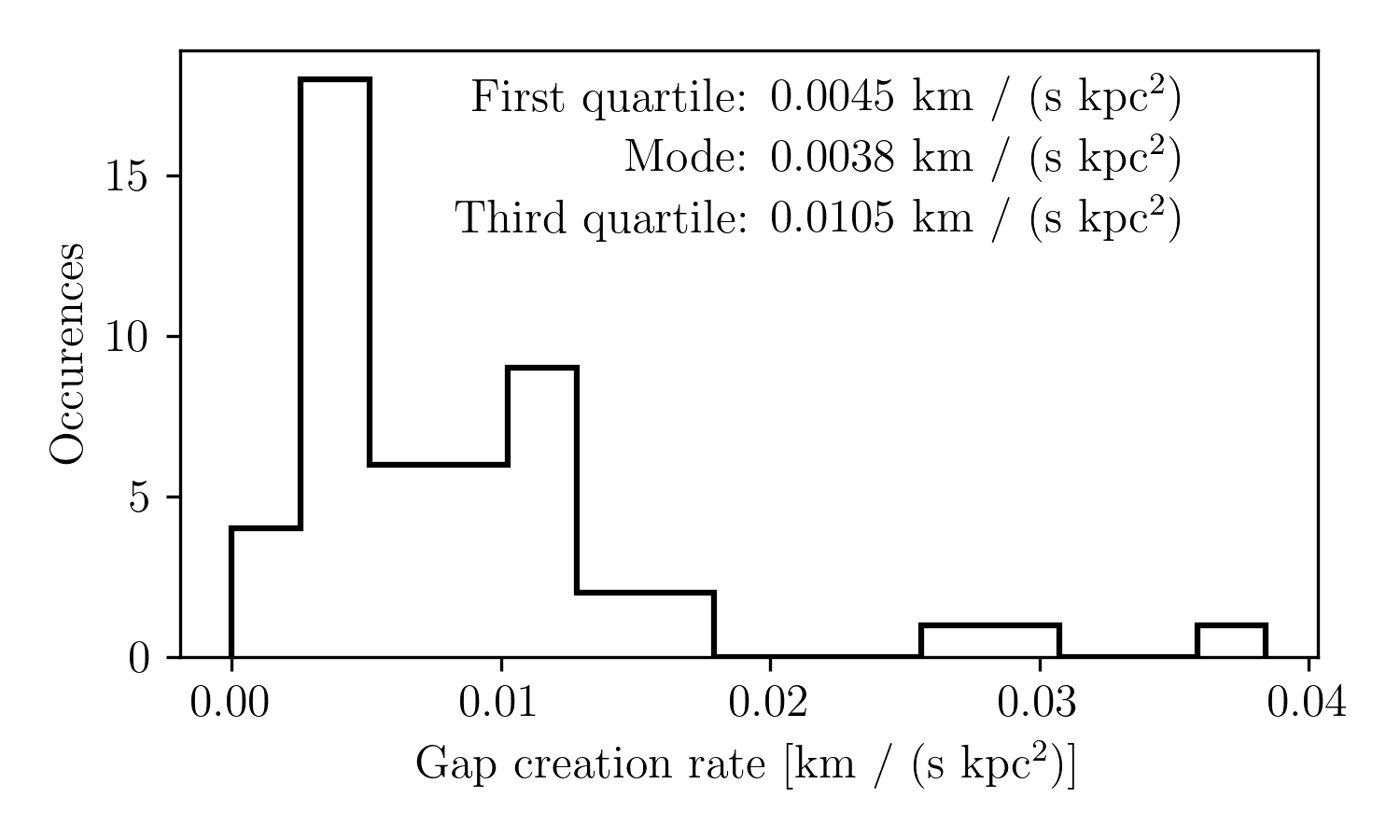}
      \caption{The distribution of the number of gaps normalized over the total integration time and unit stream length, as described by Eq.~\ref{eq:gap_creation_rate} for the whole set of 50 \texttt{full} simulations. }
      \label{fig:gapcreationrate}
    \end{figure}

    This computation allows us to analyze the distribution of gap creation rates across all simulations. Notice that since the gap creation rate \textit{adds in parallel}, naturally gaps that occur at earlier times when the stream was shorter are weighted higher than those that occur when the stream is longer. The results are shown in Fig.~\ref{fig:gapcreationrate}. This is roughly consistent with a simple estimate of the average gap creation rate: with 73 gaps over 5~Gyr of integration time for a stream about 20 kpc in length, the naive estimate is approximately 0.015~km~s$^{-1}$~kpc$^{-2} $ (which is roughly equivalent to $0.015~\rm{Gyr^{-1}kpc^{-1}}$). This is about double the weighted mean gap creation rate of 0.009~km~s$^{-1}$~kpc$^{-2}$. The naive estimate is higher because it does not account for the growth of the stream over time, unlike Eq.~\ref{eq:gap_creation_rate}.

    Lastly, we note that of the 73 observed gaps only 8 are in the trailing tail and the rest are in the leading tail. This is a surprising result. A priori, since the number of star-particles escape at similar rates from the L1 and L2 Lagrange points, each tail is of similar length and density. The main difference between the two tails is that the leading tail is closer to the Galactic center than the trailing tail by about 400~pc. Since the lengths are equal, and the offset between the tails is small compared to the galactocentric distance of about 10 kpc, we expected the number of gaps in each tail to be more or less the same. We can compute the probability of observing the unequal occurrences through the binomial distribution. First, the since the 44 gaps linked to NGC~2808 are the same fly-by, they are not independent events. We remove them from this consideration. This leaves 21 gaps in the leading tail and 8 in the trailing. The probability of observing up to 8 successes in 29 trials, given a 50\% chance of success, is 1.2\%--unlikely, but possible. Additionally, other perturbers impact at consistent times, which may also violate the assumption of independent events, as seen with NGC~2808.

  \subsection{Impact geometry and parameters of the perturbers}\label{sect:geometry}
  
    \begin{figure*}
      \centering
      \includegraphics[width=\linewidth]{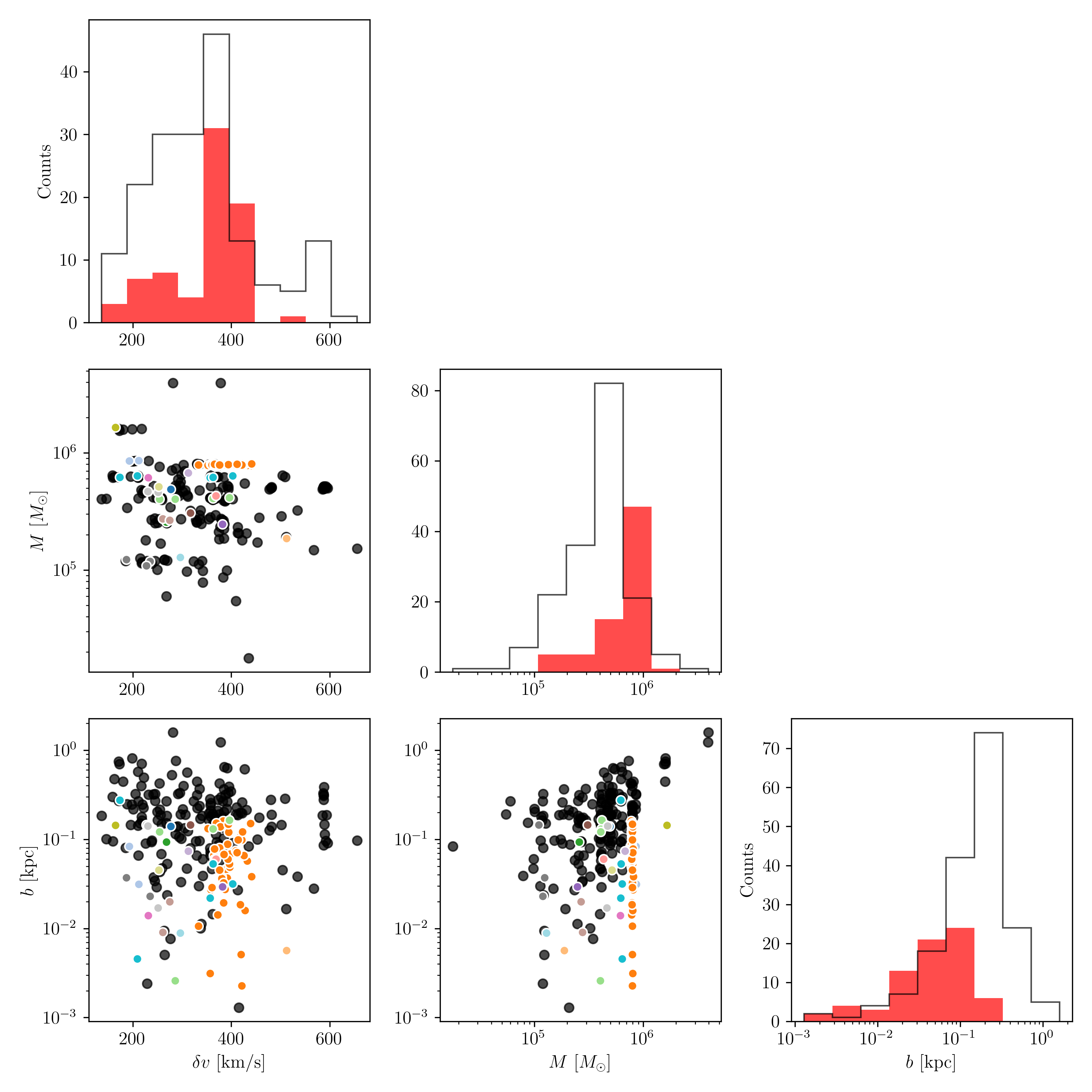}
      \caption{The distribution and relationship between the impact variables from Eq.~\ref{eq:change_in_momentum} for all close fly-bys considered. The encounters that cause gaps are labeled with the same colors as from Fig.~\ref{fig:histogram_impact_time} in the scatter plot, with white rings edges for visibility, and binned in red for the histograms. The close encounters that did not lead to gaps are shown in black. Indeed, no obvious trend delimiting these planes into gap producing or not emerges.}
      \label{fig:impact_geometry_statistics}    
    \end{figure*}

    With the perturbers identified, we perform statistics in the pursuit of understanding what conditions are necessary for a globular cluster to induce a gap on the Palomar~5 stream. We turn to impact theory, which in its simplest form is presented in works such as \citet{2008gady.book.....B}. Consider two particles: one stationary and the other moving past it. The distance between the two particles at their closest approach is known as the impact parameter. To simplify the analysis, the \textit{impulse approximation} is employed, which assumes that the velocity of the perturber remains unchanged during the interaction. This assumption simplifies the computation.

    To understand how the impacted particle is perturbed, one needs to compute its change in momentum, which is determined by integrating the force acting on the particle over the duration of the interaction. A useful approximation for this change in momentum, per unit mass, is the force at the closest approach multiplied by an estimate of the interaction time:
    \begin{equation} \label{eq:change_in_momentum} \Delta p \approx \text{Force} \times \text{interaction time} = \frac{GM}{b^2} \times 2\frac{b}{\delta v} = 2\frac{GM}{b \delta v}, \end{equation}where $M$ is the mass of the perturber, $b$ is the impact parameter, $\delta v$ is the relative velocity of the perturber with respect to the particle, and $G$ is the gravitational constant.

    This equation asserts that a more massive perturber, passing closer to the particle, and moving more slowly, will have a greater impact. It is important to note that the momentum change is inversely proportional to the velocity of the perturber. Note that this contrasts with the intuition from elastic collisions such as those between billiard balls where higher velocities result in greater impacts.

    \citet{2015MNRAS.450.1136E} extended this impact theory from one point mass impacting another to studying how an extended body impacts a stream. This quantifies the change in momentum of a given particle as a function of its distance from the point of greatest impact along the stream. In their analysis, the perturber was modeled as a Plummer sphere, like in our simulations. Since the stream is not zero-dimensional and has length, both the parallel and perpendicular components of the perturber velocity must be considered. Consequently, five parameters determine the change in velocity of a given stream particle: $M$, $r_p$, $b$, $W_\parallel$, and $W_\perp$, which are the: mass of the perturber, size of the perturber, impact parameter, parallel and perpendicular components of the relative velocity. As detailed in Appendix~\ref{sec:reconstruction}, we calculated these parameters for all our \texttt{full} simulations, by selecting -- for each of them --  the strongest 5 flybys of a perturber with the Palomar~5 stream. Thus, we compute a total of  250 impacts, and flag those that give way to gaps.

    Visual inspection of the five key impact parameters ($M$, $r_p$, $b$, $W_\parallel$, and $W_\perp$) did not reveal a clear distinction between flybys that create gaps and those that do not. Therefore, we only present the quantities from Eq.~\ref{eq:change_in_momentum} in Fig.~\ref{fig:impact_geometry_statistics}. 
    Note that in this figure, we show the total relative velocity rather than separating parallel and perpendicular components, because from a visual inspection, we could not find any specific trends when the two velocity components were plotted separately. We also exclude the characteristic cluster radius, as it showed little correlation with the results, likely due to the narrow range of globular cluster radii (see Fig.~\ref{fig:mass_size_plane}). This factor might be more significant for dark matter sub-halos, where size variation is greater.

    While Fig.~\ref{fig:impact_geometry_statistics} demonstrates that mass, relative velocity, or impact parameter alone cannot predict gap formation, one interesting result emerges: no gaps are created from interactions with impact parameters greater than 300~pc. The stream widths are roughly 200~pc, as seen in Appendix~\ref{sec:gallery_of_gaps}. This finding is even more evident when examining the $b$-$M$ plane. A series of perturbers at roughly $\sim8 \times 10^5 M_\odot$ highlights NGC~2808's flybys, where all encounters with impact parameters under 200~pc result in gaps, while those beyond this distance do not. In other words, even the most massive globulars with masses greater than about $10^6 M_\odot$ cannot cause gaps if their impact parameters are greater than about $300$~pc. It is interesting to note that for values of the impact parameter below this threshold, even fast encounters ($\delta v > 300$~km/s) can produce gaps. Perhaps, this is not surprising since the range of possible relative velocities is much less than mass and impact parameter which vary two and three orders, respectively while the relative velocities only vary by about a factor of three.

    \begin{figure}
      \centering
      \includegraphics[width=1\linewidth]{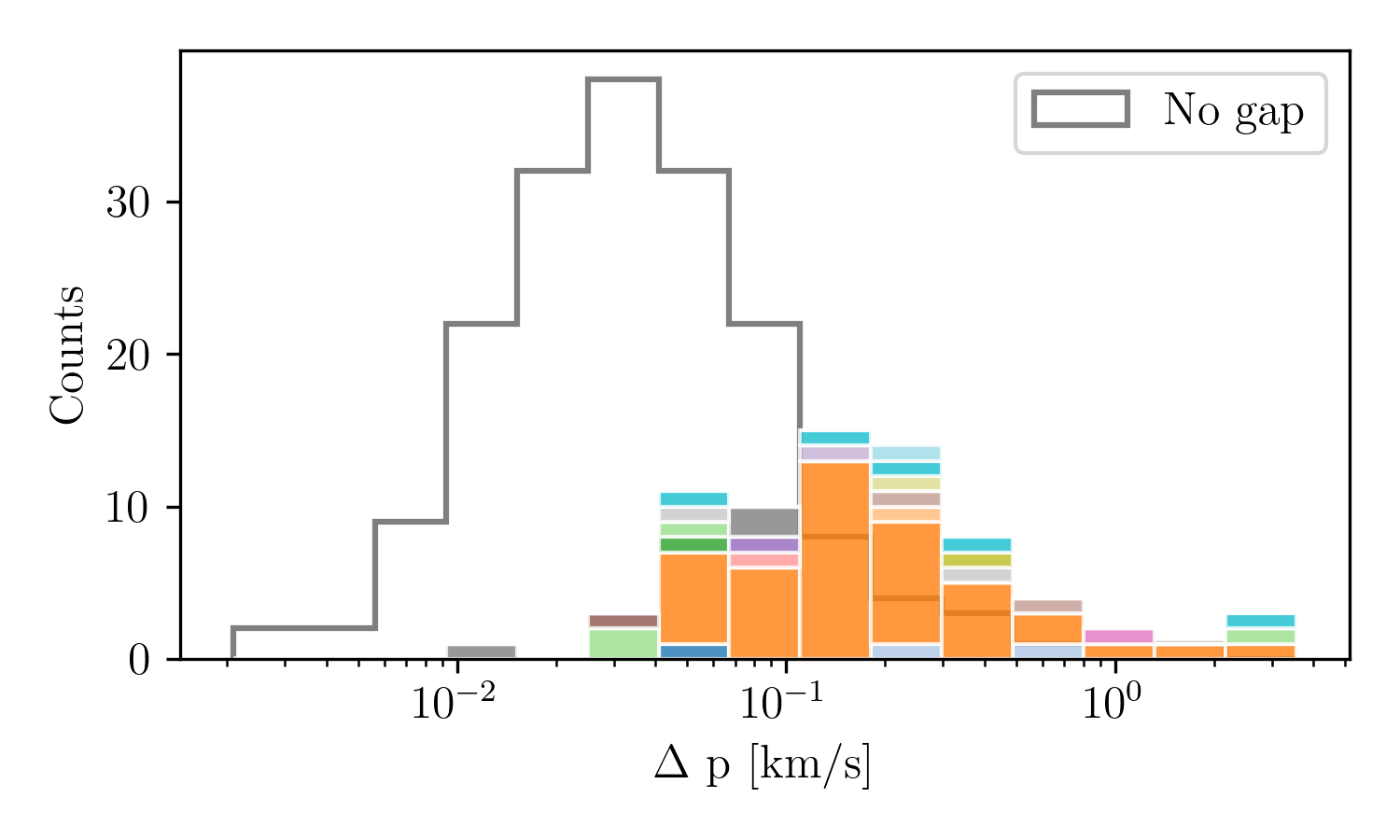}
      \caption{The distribution of imparted change in momentum (per unit mass) from a cluster flyby given by Eq.~\ref{eq:change_in_momentum}. The data set includes the top 5 strongest flybys from each simulation. Those that cause gaps are colored, stacked, and overlain atop of those that do--the distribution of which is outlined in grey. Note that the meaning of the colors is the same as in Fig.~\ref{fig:histogram_impact_time}.}
      \label{fig:deltap}
    \end{figure}
      
    \begin{figure*}
      \centering
      \includegraphics[width=0.45\linewidth]{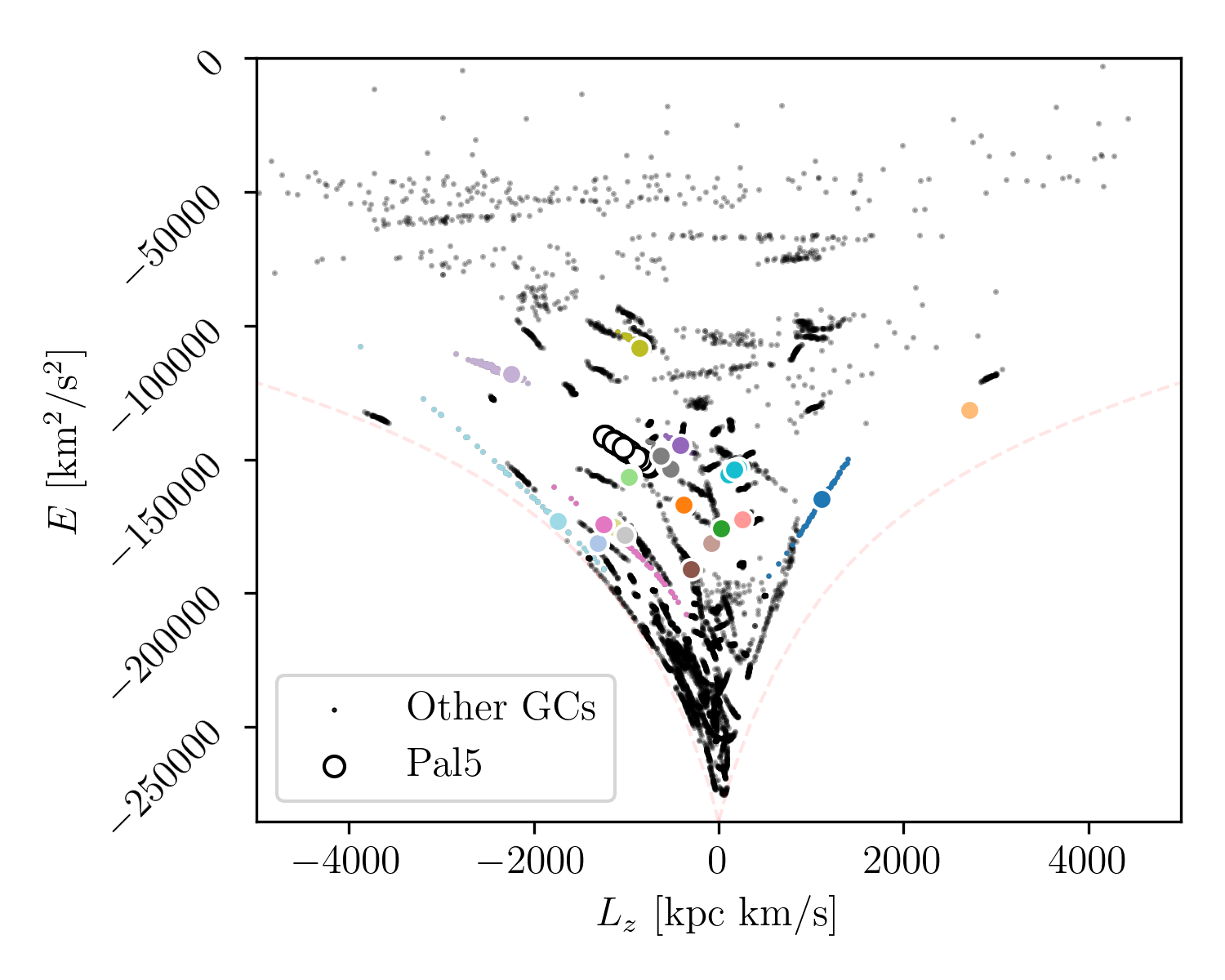}
      \includegraphics[width=0.45\linewidth]{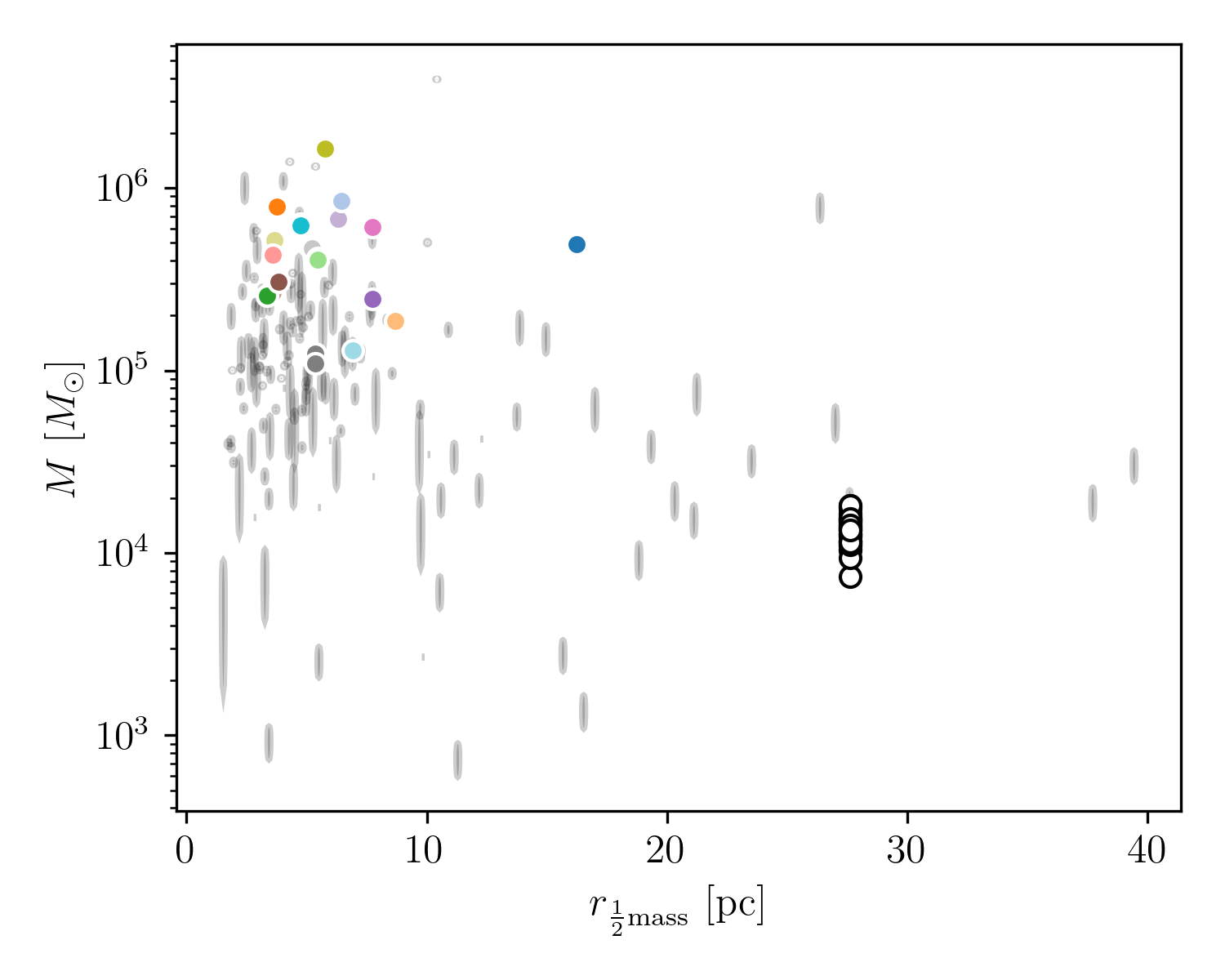}
      \includegraphics[width=0.45\linewidth]{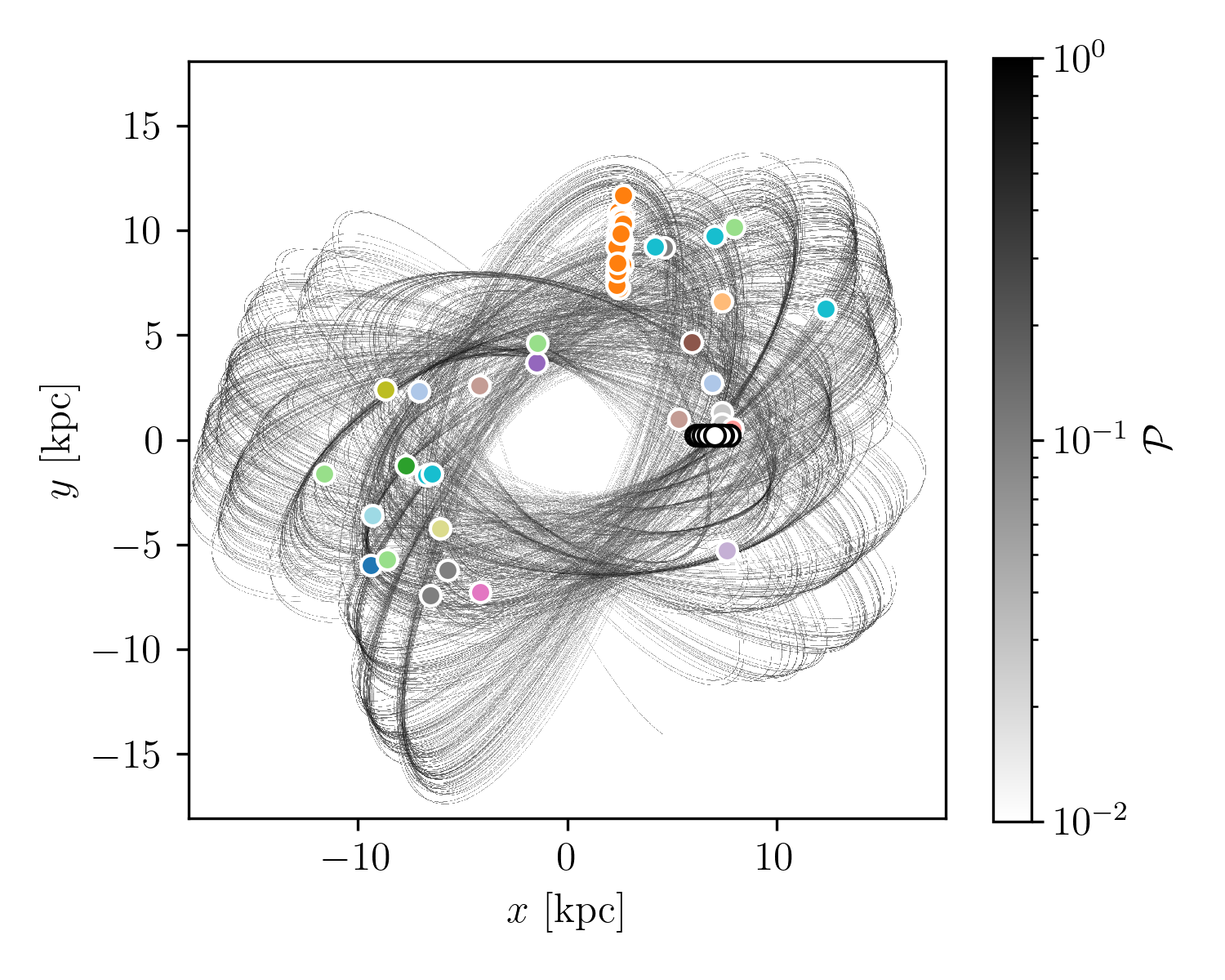}
      \includegraphics[width=0.45\linewidth]{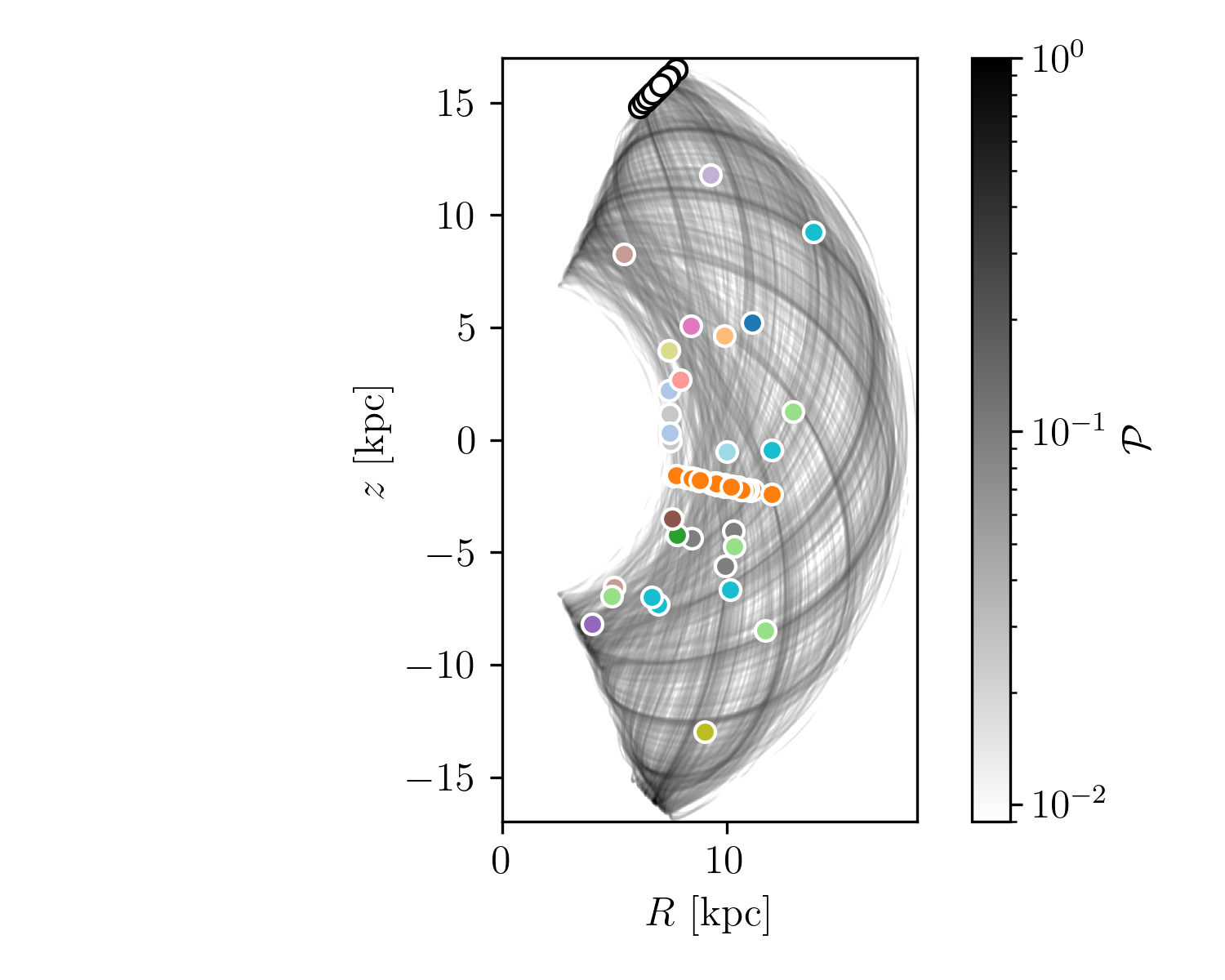}
      \caption{\textit{Characteristics of gap causing clusters}. \textit{Top Left:} Energy-angular momentum space of the globular clusters in the simulations, with 50~$\times$~165 data points representing all sampled initial conditions. Clusters impacting Palomar~5 are shown with colored markers; large for the samplings that induce a gap and small for those that do not. Non-gap-causing clusters are shown in small gray dots. The 50 white dots indicate Palomar~5's sampled initial conditions for the current day position. The circular velocity-curve is drawn with a light-pink dashed curve. \textit{Top Right:} Mass-size plane of the globular clusters in the simulations, with uncertainties on the masses indicated as vertical lines. We remind the reader that the globular cluster catalog at this time does not provide uncertainties for clusters' characteristic radii. \textit{Bottom Left:} Palomar~5's orbit in the Galactocentric xy plane. The gray-scale represents all 50 stacked orbits, with $\mathcal{P}$ indicating the probability of Palomar~5's position, normalized to $\mathcal{P}_\textrm{max}=1$. The colored markers indicate where the perturber was located when it impacted the stream and \textit{not} its present day position. \textit{Bottom Right:} same as bottom left but in the meridional plane. In all panels the colors of the markers and histogram bars correspond to specific perturber as specified in Fig.~\ref{fig:histogram_impact_time}.}
      \label{fig:mass_size_plane}
    \end{figure*}

    Once all the key impact parameters estimated, we can use Eq.~\ref{eq:change_in_momentum} and calculate $\Delta p$, the change in momentum (per unit mass) imparted by a cluster flyby on Palomar~5 stream. The result  is shown in Fig.~\ref{fig:deltap}, where the distribution of imparted change in momentum is shown for all impacts which produce a gap, and compared to those that do not produce one. On average, encounters which lead to gap creations impart a change in momentum on stream particles which is a factor of 10 higher than that of encounters which do not form gaps (but with some overlap in the low-velocity tail). Interestingly, changes in momentum which lead to gap creations extend over a large range in velocities. There is a factor of about 100 between the smallest and largest changes with NGC~2808 (orange color in the histogram) imparting changes in the velocity of stream particles which redistribute over the whole range of $\Delta p$. 

    In addition to characterizing the parameters governing cluster encounters with the stream, since we know which clusters have produced gaps on the tail of Palomar~5, we can also verify their orbital and structural properties. This is shown in Fig~\ref{fig:mass_size_plane}, where we first show the mass and size (i.e. half-mass radius) of the clusters that cause gaps on the tails of Palomar~5, dividing them from those that do not. As can be seen, no cluster with mass below $10^5 M_\odot$ causes gaps on the Palomar~5 stream, and all perturbers, except FSR~1758 have a half-mass radius below 10~pc. Even more interesting is their distribution in the E-L$_z$ plane, which shows that the clusters that cause gaps are on both direct and retrograde orbits (respectively, negative and positive values of $L_z$), but all confined to an energy interval between $-2$ and $-1 \times10^5~\textrm{km}^2\textrm{s}^{-2}$. This is because only clusters within Palomar~5's orbital space can interact with the stream: clusters with higher orbital energies tend to have larger apocenter than that of Palomar~5 and thus spend most of their time away from Palomar~5's orbital volume. 
    
    Finally, it is worth commenting on the location of impacts with the stream, whether they occur when Palomar~5 is close to its pericenter or not. As it can be seen from Fig.~\ref{fig:mass_size_plane} (bottom panel) encounters can occur at all orbital phases of Palomar~5, when it is close to pericenter, but also very far away from it, at the outskirts of its orbital space. However, when taken all together, the location of gap creating impacts shows a strong negative correlation with the galactocentric radius $r$, with the number $N$ of encounters favorable to gap formation  going as $N = -2.5r + 50$ (with a Pearson coefficient of -0.86). While there are more clusters near the Galactic center, clusters naturally spend more time near their apocenters, and the lower relative velocities in these regions should favor gap creation. However, this result suggests that cluster Galactic number density outweighs these factors in determining the number of gaps. It should be noted that these results are for Palomar~5 only. Streams along various orbits need to be studied in order for this conclusion to be generalized. 


  \subsection{Comparison to observations}

\begin{figure}
        \centering
        \includegraphics[width=\linewidth]{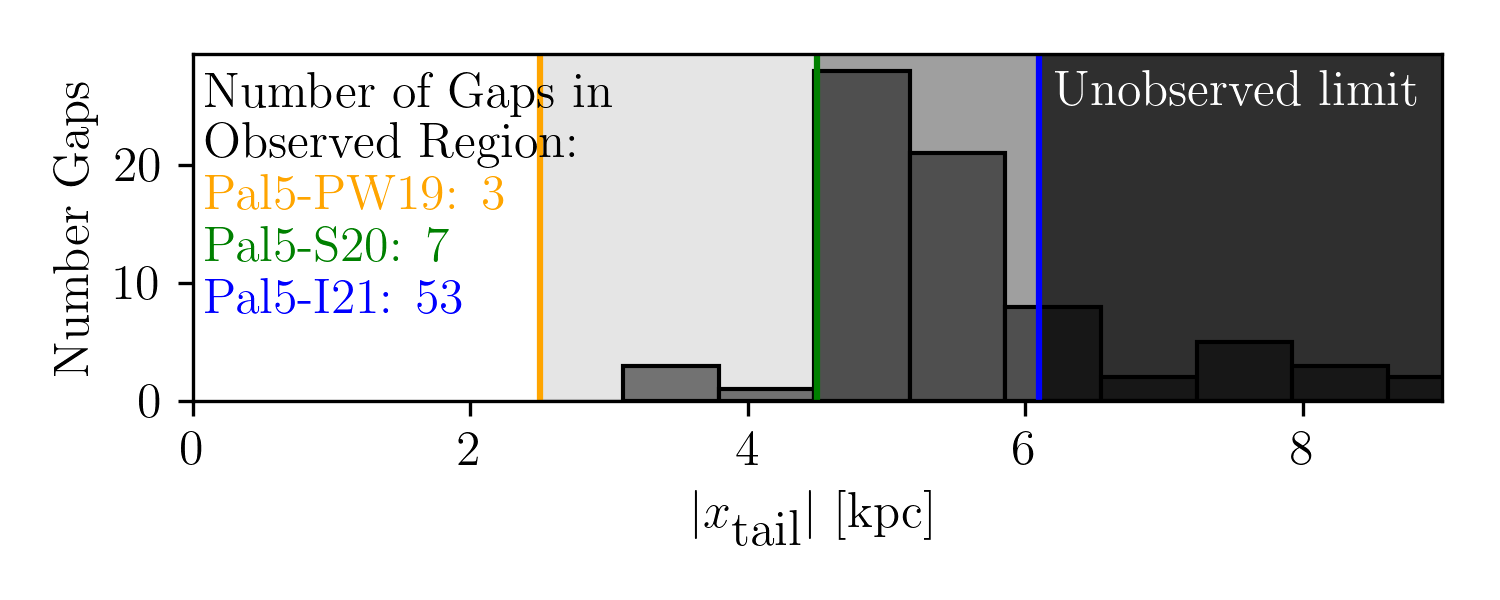}
        \caption{The distribution of gaps as a function of absolute distance from the center of mass of Palomar 5's globular cluster. The number of gaps within the observable range is evaluated on an individual track basis, with track lengths determined by the minimum and maximum $x'$-coordinates from the bottom panel of Fig.~\ref{fig:decomposition}. The ranges for \citet{2019AJ....158..223P}, \citet{2020MNRAS.495.2222S}, and \citet{2021ApJ...914..123I} are shown for their respective dimensions, and the number of gaps within these ranges is indicated in the figure. Vertical bars mark the $x' > 0$ limit of the stream--we only show the right hand side instead of both for clarity, as most gaps are located on the right-hand side of the stream in any case. }
        \label{fig:GapsWithinSight}
      \end{figure}

    We briefly compare our simulated gaps to the literature on Palomar~5. In the bottom panel of Fig.~\ref{fig:decomposition}, we compare the tracks of Palomar~5 that were compiled in \texttt{galstreams} by \citet{2023MNRAS.520.5225M}. The mid-point positions of the streams do not have the same galacto-centric positions as Palomar~5 from Baumgardt's catalog and are thus offset when projected into tail coordinates. Moreover, since we sample the distances to Palomar~5, the galactocentric position within the Baumgardt catalog varies. To combat this, we position the mid-point of the Galstream tracks at the center of mass of the cluster, which allows us to compare the length of the observed tracks to our simulated streams. We use the three tracks Pal5-PW19, Pal5-S20, and Pal5-I21 from \citet{2019AJ....158..223P}, \citet{2020MNRAS.495.2222S}, and \citet{2021ApJ...914..123I}, respectively. For each track, we find the distance from the cluster in either direction and count the number of gaps that fall within this range. This is presented in Fig.~\ref{fig:GapsWithinSight}. In the maximum limit many gaps could appear, at a rate of about one per realization. However, at the shortest reported stream-length only a few gaps occur.
    
    We observe that there are no gaps near within 3~kpc to the cluster in our simulation. There may be a few reasons for the absence of gaps in the portion of the tails closer to the cluster center. First, as \citet{2016MNRAS.457.3817S} demonstrated, the dispersion of action-frequencies in the stream plays a role. For instance, the frequency corresponding to the azimuthal action, $J_\phi$, is $\dot\theta_\phi = - \frac{\mathcal{\partial H}}{\partial J_\phi}$ and where $\theta_\phi$ is the angle describing the position of the particle phase space between momentum ($p_\phi$) and position ($\phi$) axes---where $\phi$ is the azimuthal angle between the x-y axes in physical space. Regions of the stream with well-separated frequencies are more susceptible to gap formation, while those with a wide frequency range (near the cluster) tend to erase the history of impacts. Thus, for a gap to form, the imparted change in frequency must exceed the range of frequencies in the impacted region. Thus, the strong flybys that occurred close to the cluster were inconsequential for gap formation. 
    
    Another possible explanation is the different thickness of the simulated tails at different distances from the center of the cluster. At distances of less than about 3~kpc from Palomar~5, the tails are very thin, with a typical thickness being less than 100~pc. In order to cause gaps in these regions, the clusters would have to pass close to the stream, with impact parameters similar to the thickness itself. Note that the thinness of the tails at these distances is probably a direct consequence of the initial parameters we chose for the simulation. We have assumed for Palomar~5, 5~Gyr years ago, the same internal parameters (mass and size) that the cluster has today. With such parameters, after 5~Gyr of evolution, our system has lost most of its mass, and therefore the part of the tails closest to the cluster itself, whose density depends mainly on the most recent mass loss \citep[see, for example, Fig.~A.3 in][]{2012A&A...546L...7M}, is necessarily very thin because the simulated cluster has essentially no more mass to lose. This last point has also a consequence in the gap creation rate, which we derived at the end of Sect.~\ref{sect:geometry}: with only 70\% of the tail  ($\sim$ 14~kpc over 20~kpc, excluding the innermost $\pm 3$~kpc from the cluster center) really suitable for forming gaps, the gap creation rates are about 50$\%$ higher than those estimated in the aforementioned section, where values have been derived taking into account the full tail extent. This gap creation rate would be still not high enough to reproduce the  number of gaps in Palomar~5's tails, as reported by \citet{2012ApJ...760...75C}. In their study, they indeed suggested the presence of  five gaps with a 99~\% detection confidence, leading to a gap creation rate of 0.17~Gyr$^{-1}$~kpc$^{-1}$. If correct, this rate would be too high to be explained by Galactic globular clusters only.

\section{Discussion}
  
  The simulations presented in this paper suggest that, in the last 5~Gyr of evolution, Palomar~5's stream could have experienced multiple close encounters with other Galactic globular clusters, some of which are able to create gaps -- even a few kpc wide -- in its tails. Currently, the presence -- or not -- of gaps in the observed portion of Palomar~5 tails is debated: \citet{2016ApJ...819....1I} found no statistically significant gap in Palomar~5 tails, while \citet{2017MNRAS.470...60E}, analyzing the same data-set as \citet{2016ApJ...819....1I}, suggested the presence of a few dips and gaps in the tails, at angular distances between $2^\circ$ and $9^\circ$ from the cluster center \citep[see also][]{ 2020ApJ...889...70B}. While our simulations produced 73 gaps across 50 realizations, only about 22 are beyond the current length of the observed portion of the stream. This means on average, we obtain at least one gap from a globular cluster within the past 5~Gyr. However, we did not attempt to create mock observations or simulate a full detection process accounting for the challenges of disentangling field stars from stream stars. While such an analysis would be valuable, it is beyond the scope of this study.

  Our simulations produce a stream for Palomar~5 that is longer than the currently observed extent. In simulations, stream detections are straightforward because we can use reference runs to clearly separate stream particles from the field and compare against a known ``true'' structure. In contrast, observational data are inherently more challenging due to magnitude limits, contamination from field stars, and the lack of a ground truth for comparison. 
  
  The aforementioned factors could lead one to the conclusion that the number of gaps identified in this study could represent an upper limit. However, this conclusion is incomplete. Globular clusters lose mass and evaporate over time, leading to an incomplete catalog of perturbers. Moreover, the present-day masses used in our simulations are likely lower than the historical masses of these clusters. For example, \citet{2024ApJ...976...54P} conducted a study simulating the dissolution of a realistic globular cluster population, in the aims of identifying how many stellar streams we should expect in the Milky Way and used a mock catalog of globular clusters with an initial amount with masses above 10$^4 M_\odot$ totaling at about 10,000 clusters. This implies that more perturbers could have been present in the past, potentially increasing the frequency and number of gaps in Palomar~5's stream.

  We note that our results seem to be in tension with the conclusions of \citet{2019MNRAS.484.2009B}, who presented a numerical study of Palomar~5 tails, orbiting a Milky Way-like potential, where both dark matter sub-halos and baryonic sub-structures (Galactic bar, spiral arms, giant molecular clouds, globular clusters) were taken into account to quantify the importance of these latter in density variations in Pal5 streams. While their methods are extremely similar to ours, their analysis diverges significantly. Specifically, \citet{2019MNRAS.484.2009B} focused on examining power spectra, analyzing variations in the one-dimensional stream density in stellar counts along the length of the stream. 

  Upon comparison with our simulations, we observe that the power spectra from our \texttt{full} simulations that include globular clusters and the \texttt{reference} simulations that do not, are not significantly different. We suggest that this may be due to the signal from only one or two gaps caused by globular clusters not being sufficient to produce notable differences in the overall power spectrum. Additionally, \citet{2019MNRAS.484.2009B} may not have inspected the profiles for individual gaps, which could explain why these features were not reported in their study. Further investigations would be needed to confirm this interpretation and to thoroughly assess the impact of globular clusters on stream density profiles. \\

  The impact of globular clusters close interactions with stellar streams has also been the object of another recent paper, by \citet{2022ApJ...941..129D}, who concluded that the chance for GD-1 gaps, which have been reported in a number of works \citep[see, for example, ][]{2019ApJ...880...38B,2018MNRAS.477.1893D,2020AAS...23533607D} to be produced by globular clusters is very low. This result is not necessarily in contradiction with ours, since GD-1 has a pericenter which is almost twice that of Palomar~5 \citep[see, for example][]{2019MNRAS.486.2995M}. As we have discussed in Sect.~\ref{sect:geometry}, the number, $N$, of close encounters which lead to gap creation is anti-correlated with the distance $r$ to the Galactic center. If we naively use the same radial dependence of gaps from Palomar~5 for GD-1 by swapping a pericenter from 6~kpc to 16~kpc, we would reduce the number of gap-favorable impacts of more than a factor of 2. Moreover, we note that \citet{2022ApJ...941..129D} pre-select the globular clusters which could have experienced a close encounter with the GD-1 stream, by selecting only clusters which pass at a distance less than 0.5 kpc from the stream, having a relative velocity smaller than  300 km/s. As shown in our Fig.~\ref{fig:impact_geometry_statistics}, bottom-left panel, in the case of Palomar~5 this choice would lead to exclude most of the encounters favorable to gap creation, which turn out to have relative velocities above 300 km/s. It would be interesting to repeat a similar study as the one made by \citet{2022ApJ...941..129D} for GD-1, this time imposing no selection on possible candidate clusters.\\
  
  As already suggested in previous works which have studied the impact of baryonic structures on Palomar~5 tails \citep{2017NatAs...1..633P, 2019MNRAS.484.2009B}, it is possible that this cluster lies in a region of the phase-space which is not favorable to distinguish gaps created by dark matter sub-halos from gaps created by baryonic structures, such as the Galactic bar and giant molecular clouds. With our study, we show that close encounters with globular clusters constitute a further element that confuses a simple interpretation of the observed Palomar~5 stream gaps (if any). Other clusters and streams in the orbital energy range of Palomar~5 may suffer of the same difficulty. Going to lower orbital energies make the situation even worse, because the effect of the Galactic bar, giant molecular clouds and interactions with globular clusters becomes even more efficient. In the innermost regions of the Galaxy, where the density of dark matter sub-halos is expected to be maximal, globular cluster streams are intrinsically more difficult to find, in addition to the fact that the dynamical times in these regions become so small that the stars lost from globular clusters do not redistribute themselves for the most part into thin structures \citep[see][]{2023A&A...673A..44F}. It is thus probably only at larger distances from the Galactic center than those spanned by Palomar~5's orbit that the impact of dark matter sub-halos may become dominant, but it should be borne in mind that at these distances the number density of sub-halos also decreases. In the future, it will be interesting to apply the type of study conducted here to the whole set of clusters for which streams have been found to quantify the regime in which they are found, whether favorable to the creation of gaps from baryonic structures or not.

\section{Conclusions}
  Our study demonstrates that globular cluster flybys can produce density gaps in the stellar streams of Palomar~5. The occurrence and characteristics of these gaps depend on the dynamical and structural properties of the stream, the mass of the perturber, and the impact parameter. While our simulations predict the formation of gaps in Palomar~5's tails, the predicted gaps do not align with the regions of the stream currently observed. Several factors contribute to the absence of simulated gaps in the observed portions of Palomar~5's tails, including the stream's varying thickness, the dispersion of action-frequencies near the cluster, and the initial conditions adopted in our simulations. 

  The broader implications of our findings indicate that globular cluster interactions add a layer of complexity to interpreting stellar stream substructures, complicating efforts to distinguish between baryonic and dark matter-induced gaps. While Palomar~5's phase-space region may not be ideal for isolating the impact of dark matter sub-halos, extending this analysis to other streams, particularly those at larger galactocentric distances, could provide clearer insights. 

  Future work should involve a systematic comparison of gap creation rates across various Galactic potentials and extend the scope to a wider range of streams and globular clusters. This would help identify regimes where baryonic and dark matter-induced substructures can be disentangled, shedding light on the elusive influence of dark matter sub-halos on stellar streams. Additionally, we can continue to add realistic physics to the Galactic environment such as: the merger of the Sagittarius dwarf galaxy, the passages of the Large and Small Magellanic clouds, as well as a time evolving globular cluster population that compensates for the clusters' mass loss, evaporation, and similarly add members from the current incomplete census.

\section*{Data Availability}

  The code used to produce the data in these simulations is publicly available at https://github.com/salvatore-ferrone/tstrippy. Examples on the codes installation and use are provided at https://tstrippy.readthedocs.io. The simulations will also be made publicly available at etidal-project.obspm.fr.

\begin{acknowledgements}
  This work has made use of the computational resources available at the Paris Observatory, as well as those obtained through the DARI grant A1020410154 (PI: P. Di Matteo), and of the Astropy, Numpy and Matplotlib libraries (Astropy Collaboration 2013, 2018; Harris et al. 2020; Hunter 2007). SF, PDM and MM would like to thank the Ecole Doctorale 127 for its financial support. MM thanks the CNR STM program, thanks to which the collaboration on this research topic began. 
\end{acknowledgements}

\bibliographystyle{aa} 
\bibliography{bibliography} 

\begin{appendix}

  \section{Gap Detection} \label{sec:gap_detection}
\begin{figure}
      \centering
      \includegraphics[width=\linewidth]{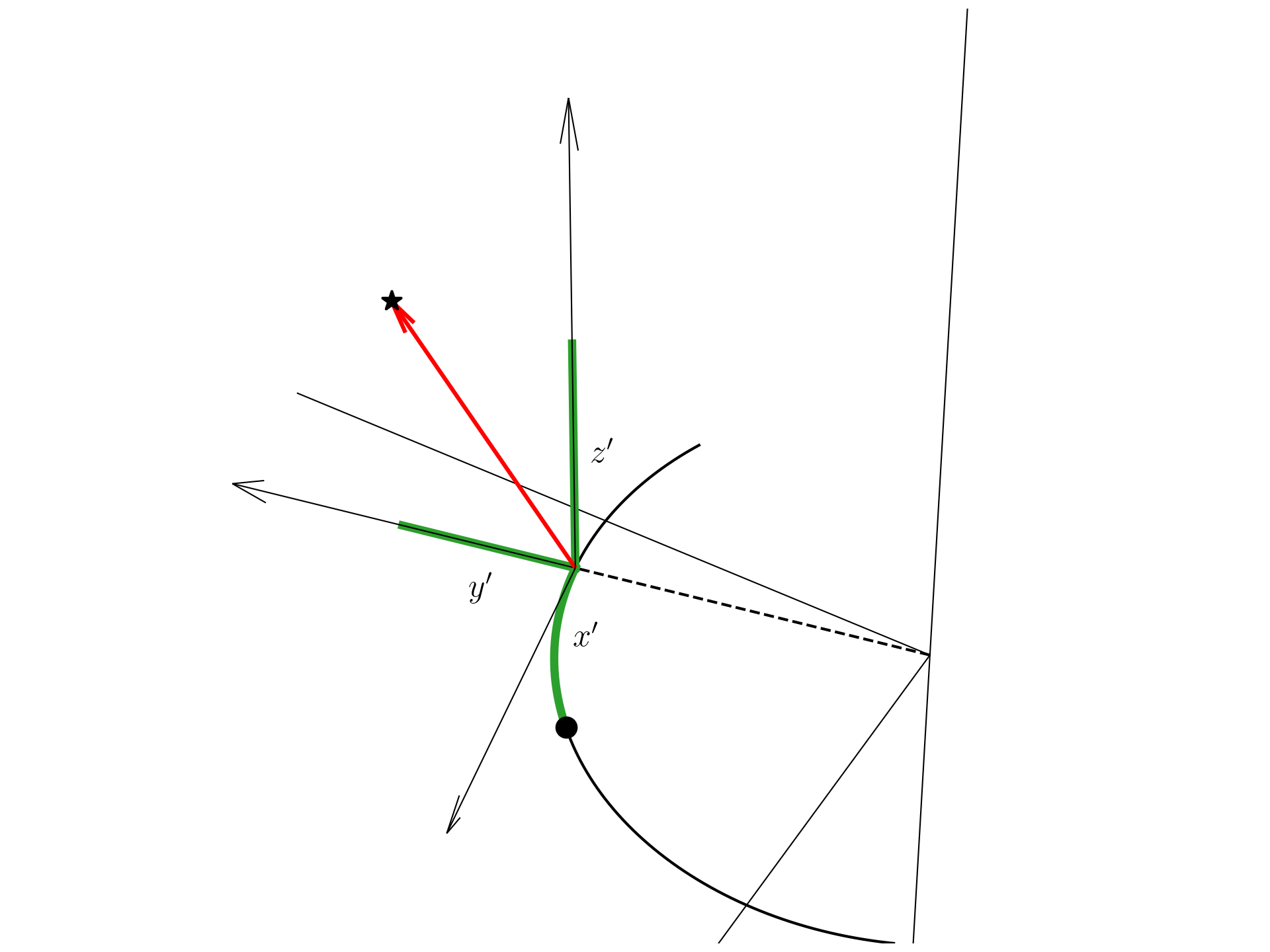}
      \caption{A demonstration of the transformation from galactic coordinates to tail coordinates. The black dot represents the globular cluster, while the thick black line shows a segment of its orbit. A star particle is marked with a star symbol. The red vector points from the nearest point on the cluster's orbit to the star particle. Its projection onto the galactocentric unit vector is shown in green and labeled $y'$. The red vector is also projected onto the vector perpendicular to both the orbital tangent and the galactocentric vector, which defines the local orbital plane and is labeled $z'$. The path length along the orbit from the cluster to the nearest point is denoted as $x'$, with $x'$ being positive if the particle is ahead of the cluster.}
      \label{fig:TailCoordinates}
    \end{figure}

\begin{figure*}
      \centering
      \includegraphics[width=\linewidth, trim=20 0 15 0]{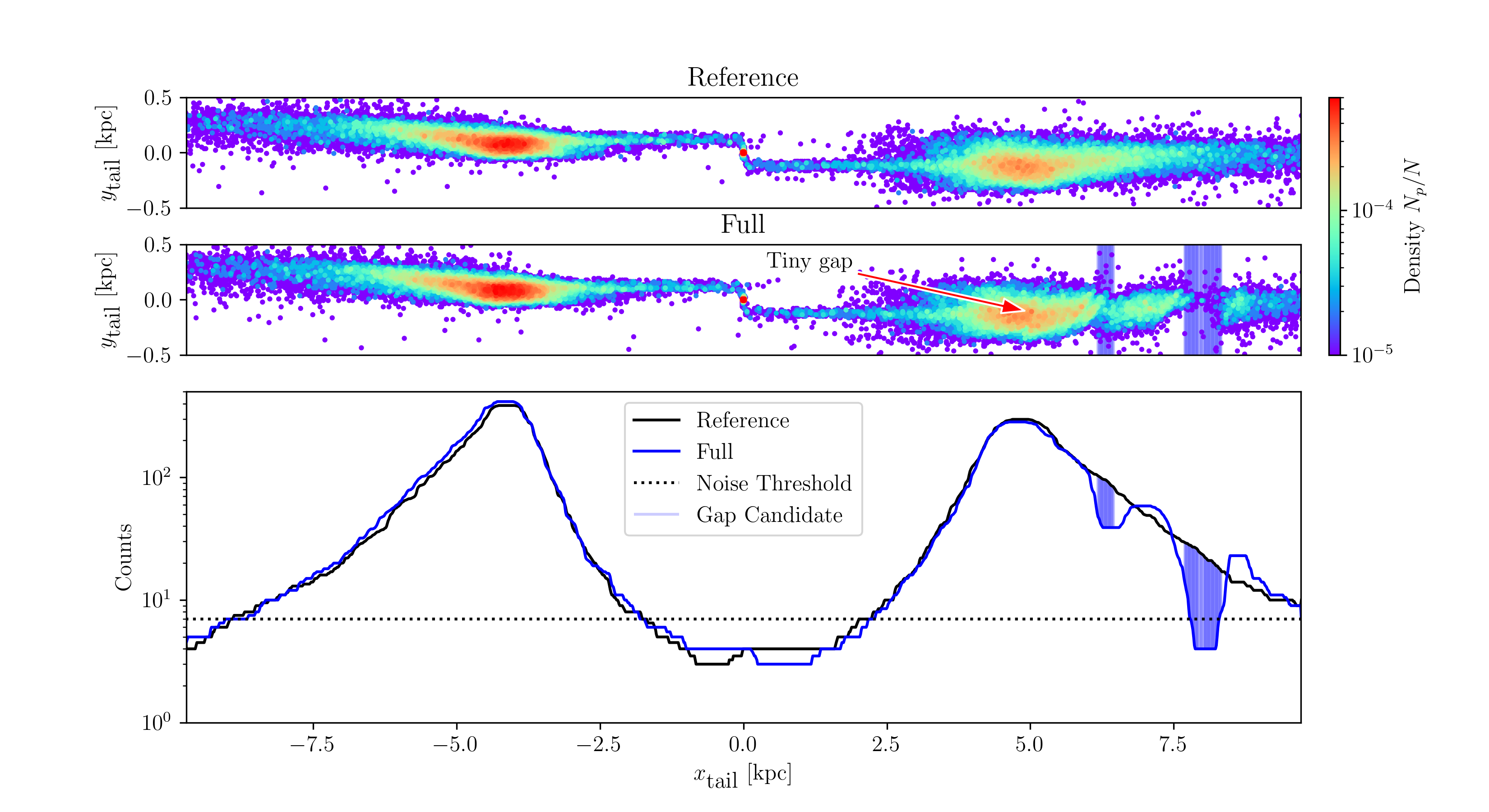}
      \caption{A comparison between the density maps and profiles of the \texttt{full} and \texttt{reference} simulations is presented. The locations where the \texttt{full} simulation is less dense than the \texttt{reference} simulation by 2-$\sigma$ is highlighted by blue vertical bars. Bins below the noise threshold are not considered when measuring the differences between the two 1D profiles. The 1D profiles have been smoothed with a median box-car filter.}
      \label{fig:profiles}
      \end{figure*}

    For each of the 50 simulations, we compare the final snapshots between the \texttt{reference} and \texttt{full} simulations using the tail coordinate system, as shown in Fig.~\ref{fig:TailCoordinates}. The \texttt{reference} simulations were only saved at the final time-stamps, thus from Eq.~\ref{eq:data_volume_estimate} where $N_{ts}$=1, leading to a data volume of about one hundred measly megabytes. We inspect these differences by generating 2D density maps of the streams. Additionally, we marginalize over the $y'$-coordinate to produce 1D density profiles along the $x'$-axis. These comparisons are shown in the results section (see Fig.~\ref{fig:profiles}).
  
    To construct the 1D density profiles, we bin the data using the $\sqrt{N}$ rule, where $N$ is the number of data points ($N_p = 10^5$). After binning the 1D profiles, we apply a boxcar smoothing technique. This raw profile density curve is smoothed with the median box-car filter. At each bin, a selected number of adjacent data points from both sides are placed in a list. The bin is then replaced with the median. We choose to use 10 adjacent points per side, which correspond to a smoothing length of approximately 1~kpc. This procedure reduces high-frequency noise and smooths the profiles. For instance, notice the absence of a high mass peak indicating for the center of mass in the bottom panel of Fig.~\ref{fig:profiles}.

    With the smoothed 1D density profiles in hand, we search for regions where the \texttt{full} simulations are significantly under-dense compared to \texttt{reference} simulations, surpassing stochastic fluctuations. To accomplish this, we first impose a signal-to-noise ratio threshold, $\mathcal{SNR}$. The signal is defined as the log of the counts per bin from the \texttt{reference} 1D density profile, and errors are propagated assuming a Poisson distribution. We then compute a threshold for the number of counts in the \texttt{reference} simulation, $N_v$, using the transcendental equation:
    \begin{equation}
        \mathcal{SNR} = \ln(10) \log_{10}\left(N_v\right) \sqrt{N_v}.
      \end{equation} \label{eq:density_threshold}
    By setting $\mathcal{SNR} = 5$, we solve for $N$ using \texttt{scipy.optimize.fsolve}, finding that $N$ must be greater than 7. After discarding insignificant bins (i.e., those with counts below the threshold), we compute the log ratio of the counts between the \texttt{reference} and \texttt{full} simulations:
    \begin{equation}
        \mathcal{R}_i = \log_{10}\left(\frac{N_{f,i}}{N_{v,i}}\right),
      \end{equation}
    where $\mathcal{R}_i$ is the log ratio, $N_{f,i}$ are the counts from the \texttt{full} simulation, and $N_{v,i}$ are the counts from the \texttt{reference} simulation for each bin $i$. We then analyze the distribution of $\mathcal{R}_i$. If the differences between the density profiles are primarily due to stochastic processes of similar magnitude, this distribution should resemble a Gaussian as expected from the central limit theorem. Thus, we flag all regions where the density is under-dense by more than two standard deviations, which should highlight regions whose under-density is unlikely to be the result of the sum of stochastic processes but rather the passage of another globular cluster. 

    However, this method has its limitations, especially when detecting smaller gaps. As outlined by \citet{2015MNRAS.450.1136E}, a small gap is not indicative of a weak impact, but a recent one. This is because gap growth is a dispersion phenomenon. Additionally, since our streams have finite width, some gaps are oblique with respect to the stream axis. In such cases, marginalizing over $y'$ erases the gap's signal, making it impossible to detect in a 1D profile. This limitation is particularly evident in gaps caused by NGC~2808, as discussed in the results. Therefore, this quantitative analysis serves as an \textit{aid} to visual inspection rather than a complete substitute for it. This method helps particularly with large subtle gaps that the eye does not notice in the 2D maps. These profiles are included in the Appendix~\ref{sec:gallery_of_gaps}.

  \section{Perturber Identification} \label{sec:Perturber_Identification}

    A key question we seek to answer is: from the gaps present at the end of the simulation, who caused them? To address this, we examine the evolution of stream density over time. Instead of using the x'-coordinate, we introduce $\tau$, which represents time rather than distance. Specifically, $\tau$ indicates how long it will take for a cluster to reach, or how long ago it passed, a given point on its orbit. This is advantageous because the growth of the stream is approximately linear in $\tau$ whereas in physical space, streams on eccentric orbits expand and contract depending on the orbital phase.

    \citet{2016MNRAS.457.3817S} extended the analysis of \citet{2015MNRAS.450.1136E}, demonstrating that action-angle variables provide a useful coordinate system for analyzing stream evolution, as actions are conserved and the angles associated with the stream's growth evolve linearly over time. Although we became aware of this work only after completing our analysis, we note that $\tau$ is a suitable approximation and behaves similarly to the angle variable corresponding to the azimuthal action: $\tau \approx \theta_{\phi,i} - \theta_{\phi,\text{GC}}$.

\begin{figure*}
      \centering
      \includegraphics[width=\linewidth]{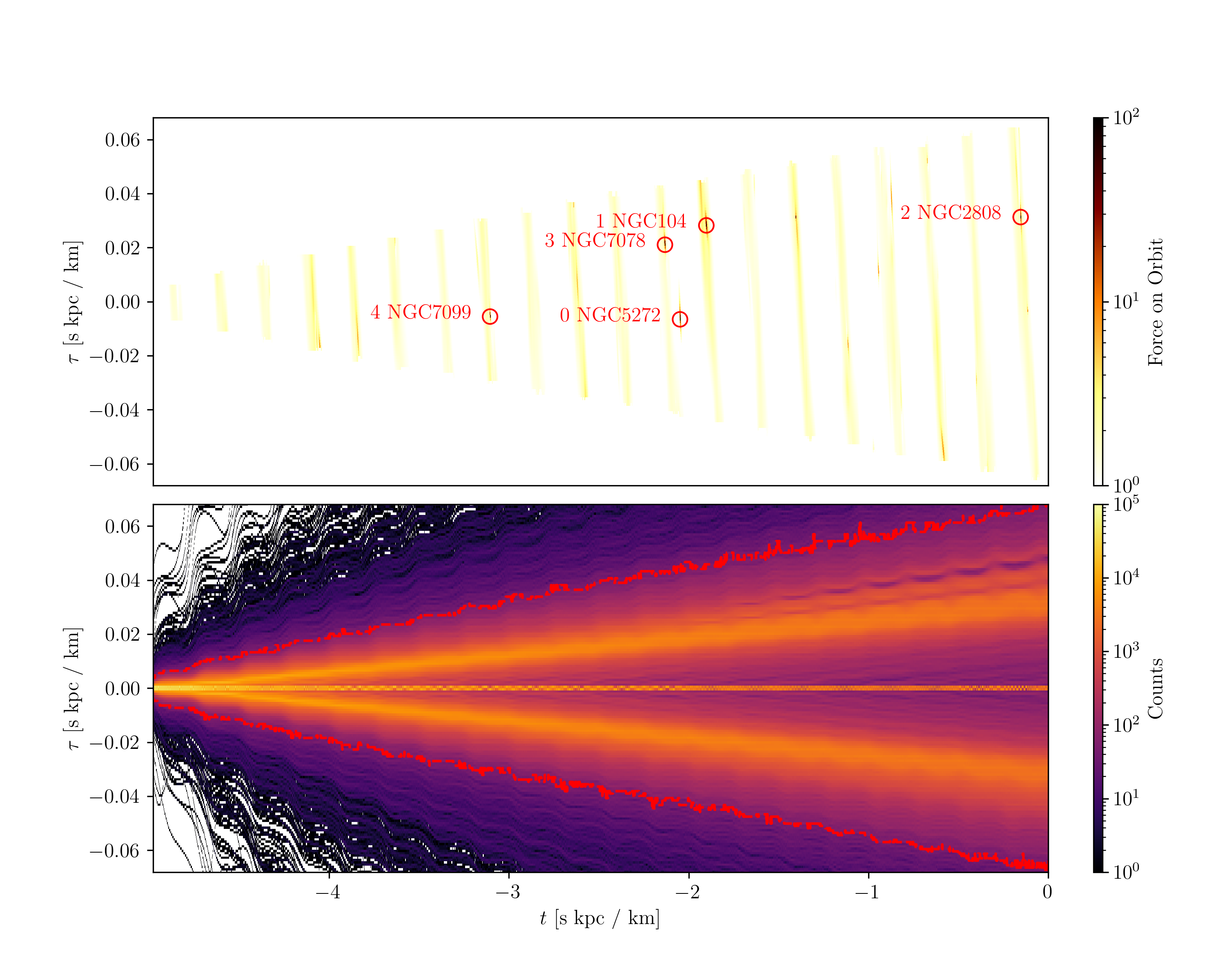}
      \caption{This figure demonstrates how we determined which globular clusters were responsible for the gaps. The y-axis is $\tau$, which is a coordinate in units of time that indicates how far ahead or behind it is from a globular cluster. The x-axis is the simulation time, $t$, where 0~s km~kpc$^{-1}$ indicates present time.  The bottom plot showcases the evolution of the stream density in simulation time. The density was used to determine a suitable length of the stream at a given function. This length was then used to extract a piece of Palomar~5's orbit. This orbital segment is used to approximate the stream. Then, the gravitational force from all other clusters was computed on the orbit. This is shown in the top plot, where the gravitational force is in measured in acceleration and is given in integration units: km$^2$~kpc$^{-1}$~s$^{-2}$. Moments of high acceleration indicate the passage of another cluster. The top 5 strongest passages are labeled with red circles as well as the name of the clusters. The example shown in this plot is the same simulation as Fig.~\ref{fig:stream_on_sky}.}
      \label{fig:force-on-orbit}
    \end{figure*}

    The core of our analysis is presented in Fig.~\ref{fig:force-on-orbit}. The bottom panel shows the evolution of the stream density over time. To avoid extreme low-density regions at the stream's edges, we applied the same density threshold as from Eq.~\ref{eq:density_threshold} to focus on the more significant areas of the stream. Next, we modeled Palomar~5's orbit as a proxy for its stream and sampled points along the orbit to measure the gravitational force exerted by other globular clusters. This force data is displayed in the top panel of Fig.~\ref{fig:force-on-orbit}, showing how the total gravitational acceleration on Palomar~5's stream evolves over its length throughout the simulation.

    We then used \texttt{scipy}'s \texttt{ndimage} \citep{2020NatMe..17..261V} package to identify the top five local maxima in the data space of gravitational acceleration $\vec{g}$ as a function of time $t$ and the stream coordinate $\tau$. This is done by first smoothing the image by taking a 5 point moving average kernel. Secondly, we use maximum filter to locate coordinates in the ($t,\tau$) data plane who are local maxima to at least 10-adjacent data points. These locations are then ordered and the top five strongest interactions are saved. Once these local maxima were identified, we iterated over the contributions of individual globular clusters to determine which cluster contributed the most to each peak in $\vec{g}$. Each significant peak was labeled with the corresponding globular cluster.

    Afterward, we cross-referenced these peaks with the locations of the gaps identified by studying the density maps and profiles from Fig.~\ref{fig:profiles}. For large gaps resulting from strong interactions, we observed that after an impact, a low-density wake is left behind in the ($t,\tau$) plane, which can be seen corresponding to the impacts of NGC~104 and NGC~7808 in the bottom panel of Fig.~\ref{fig:force-on-orbit}.

    Fig.~\ref{fig:force-on-orbit} contains some interesting information. Notice the periodic ribbons of force in the $(t,\tau)$ plane. This is due to pericenter passages where Palomar~5 is getting closer to the center of mass of the globular cluster system. Additionally, for the impacts of NGC~104 and NGC~C7078, wakes can be observed in the density map. Another important aspect is that the strongest peak in gravitational force does not necessarily create a gap. Notice how NGC~5272, which was labeled with 0 to indicate that it has the greatest local maxima, does not have a gap. The reason for this is manifold, for instance, the force needs to be modulated with the time since the change in momentum is the determining factor and not the peak magnitude of the force.  Additionally, there is an offset of about 200~pc between the stream and the orbit, as seen when viewing the stream in tail coordinates, so peaks upon the orbit are good proxies for the stream but are not definitive. We found that the top five greatest impacts accounted for all gaps, except for Sampling~014 as shown in Fig.~\ref{fig:gallery3}, whose gap from NGC~6584 corresponded to the 7th peak. 

    The results of this analysis were compiled into a table. If a gap was attributed to a particular globular cluster, it was labeled as \texttt{TRUE}; otherwise, it was marked as \texttt{FALSE}. For a handful of simulations, to double check that the verdict made correctly when passing from suspect to culprit, we re-compute the simulations yet individually adding one globular cluster at a time. As a result, we are able to confirm that singular gaps arise from the suspected clusters, an example of which is shown in Fig.~\ref{fig:decomposition}.

    \section{Reconstruction of the impact geometry} \label{sec:reconstruction}

    As discussed in Sect~\ref{sect:geometry},  five parameters determine the change in velocity of a given stream particle: $M$, $r_p$, $b$, $W_\parallel$, and $W_\perp$. In the following, we describe how we estimated these parameters during impacts in our simulations.
    To achieve this, we begin by identifying the approximate moments of impact from the most significant clusters, as determined in the previous analysis in Sec.~\ref{sec:Perturber_Identification}. Then, we refine these estimates in order to to pinpoint both the exact location of the impact along the stream and the precise moment it occurred. To do so, we fit a third-order parametric polynomial to the stream, using the saved snapshots from our simulations:
      \begin{equation}
        \vec{s}(\tau) = 
        \left\{
          \begin{aligned}
            x(\tau) &= a_0 + a_1 \tau + a_2 \tau^2 + a_3 \tau^3 \\ 
            y(\tau) &= b_0 + b_1 \tau + b_2 \tau^2 + b_3 \tau^3 \\
            z(\tau) &= c_0 + c_1 \tau + c_2 \tau^2 + c_3 \tau^3
          \end{aligned}
        \right.
        \end{equation}  
      where $x$, $y$, and $z$ represent the parametric line describing the stream in galactocentric coordinates, $\tau$ is the stream coordinate in time as described in the Appendix~\ref{sec:Perturber_Identification}, and is used as the independent variable to parameterize the position along the stream. The coefficients $a_i$, $b_i$, and $c_i$ are the polynomial coefficients. We found that a second-order polynomial was insufficient to capture the curvature along the full length of the stream, with divergence at the ends of the tails. A third-order polynomial was sufficient and desirable, as it is the lowest order that adequately captures the path of the stream over the entire length under consideration.

      In this analysis, only one side of the stream is considered. For instance, if the impact candidate was in the leading tail, only the star particles with $\tau > 0$ are used to constrain the stream track. The polynomial coefficients were determined through a minimization method using the Nelder-Mead algorithm from \texttt{scipy}'s optimization package.

      Since the simulation snapshots were saved at a temporal resolution of 1 Myr--—rather than at the integration time-step, which would have generated an excessive amount of data—--we interpolate between snapshots to more precisely estimate the impact geometry. This is achieved by fitting the stream at five time-steps surrounding the approximate impact time, a period of 5 Myr, which sufficiently covers the interaction time. The interaction time can be estimated as $t \approx \frac{100~\text{pc}}{300~\frac{\text{km}}{\text{s}}} \approx 0.3~\text{Myr}$.

      As a result, each polynomial coefficient can now be expressed as a function of time. Consequently, we can parameterize the stream as a function of both simulation time and position along the stream:
      \begin{equation}
        \vec{s}(t,\tau) = 
        \left\{
        \begin{aligned}
          x(t,\tau) &= a_0(t) + a_1(t)\tau + a_2(t) \tau^2 + a_3(t)\tau^3 \\ 
          y(t,\tau) &= b_0(t) + b_1(t)\tau + b_2(t) \tau^2 + b_3(t)\tau^3 \\
          z(t,\tau) &= c_0(t) + c_1(t)\tau + c_2(t) \tau^2 + c_3(t)\tau^3.
          \end{aligned}
        \right.
      \end{equation}
      The values of the coefficients as a function of time are obtained through linear interpolation, ensuring that the coefficients at the snapshot times match the values constrained by the simulation data.

      Next, we fit the trajectory of the perturber with a second-order polynomial. With equations for both the stream and the perturber as functions of time, we identify the time and location of impact by minimizing a cost function, defined as the distance between the stream and the perturber:
      \begin{equation} 
        b(t, \tau) = \left\lVert \vec{s}(t, \tau) - \vec{p}(t) \right\rVert, 
        \end{equation}
      where $\vec{s}(t, \tau)$ is the galactocentric position of a point on the stream, $\vec{p}(t)$ is the position of the perturber, and $b$ is the distance between the two. The minimum value of $b$, denoted as $\text{min}(b)$, represents the impact parameter. The minimization is carried out using \texttt{scipy}'s optimization package with the \textit{L-BFGS-B} method, which allows us to place bounds on $t$ and $\tau$, ensuring no extrapolation occurs \citep{davidon1991variable}.

      Once this minimization is performed, determining the relative velocity becomes straightforward. Since the minimization provides the impact parameter, time of impact, and corresponding value of $\tau$, we can compute the derivatives of the parametric equations at $t_{\text{min}}$ and $\tau_{\text{min}}$. The parallel and perpendicular components of the perturber's velocity relative to the stream are given by:
      \begin{equation}
        \begin{aligned}
          \delta \vec{v} &=\vec{v_p} - \vec{v_s} \\
          w_\parallel &= \left(\delta \vec{v}\right)\cdot \hat{v_s}\\  
          w_\perp &=  \sqrt{\Delta v ^2 - w_\parallel ^ 2}
          \end{aligned}
        \end{equation}
      where $\vec{v}_p$ and $\vec{v}_s$ are the velocities of the perturber and the stream, respectively. For each of the 50 simulations, this information was computed for the strongest 5 flybys of a perturber with the stream. Thus, we compute a sample of 250 impacts, and flag those that give rise to gaps. We refer the reader to Sect.~\ref{sect:geometry} for the presentation and discussion of the results.

  \section{Gallery of Gaps} \label{sec:gallery_of_gaps}
\begin{figure*}
      \centering
      \includegraphics[width=\linewidth]{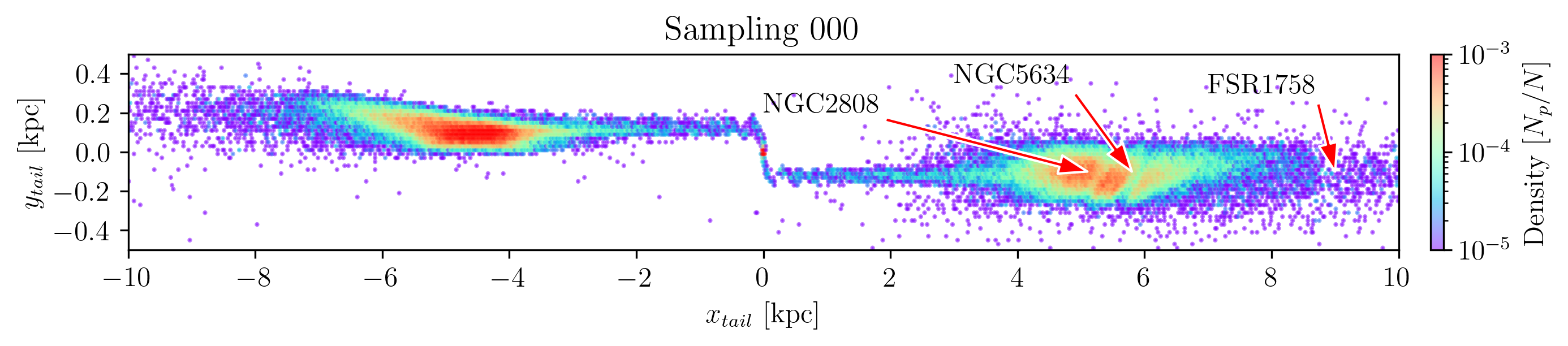}
      \includegraphics[width=\linewidth]{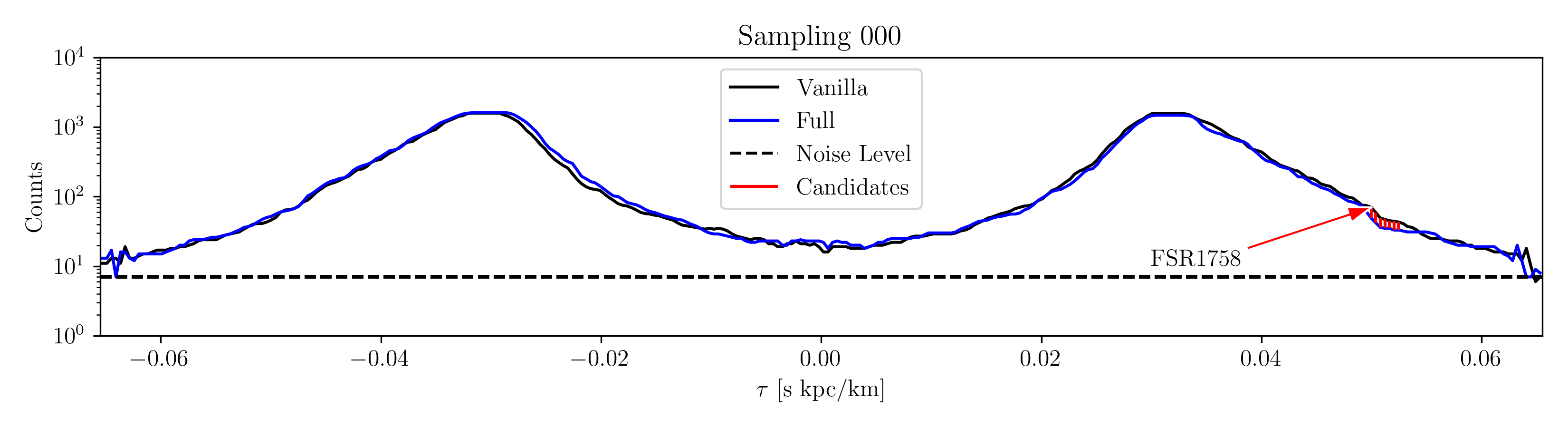}
      \includegraphics[width=\linewidth]{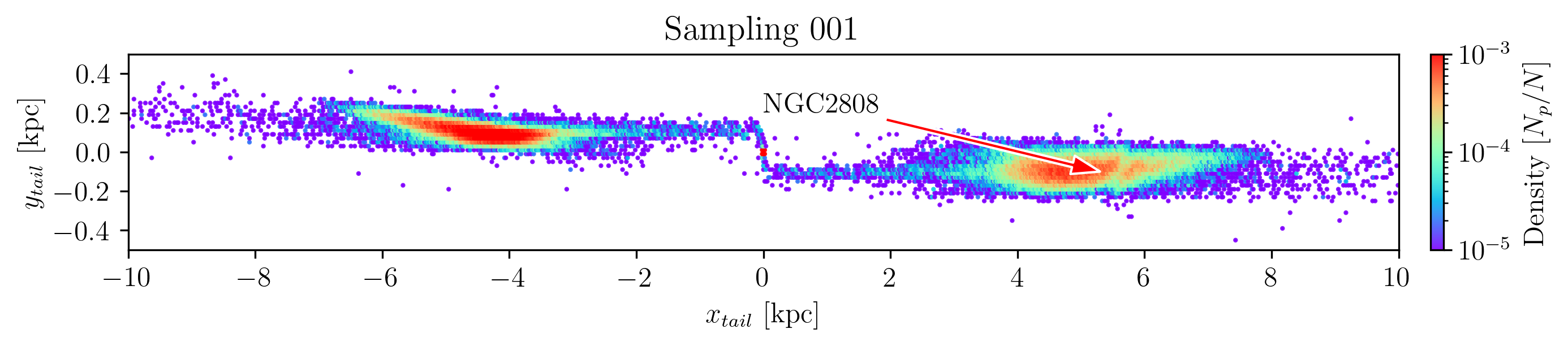}
      \includegraphics[width=\linewidth]{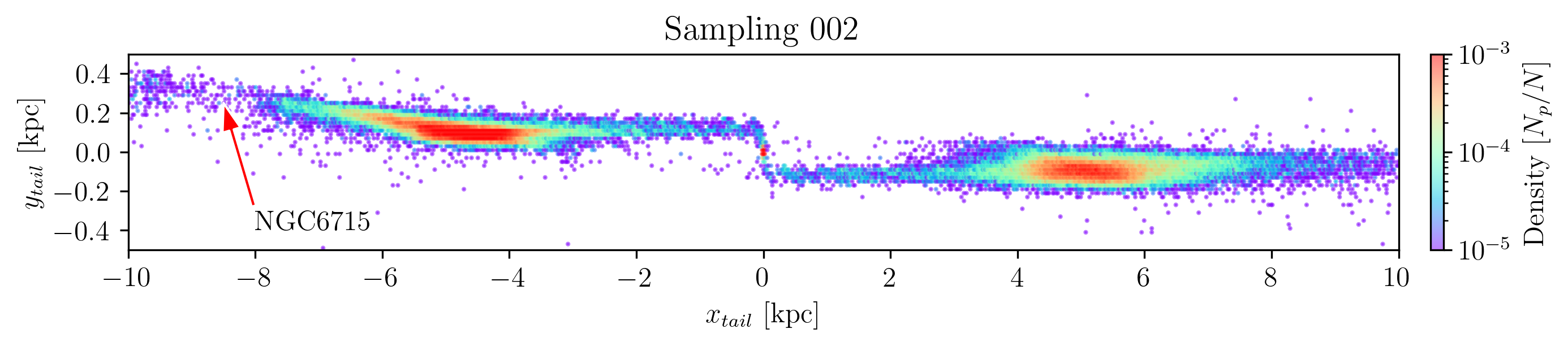}
      \includegraphics[width=\linewidth]{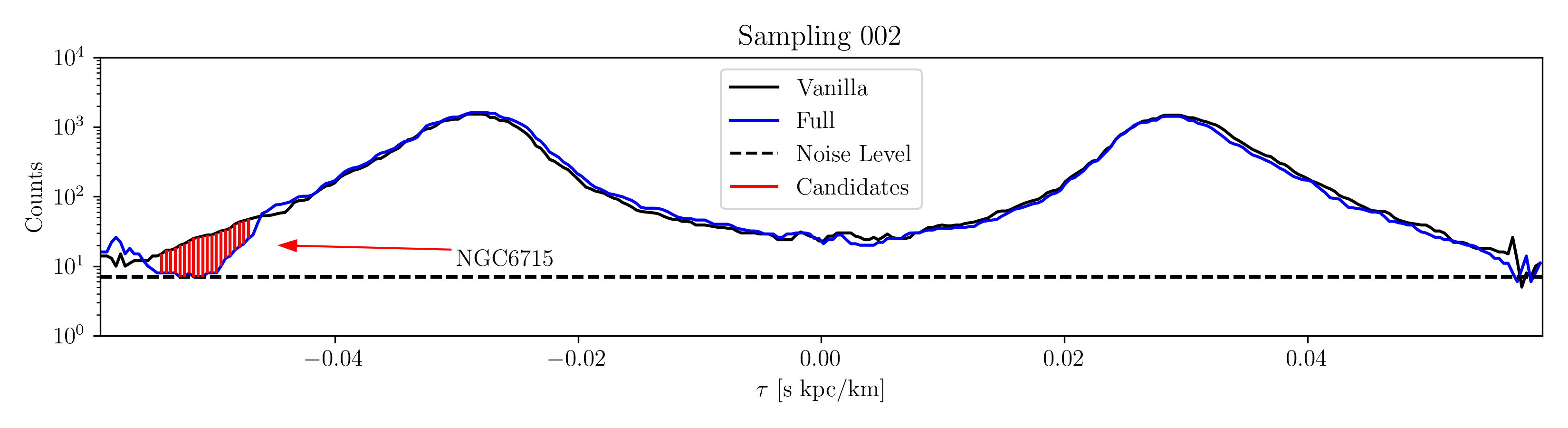}
      \caption{First gap gallery. See the text in Appendix~\ref{sec:gallery_of_gaps} for a description.}
      \label{fig:gallery0}
      \end{figure*}

\begin{figure*}
      \centering
      \includegraphics[width=\linewidth]{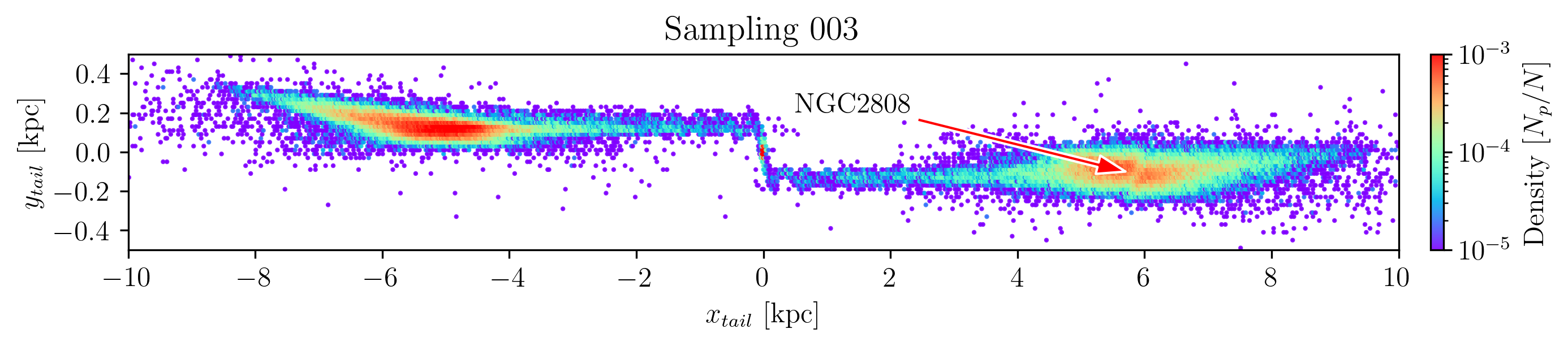}      
      \includegraphics[width=\linewidth]{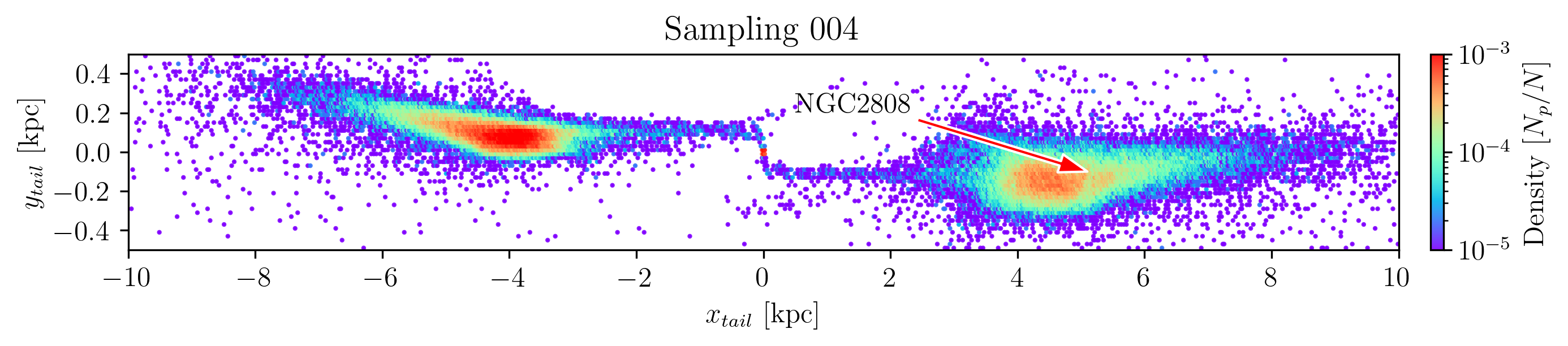}
      \includegraphics[width=\linewidth]{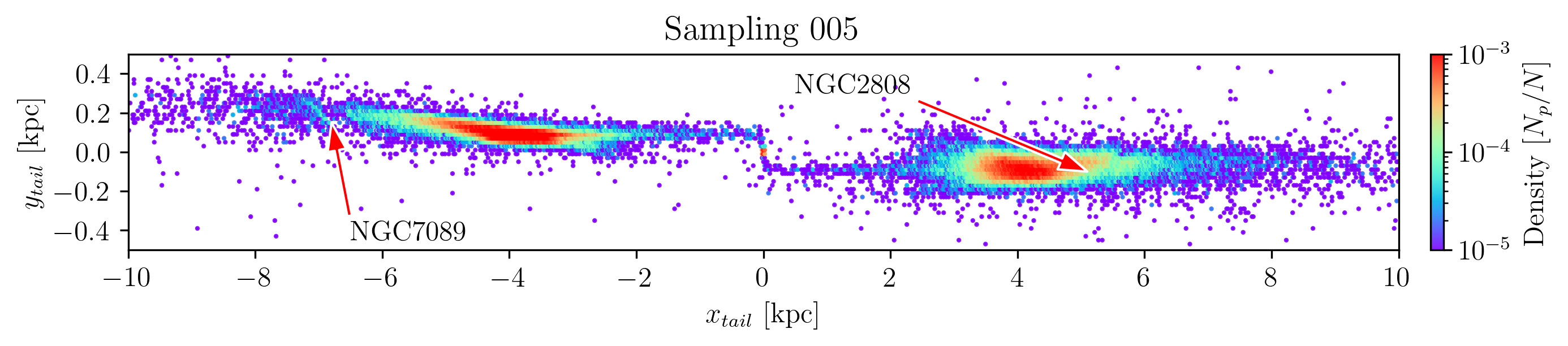}
      \includegraphics[width=\linewidth]{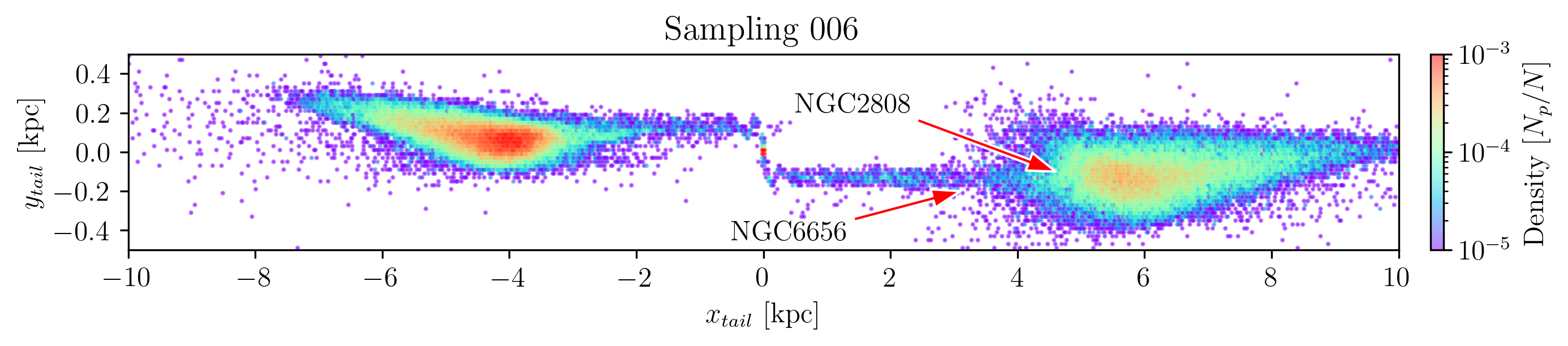}
      \includegraphics[width=\linewidth]{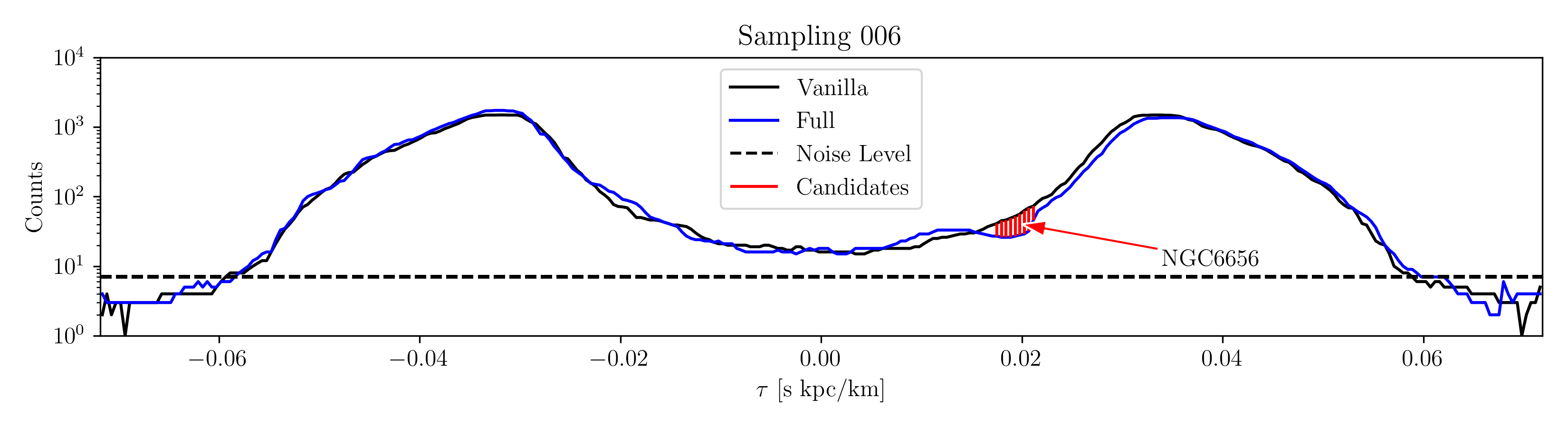}
      \caption{Second gap gallery}
      \label{fig:gallery1}
      \end{figure*}

\begin{figure*}
      \centering
      \includegraphics[width=\linewidth]{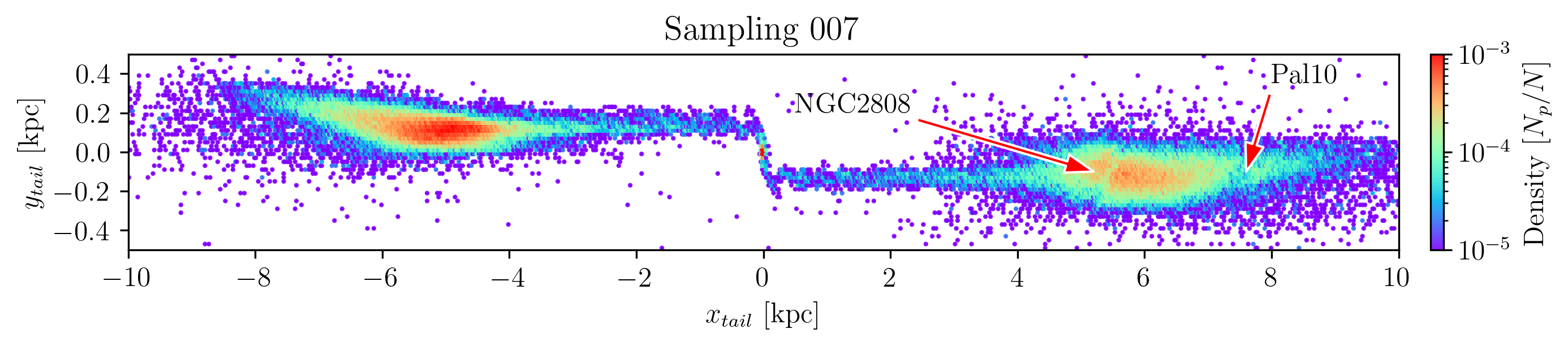}
      \includegraphics[width=\linewidth]{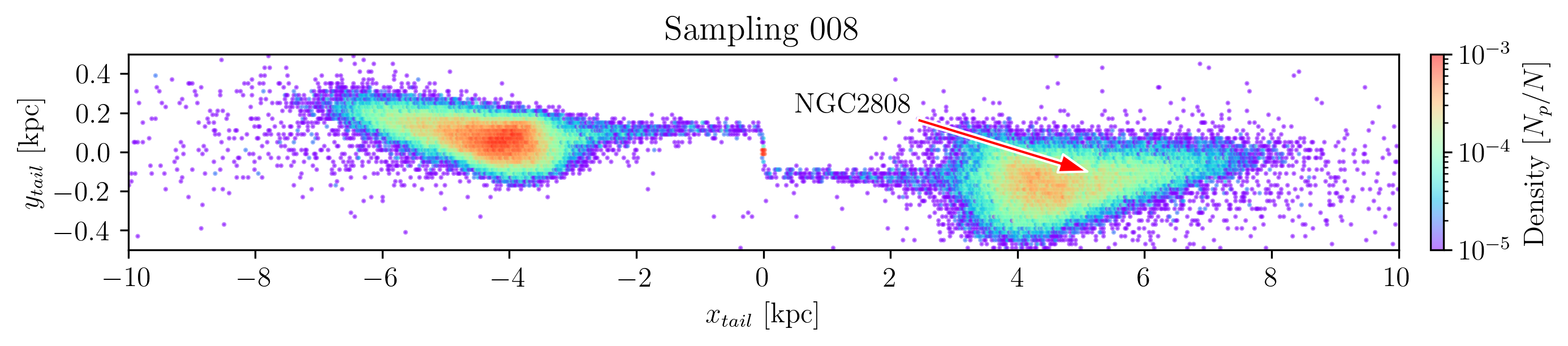}
      \includegraphics[width=\linewidth]{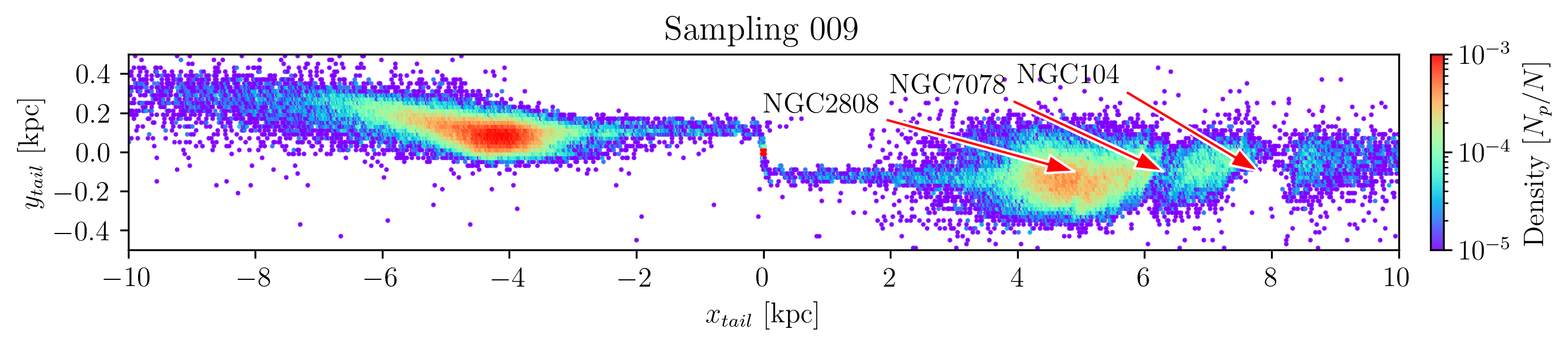}      
      \includegraphics[width=\linewidth]{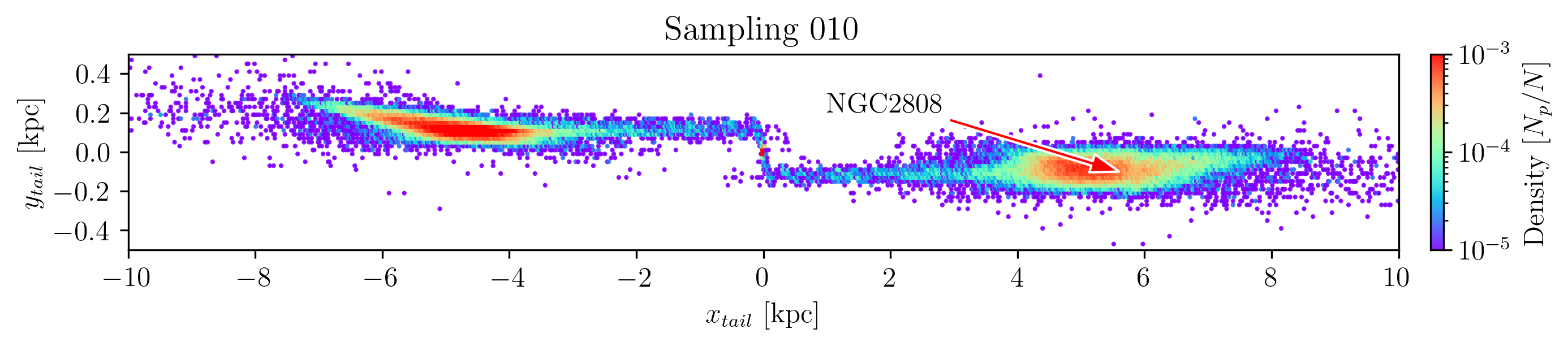}
      \includegraphics[width=\linewidth]{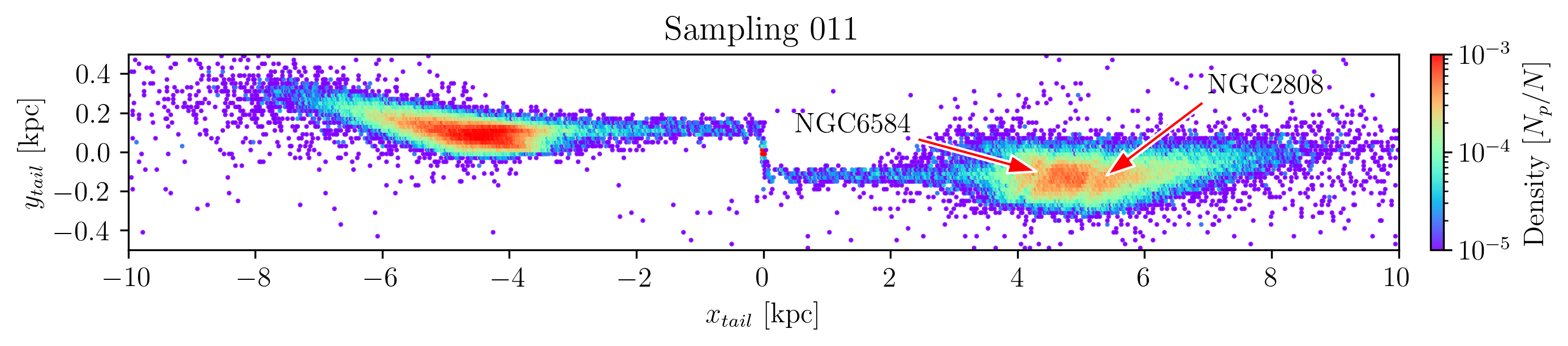}
      \caption{Third gap gallery}
      \label{fig:gallery2}
      \end{figure*}

\begin{figure*}
      \centering
      \includegraphics[width=\linewidth]{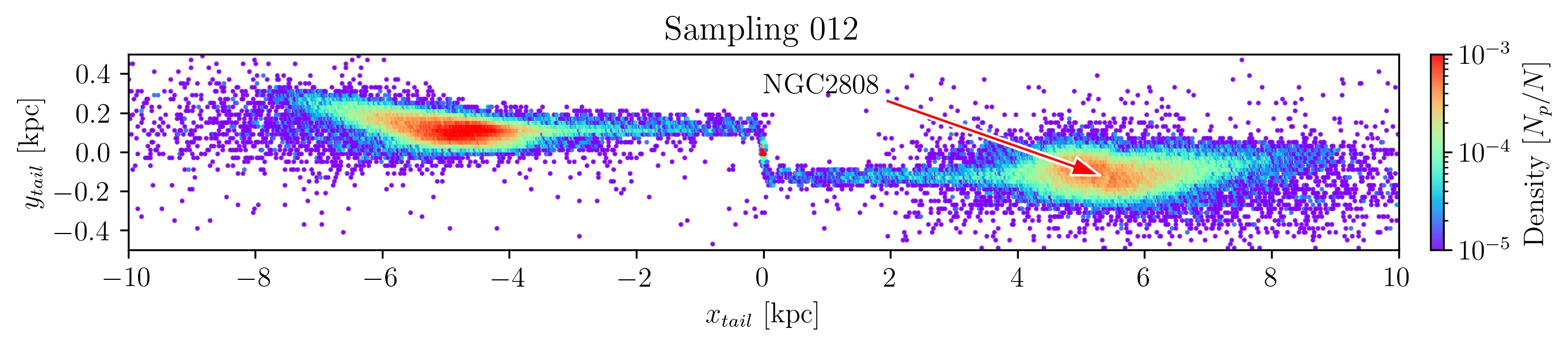}
      \includegraphics[width=\linewidth]{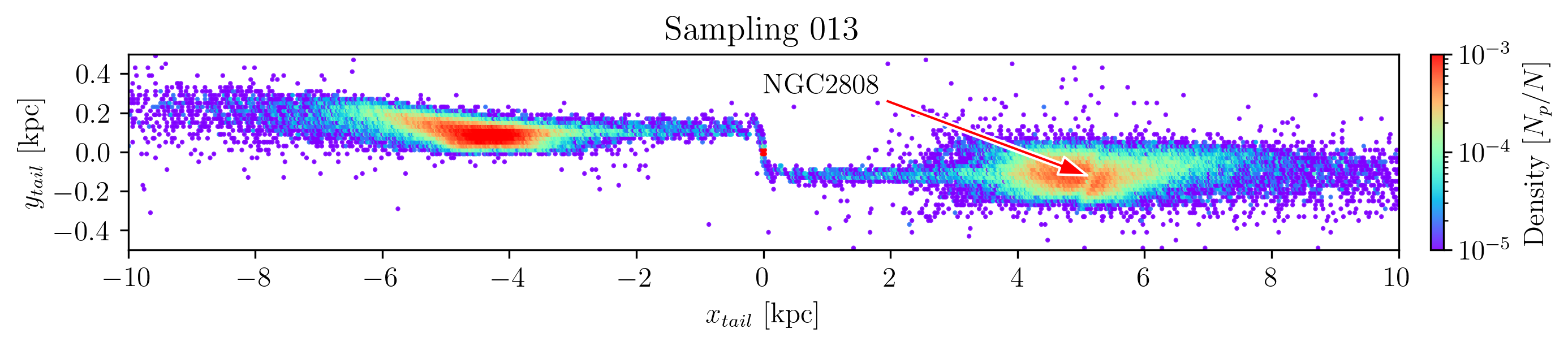}
      \includegraphics[width=\linewidth]{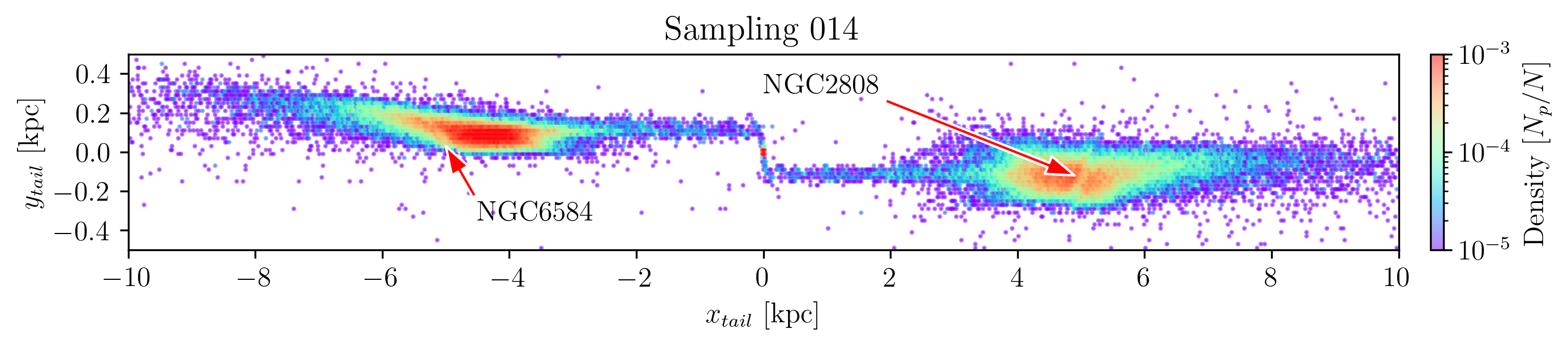}      
      \includegraphics[width=\linewidth]{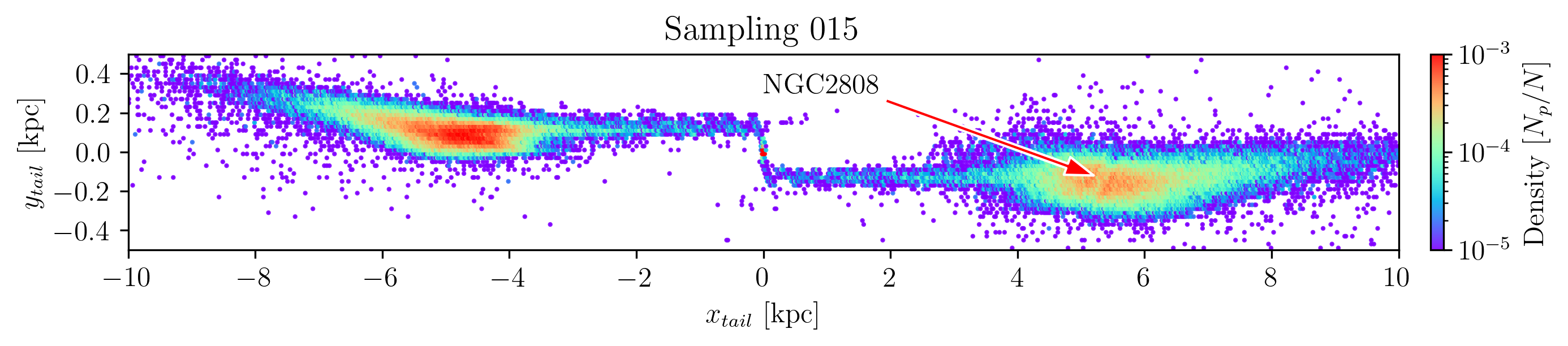}
      \includegraphics[width=\linewidth]{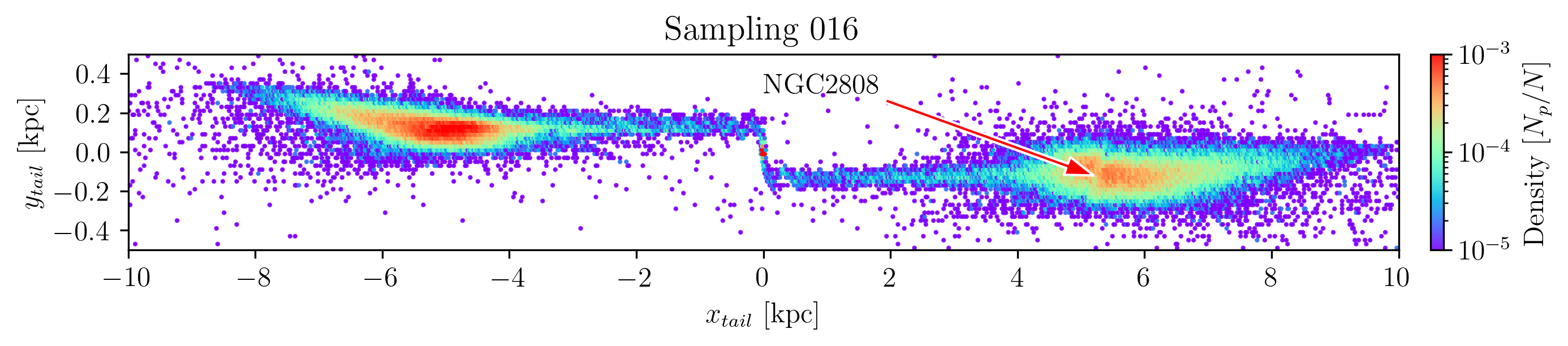}
      \caption{Fourth gap gallery}
      \label{fig:gallery3}
      \end{figure*}

\begin{figure*}
      \centering
      \includegraphics[width=\linewidth]{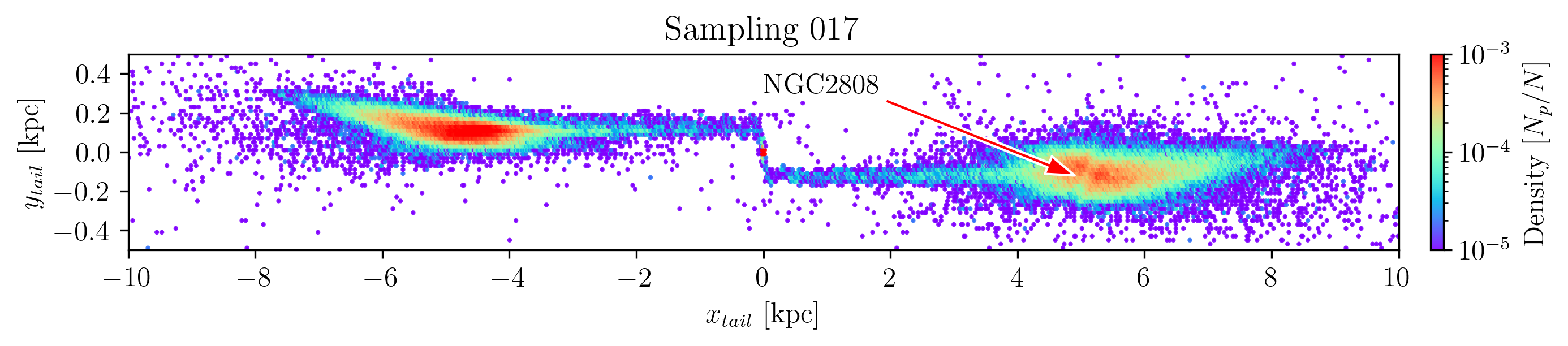}
      \includegraphics[width=\linewidth]{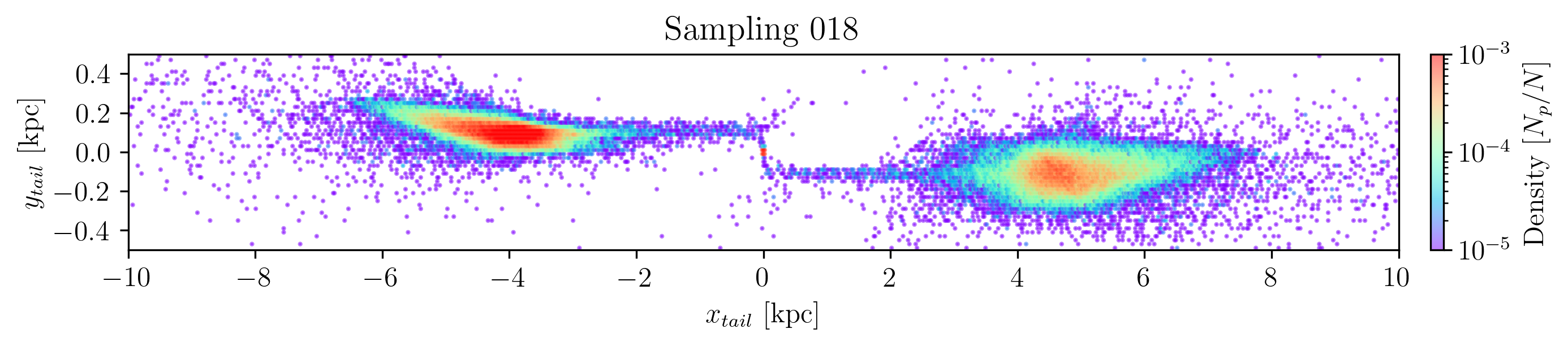}
      \includegraphics[width=\linewidth]{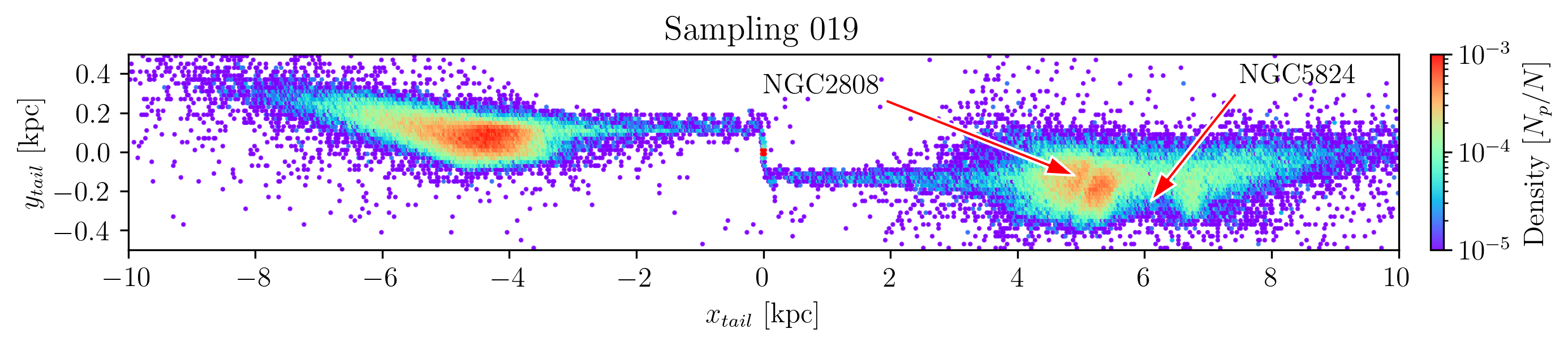}      
      \includegraphics[width=\linewidth]{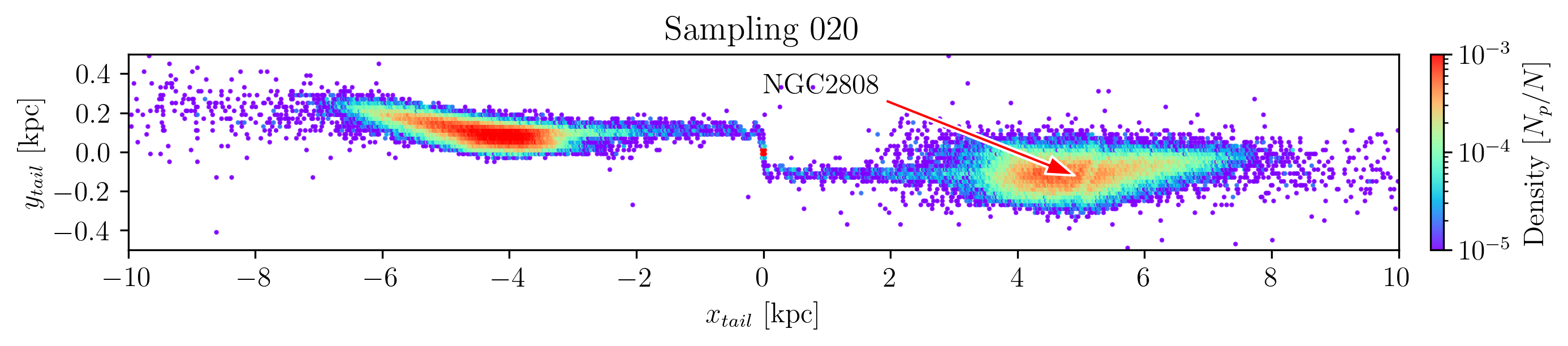}
      \includegraphics[width=\linewidth]{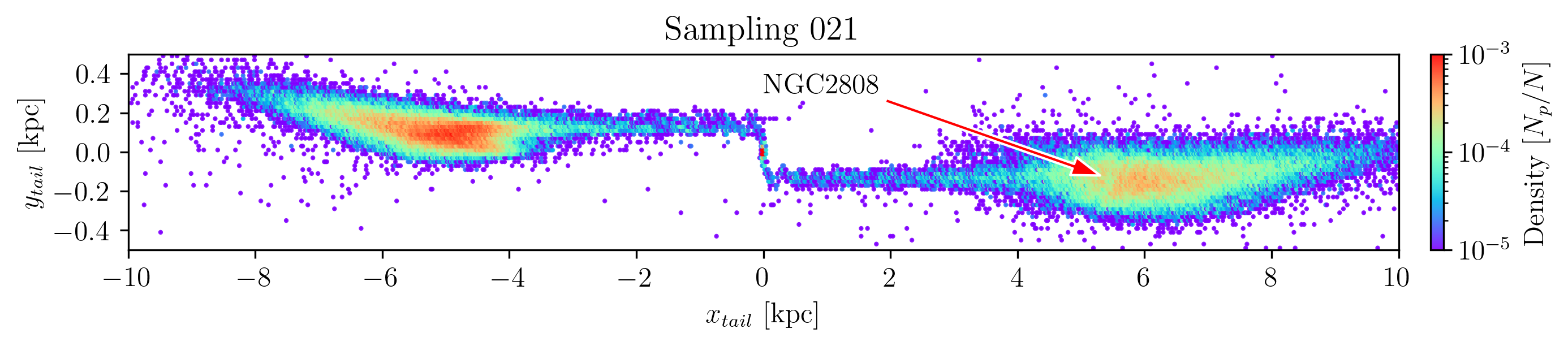}
      \caption{Fifth gap gallery}
      \label{fig:gallery4}
      \end{figure*}

\begin{figure*}
      \centering
      \includegraphics[width=\linewidth]{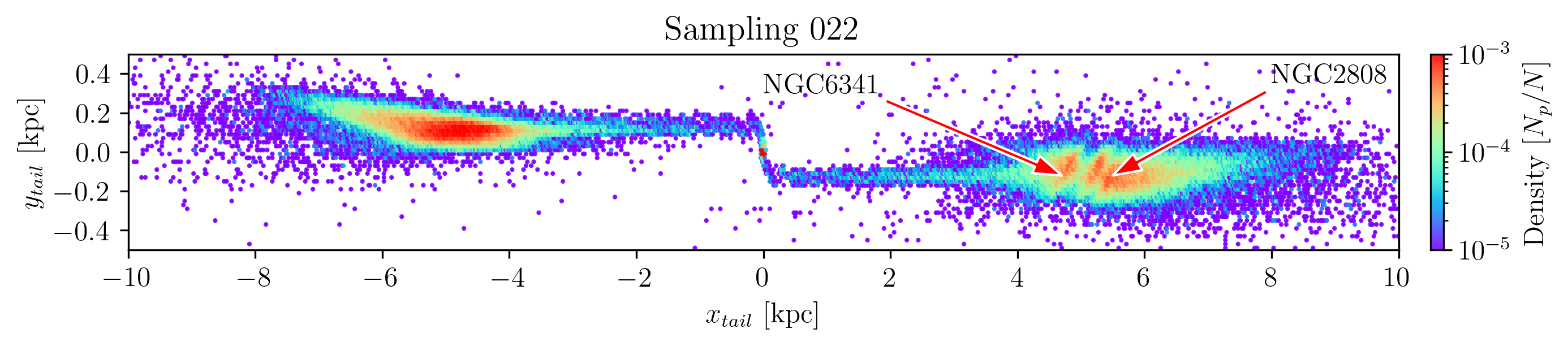}
      \includegraphics[width=\linewidth]{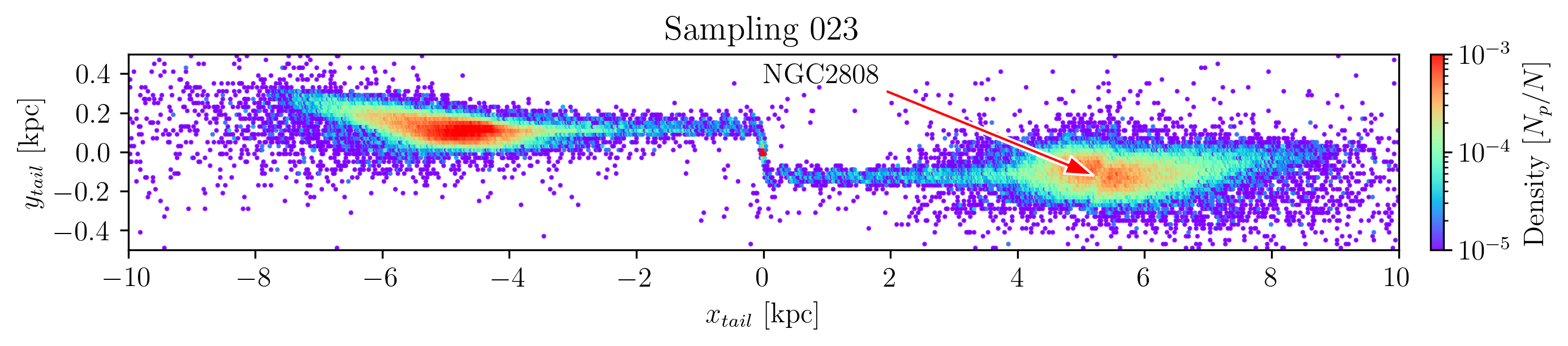}
      \includegraphics[width=\linewidth]{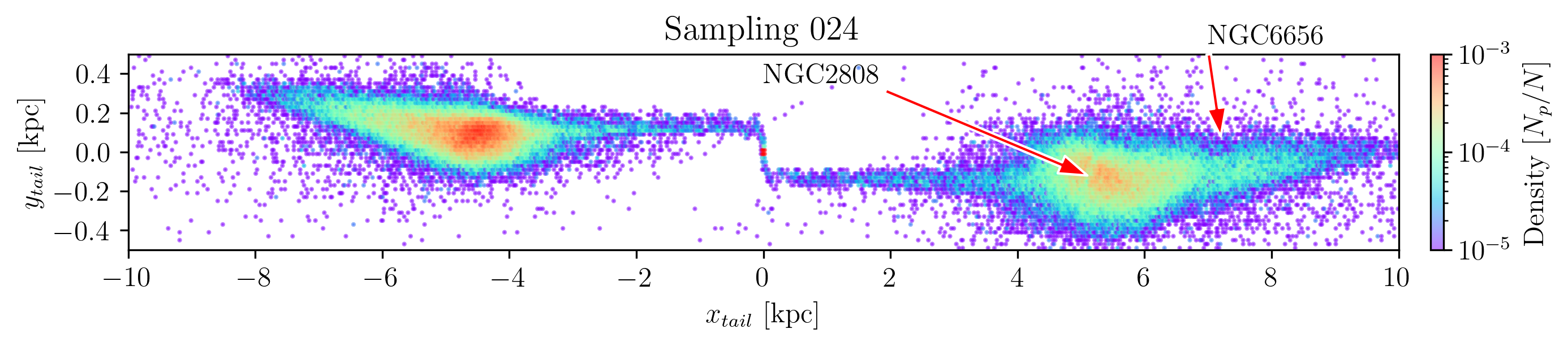}   
      \includegraphics[width=\linewidth]{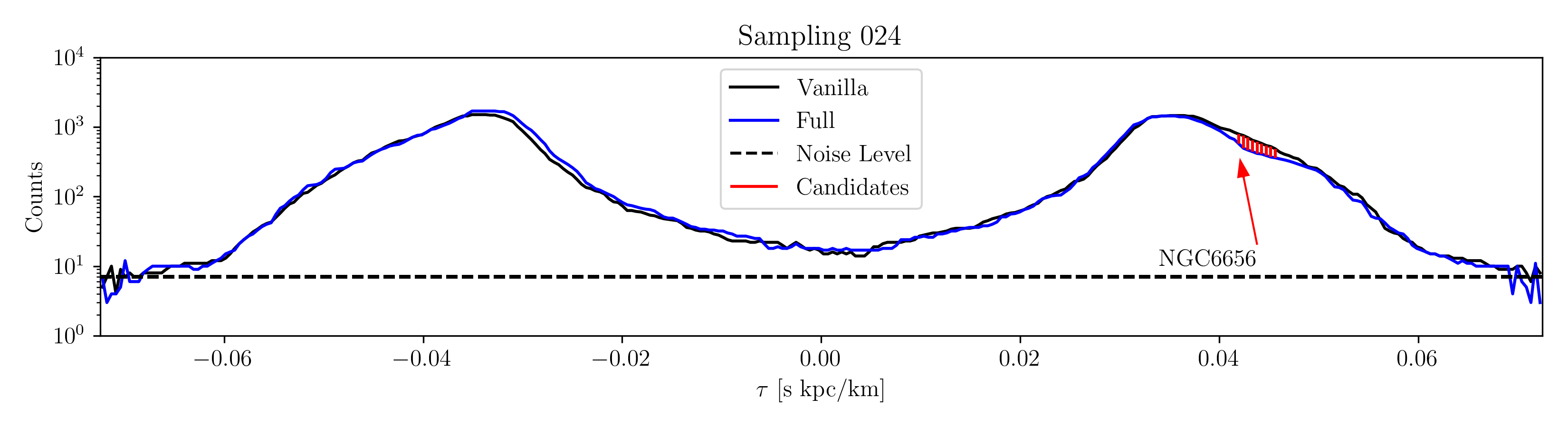}
      \includegraphics[width=\linewidth]{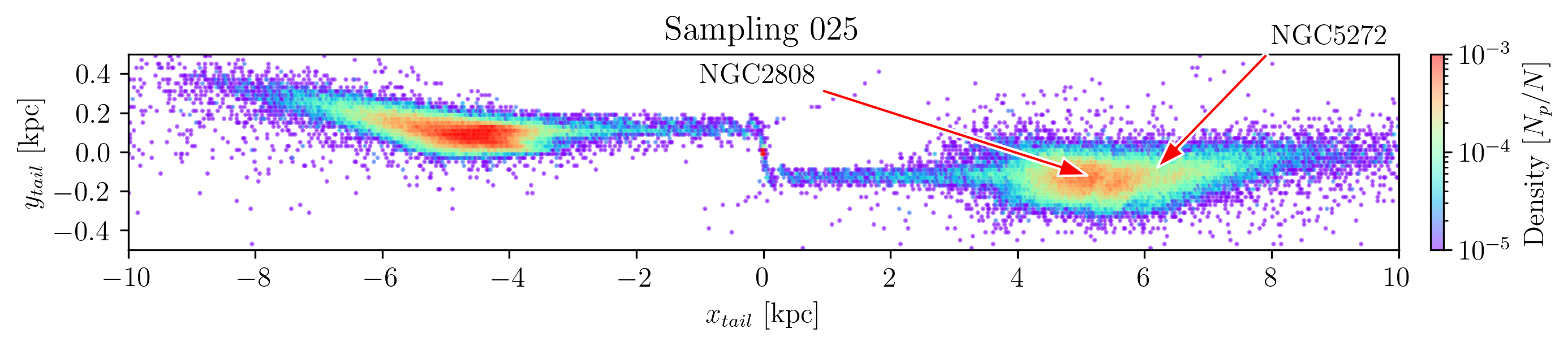}
      \caption{Sixth gap gallery}
      \label{fig:gallery5}
      \end{figure*}

\begin{figure*}
      \centering
      \includegraphics[width=\linewidth]{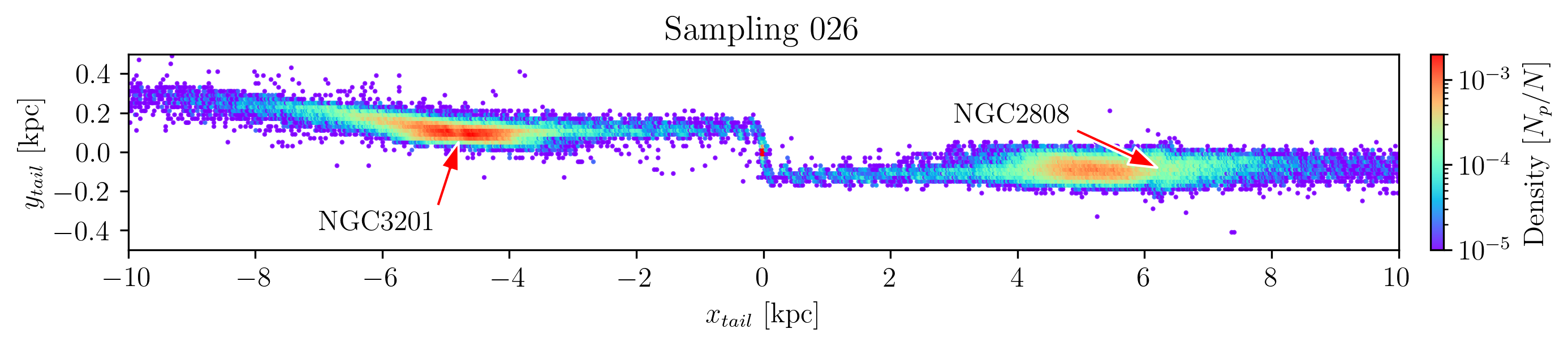}
      \includegraphics[width=\linewidth]{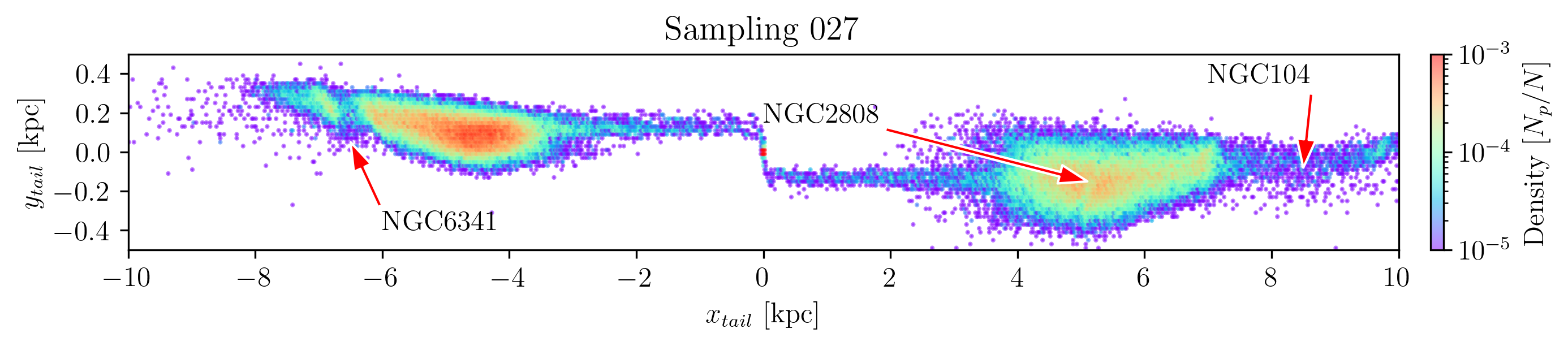}
      \includegraphics[width=\linewidth]{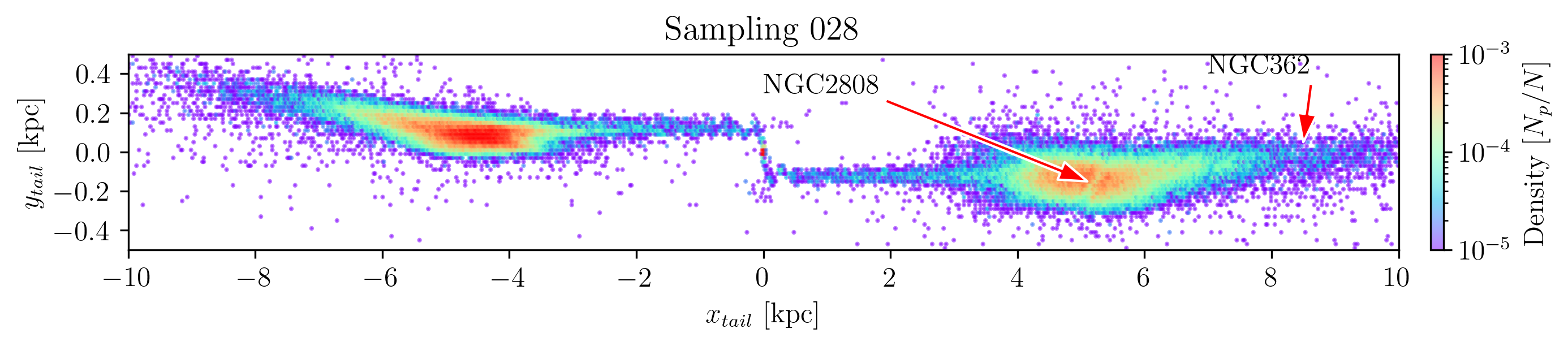}
      \includegraphics[width=\linewidth]{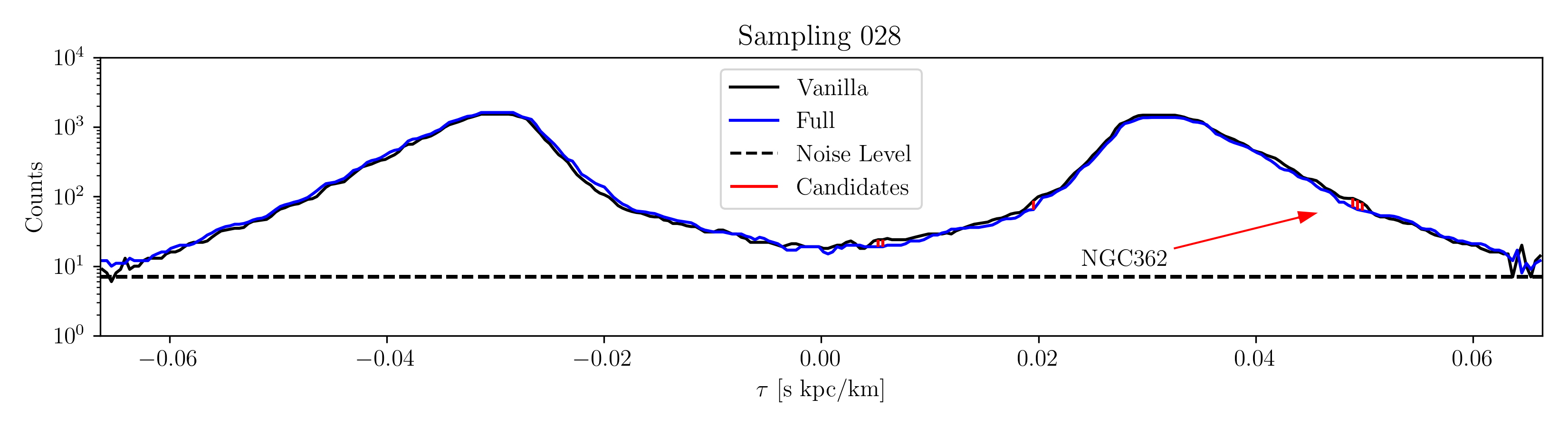}
      \caption{Seventh gap gallery}
      \label{fig:gallery6}
      \end{figure*}

\begin{figure*}
      \centering      
      \includegraphics[width=\linewidth]{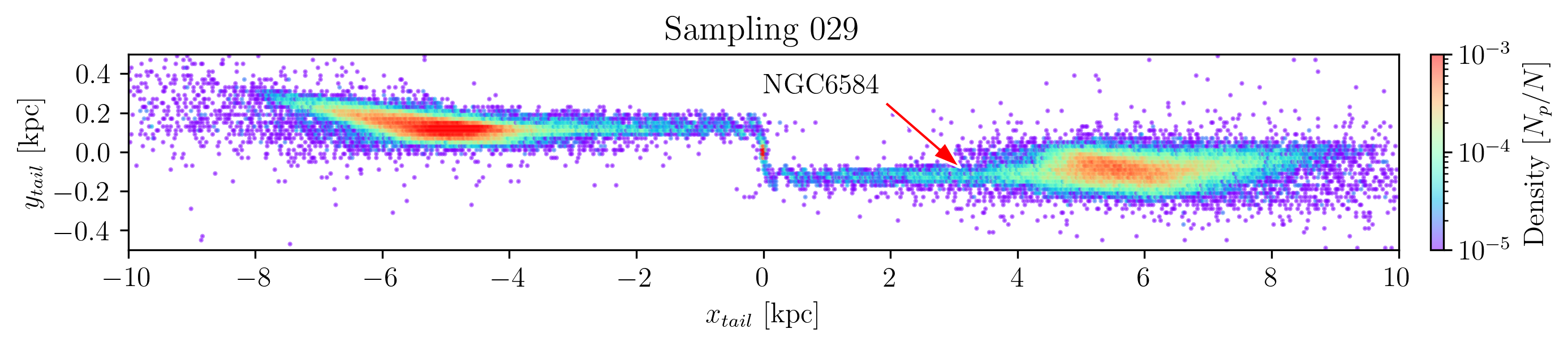}    
      \includegraphics[width=\linewidth]{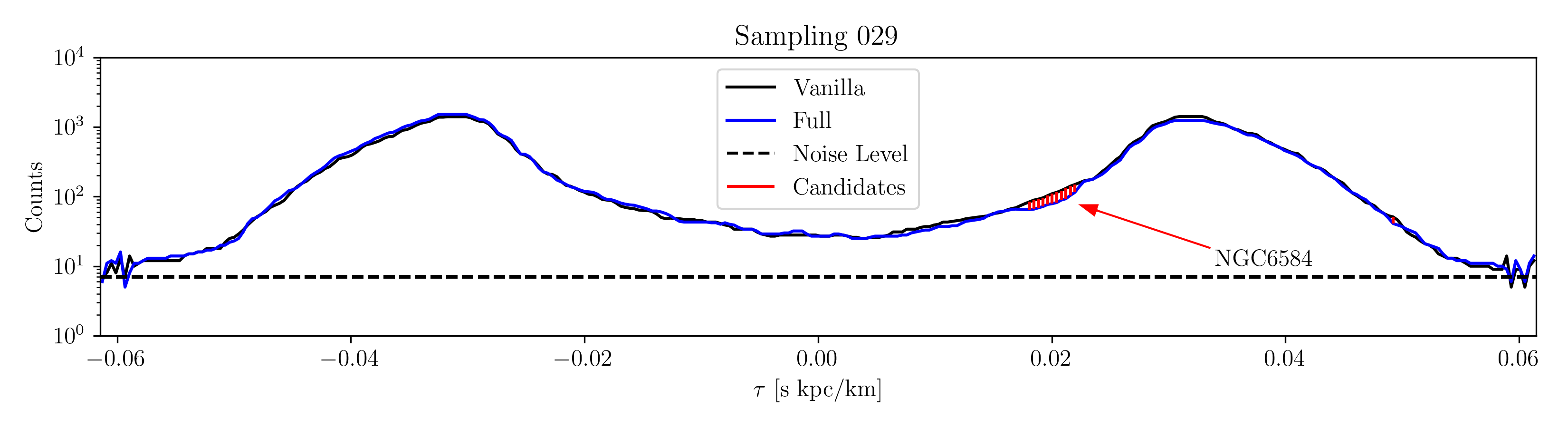}  
      \includegraphics[width=\linewidth]{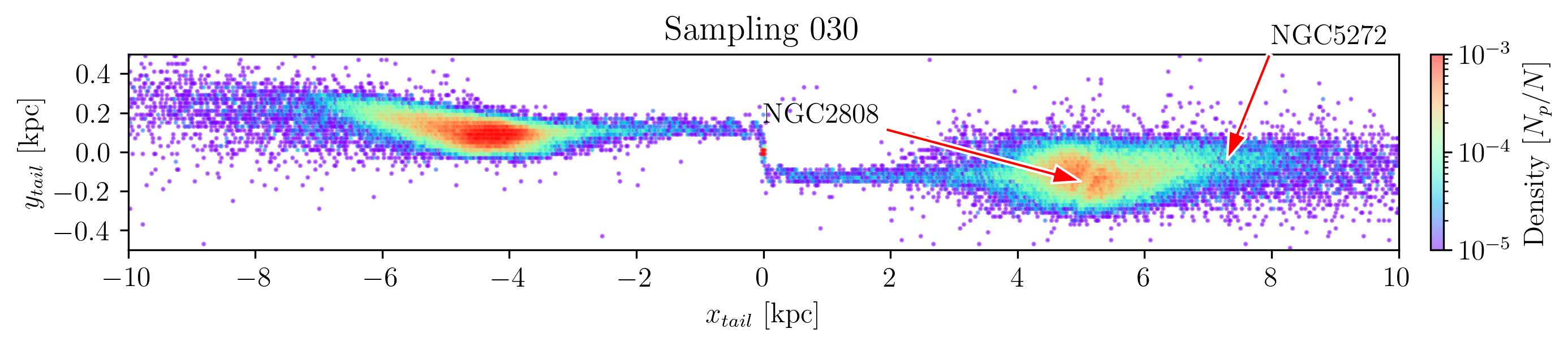}
      \includegraphics[width=\linewidth]{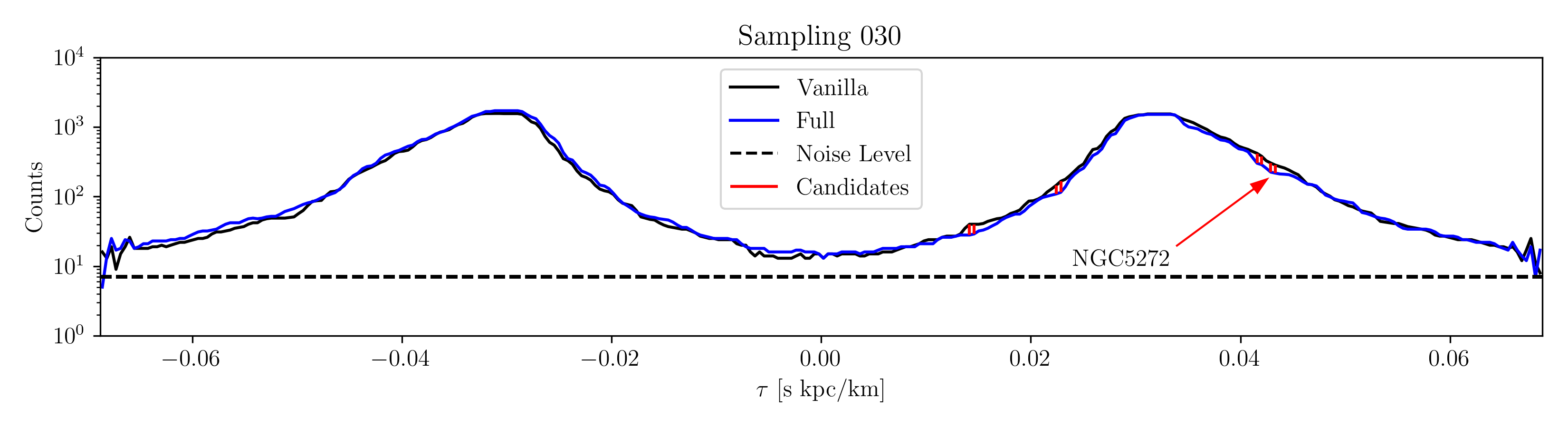}
      \includegraphics[width=\linewidth]{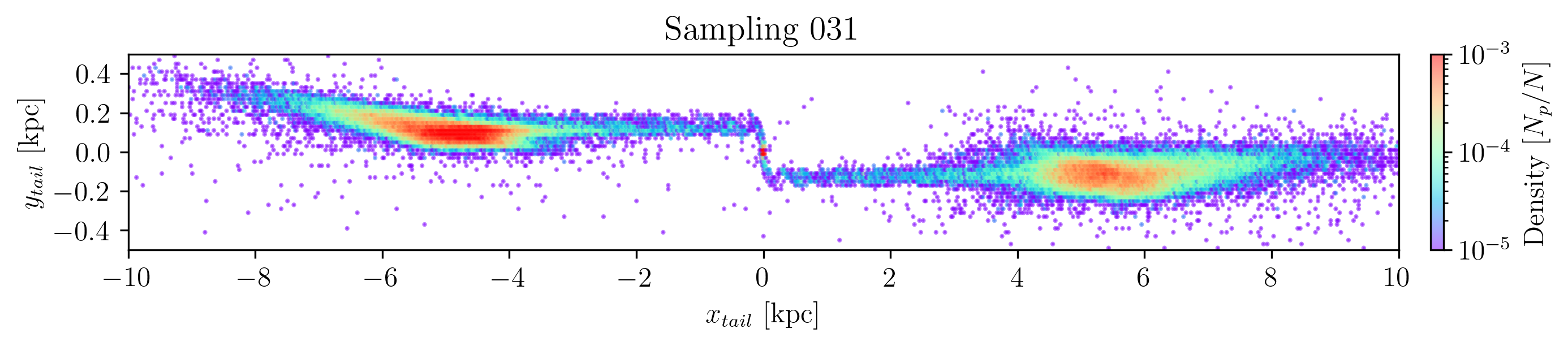}
      \caption{Eighth gap gallery}
      \label{fig:gallery7}
      \end{figure*}

\begin{figure*}
      \centering
      \includegraphics[width=\linewidth]{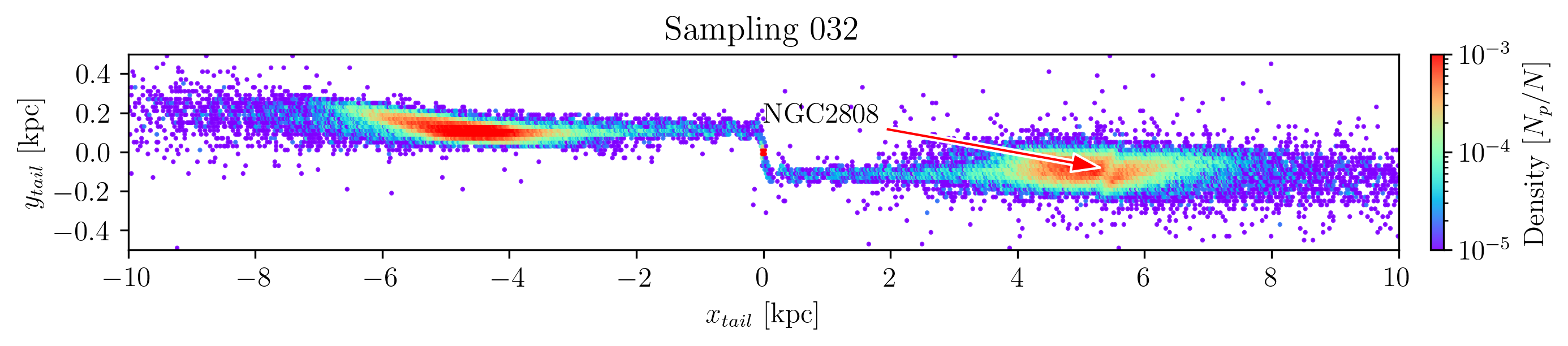}
      \includegraphics[width=\linewidth]{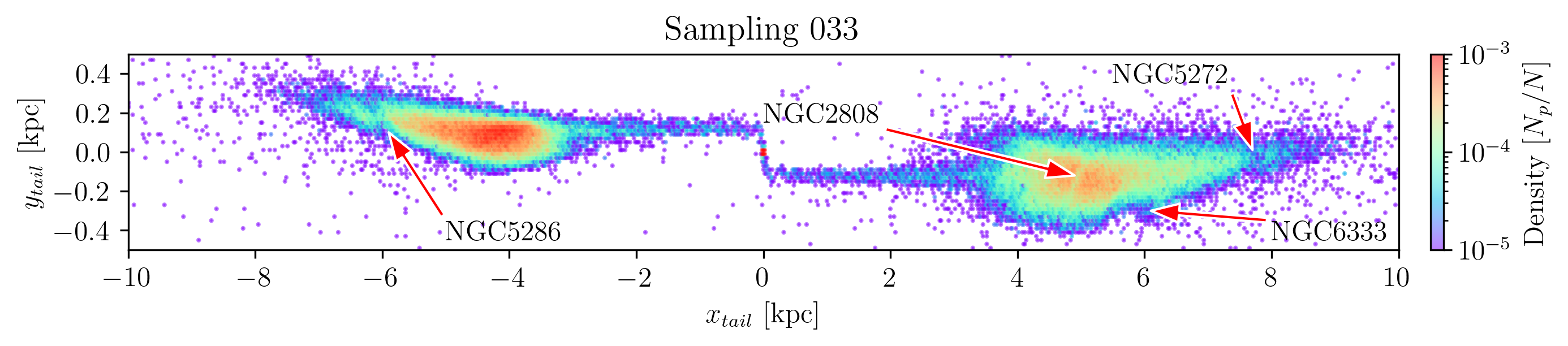}
      \includegraphics[width=\linewidth]{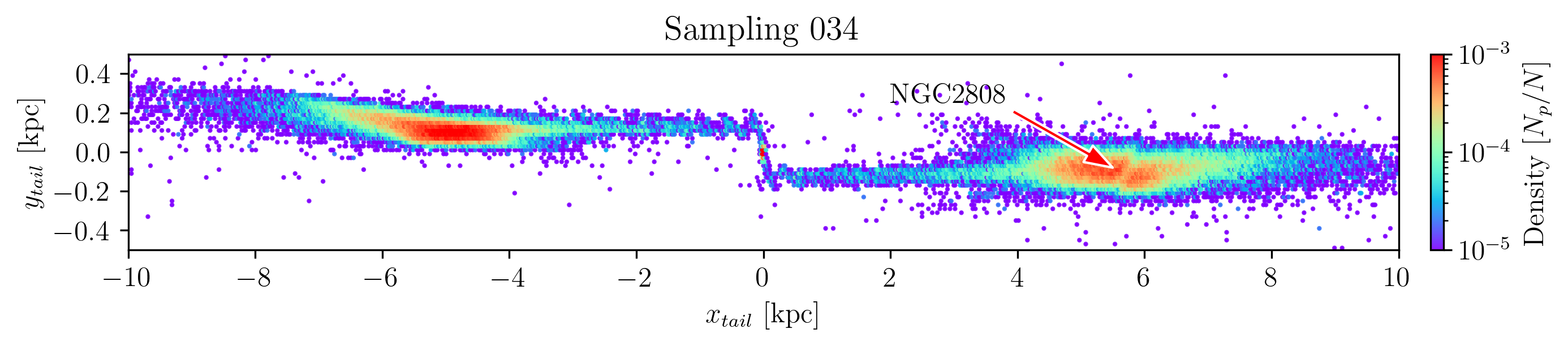}      
      \includegraphics[width=\linewidth]{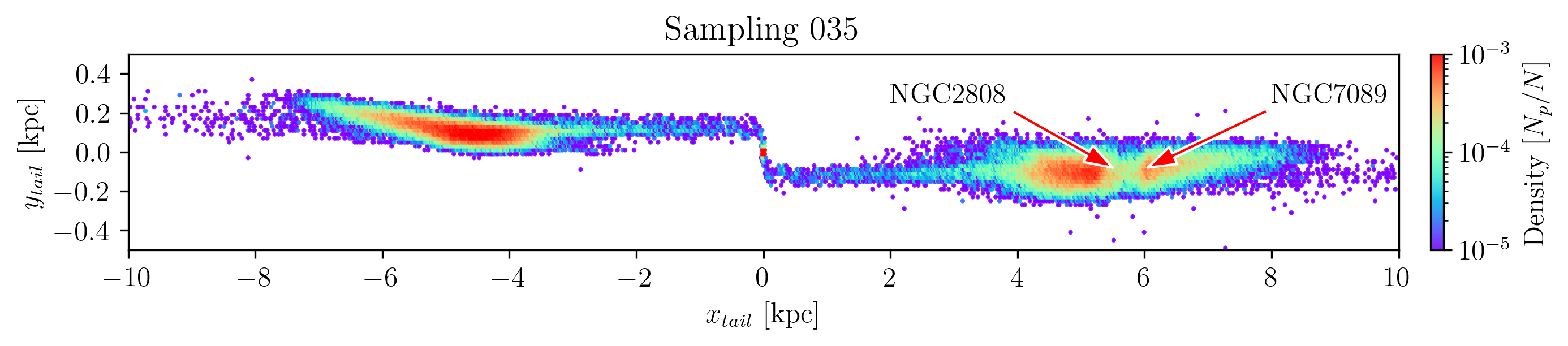}
      \includegraphics[width=\linewidth]{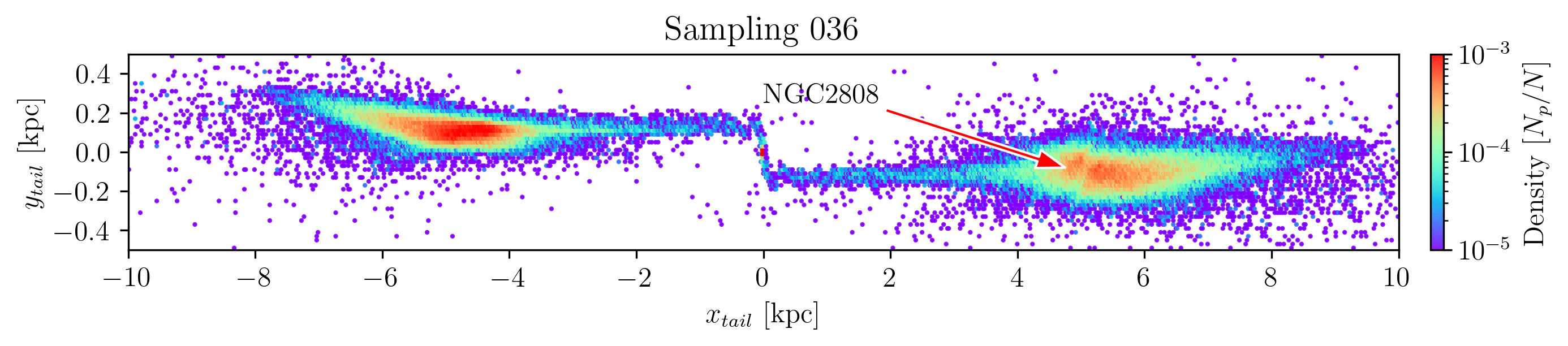}      
      \caption{Ninth gap gallery}
      \label{fig:gallery8}
      \end{figure*}

\begin{figure*}
      \centering
      \includegraphics[width=\linewidth]{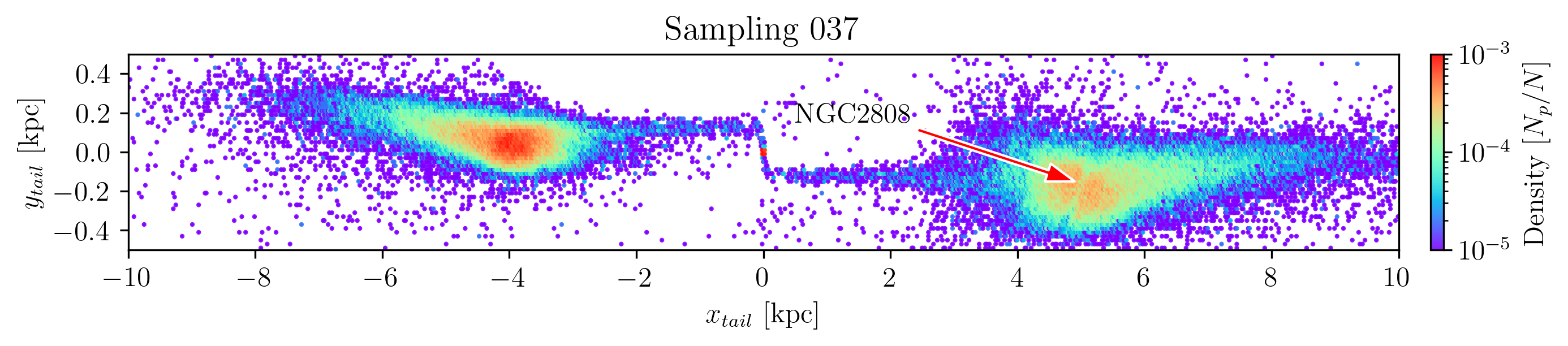}
      \includegraphics[width=\linewidth]{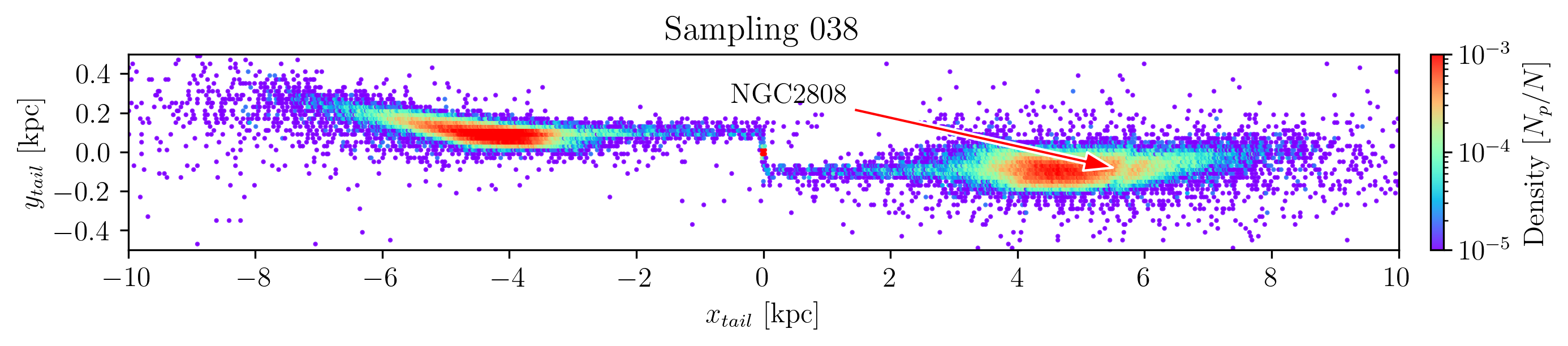}
      \includegraphics[width=\linewidth]{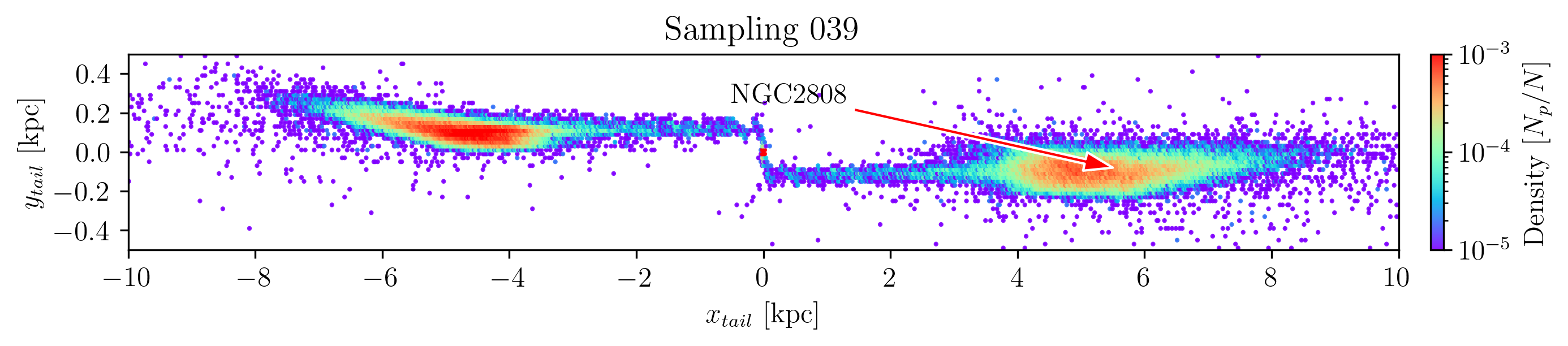}
      \includegraphics[width=\linewidth]{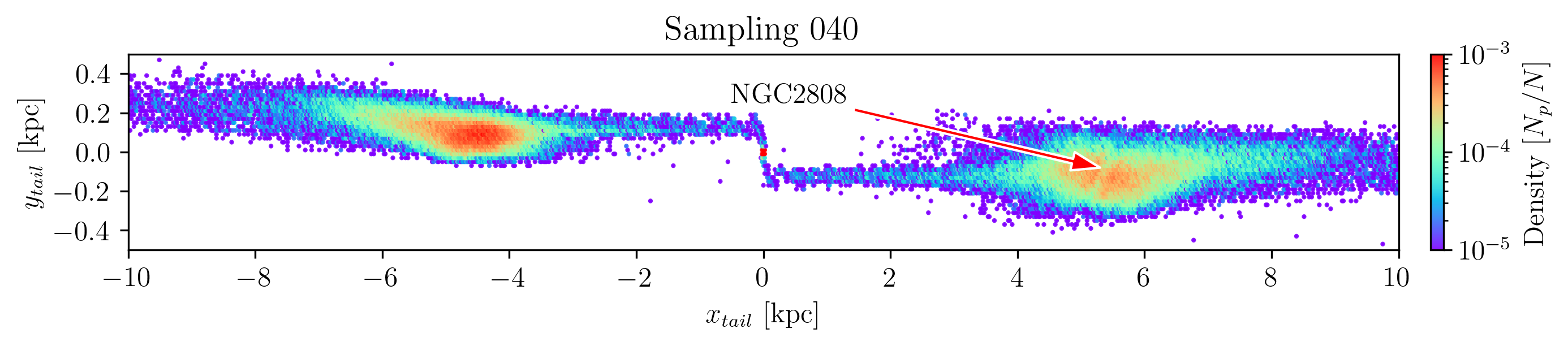}
      \includegraphics[width=\linewidth]{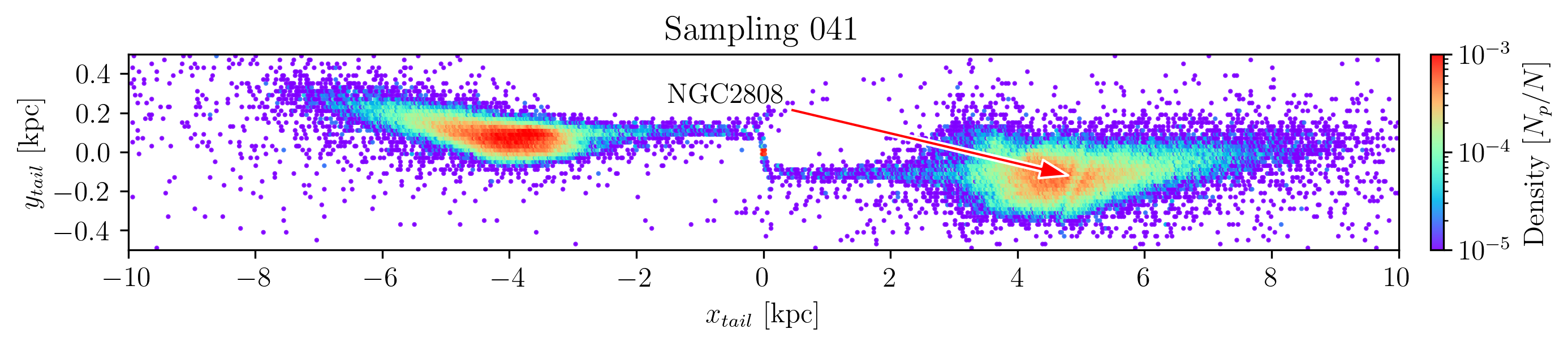}      
      \caption{Tenth gap gallery}
      \label{fig:gallery9}
    \end{figure*}

\begin{figure*}
      \centering
      \includegraphics[width=\linewidth]{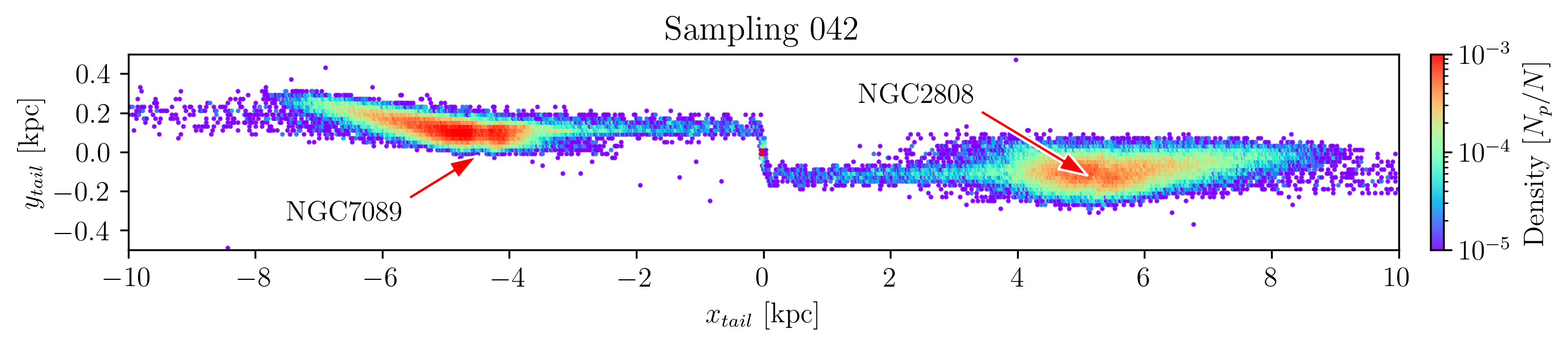}
      \includegraphics[width=\linewidth]{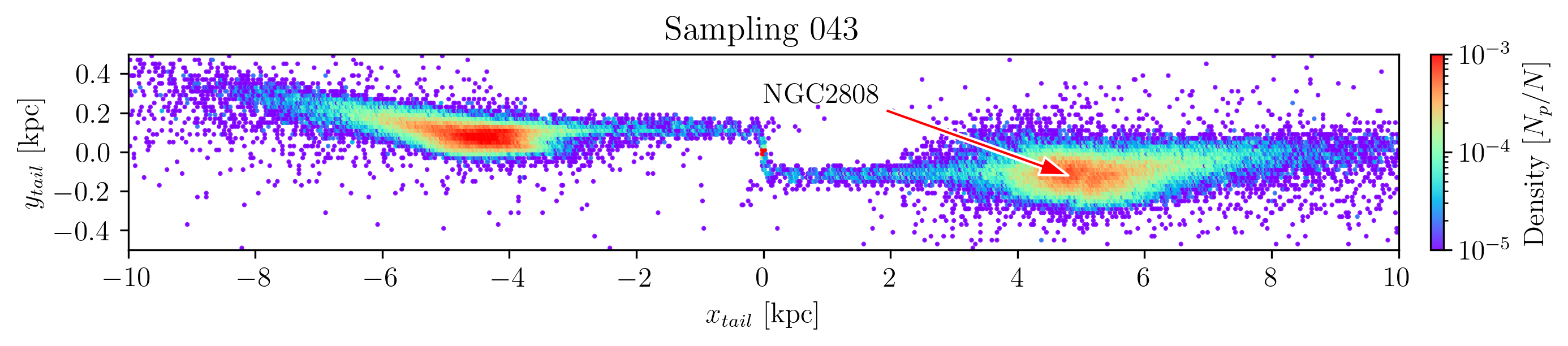}
      \includegraphics[width=\linewidth]{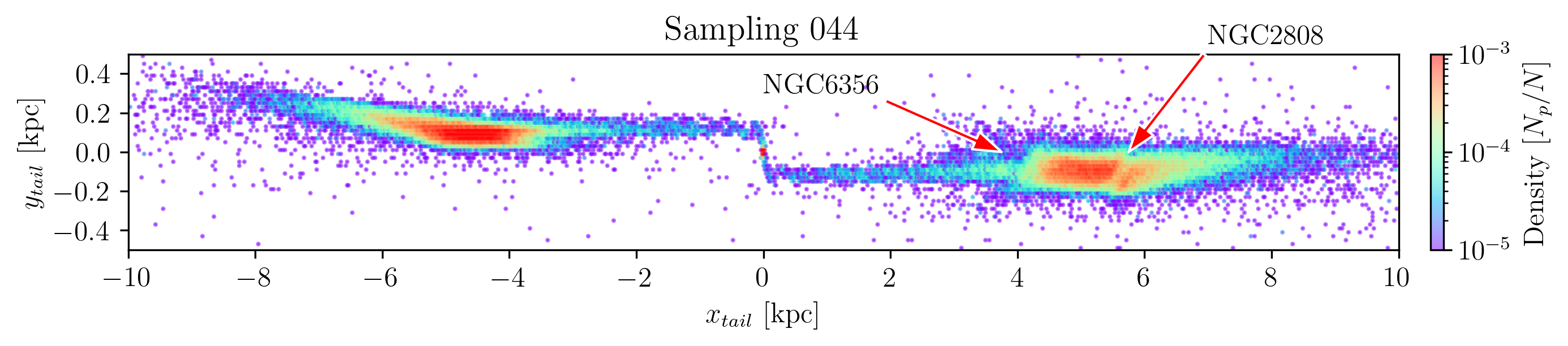}
      \includegraphics[width=\linewidth]{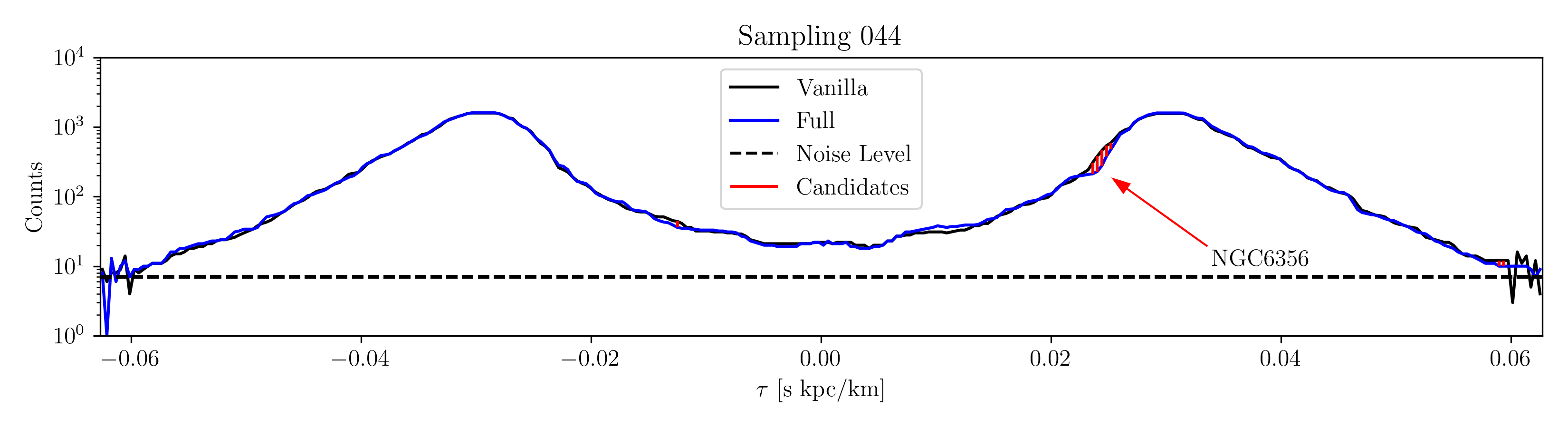}
      \includegraphics[width=\linewidth]{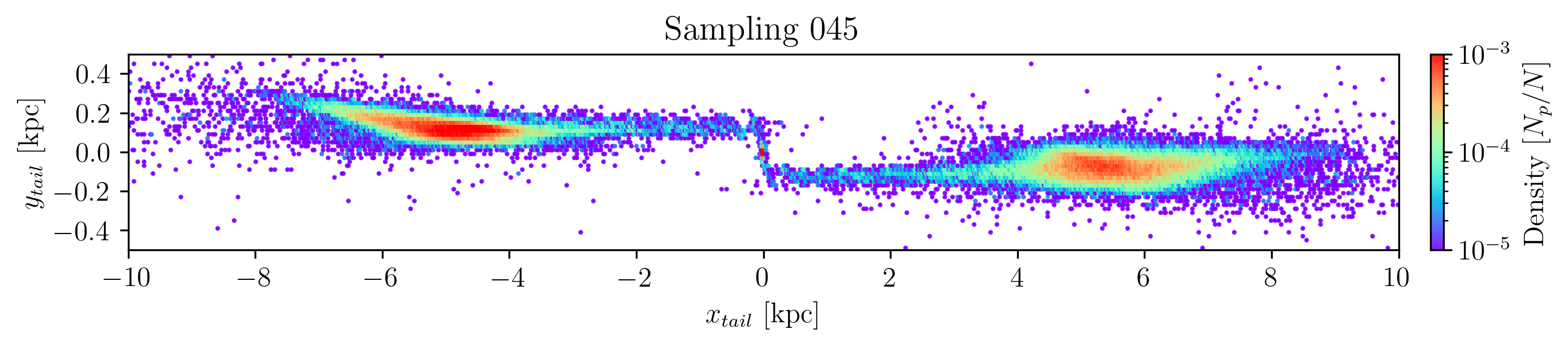}
      \caption{Eleventh gap gallery}
      \label{fig:gallery010}
    \end{figure*}

\begin{figure*}
      \centering
      \includegraphics[width=\linewidth]{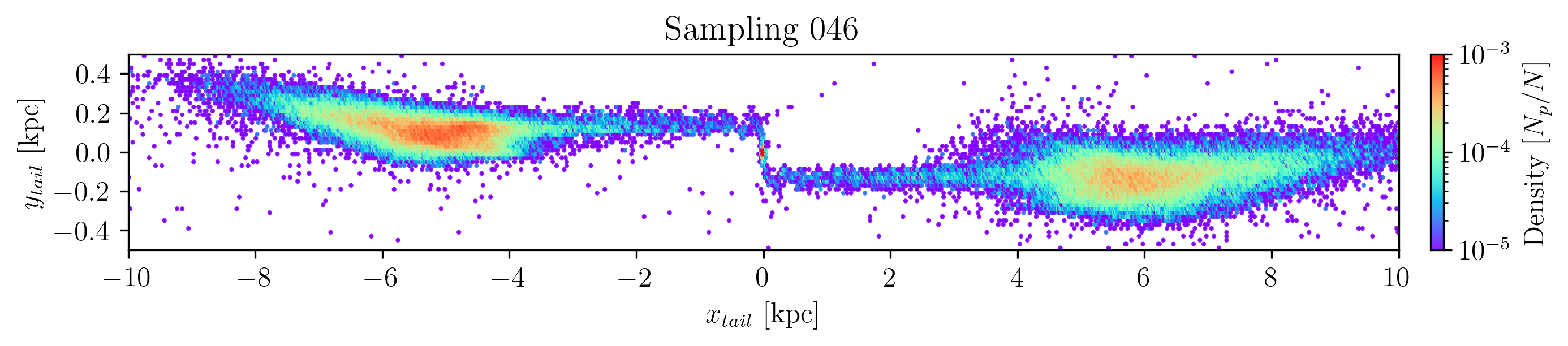}
      \includegraphics[width=\linewidth]{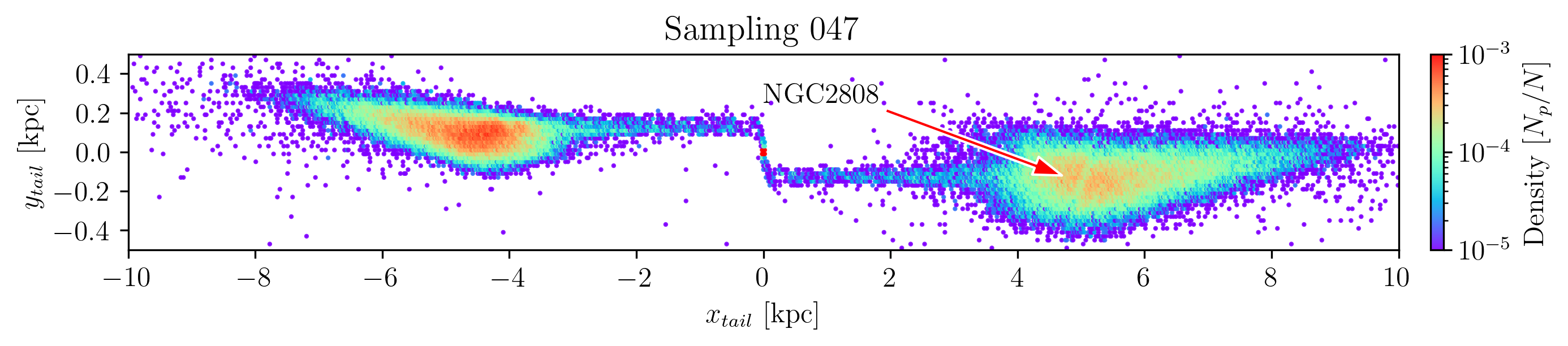}
      \includegraphics[width=\linewidth]{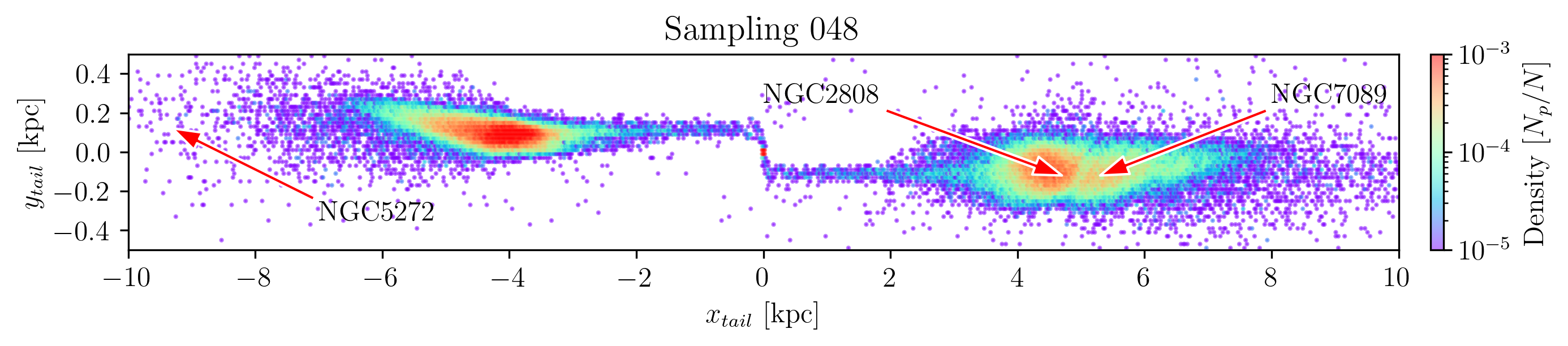}
      \includegraphics[width=\linewidth]{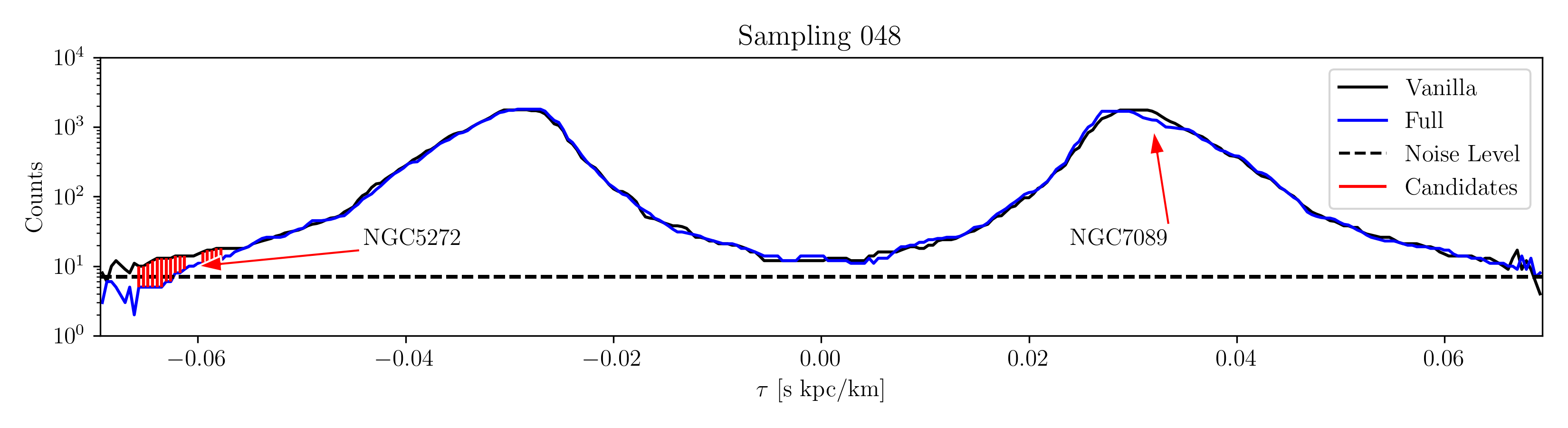}
      \includegraphics[width=\linewidth]{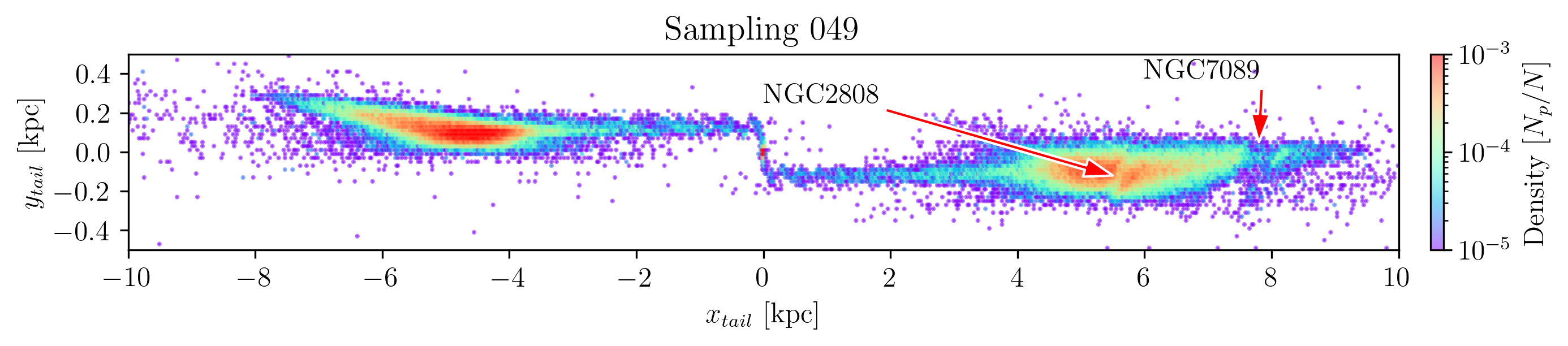}
      \caption{Twelfth gap gallery}
      \label{fig:gallery11}
    \end{figure*}

    Figures~\ref{fig:gallery0}-\ref{fig:gallery11} include all of the realizations of the \texttt{full} simulations of Palomar~5 in tail coordinates. If a gap is present, it is indicated by a red arrow with the responsible globular cluster. For the gaps that are not easily visible, we show the 1D profiles  showing the significant under-densities compared to the \texttt{reference} simulations. For instance, the second and last panels of Fig.~\ref{fig:gallery0}. However, maginalizing the 2D distribution to 1D can erase the signal of oblique gaps. Thus, not all gaps that are visible in the 2D space are flagged in the density profiles. For example, in \texttt{Sampling 000}, the gaps from NGC~2808 and NGC~5634 are not flagged present in the marginalized density profile.

\end{appendix}

\end{document}